\documentclass[11pt, lettersize]{article}
\usepackage{fullpage}
\usepackage{graphicx} 
\usepackage{comment}
\usepackage[margin=2.5cm]{geometry}
\usepackage{physics}
\usepackage{thmtools}
\usepackage{amsmath, amsfonts, amssymb}
\usepackage{amsthm}
\usepackage{tikz}
\usepackage{mathdots}
\usepackage{cancel}
\usepackage{color}
\usepackage{url}
\usepackage{hyperref}
\usepackage{cleveref}
\usepackage{siunitx}
\usepackage{array}
\usepackage{multirow}
\usepackage{gensymb}
\usepackage{tabularx}
\usepackage{extarrows}
\usepackage{booktabs}
\usepackage{subcaption}
\usepackage{todonotes}
\usetikzlibrary{fadings}
\usetikzlibrary{patterns}
\usetikzlibrary{shadows.blur}
\usetikzlibrary{shapes}
\usepackage[skins]{tcolorbox}
\usepackage{enumerate}
\usepackage{enumitem}
\usepackage{algorithm}
\usepackage{algpseudocode}

\usepackage{xcolor} 
\usepackage{tcolorbox}

\usepackage{amsthm}
\newcommand{\R}{\mathbb{R}}

\newcommand{\Q}{\mathbb{Q}}
\newcommand{\N}{\mathbb{N}}

\newcommand{\id}{\textsc{id}}

\newcommand{\cP}{\mathcal{P}}
\newcommand{\cC}{\mathcal{C}}

\newcommand{\CONGEST}{\textsf{CONGEST}}
\newcommand{\LOCAL}{\textsf{LOCAL}}
\newcommand{\biglo}[1]{\Tilde{\mathcal{O}}(#1)}

\newcommand{\bigo}[1]{\mathcal{O}(#1)}

\newcommand{\tw}{\texttt{tw}}
\newcommand{\nbsol}{\sharp\textrm{sol}}

\newtheorem{theorem}{Theorem}[section]

\newtheorem{definition}[theorem]{Definition}
\newtheorem{proposition}[theorem]{Proposition}
\newtheorem{lemma}[theorem]{Lemma}
\newtheorem{claim}[theorem]{Claim}

\newtheorem{corollary}[theorem]{Corollary}
\newtheorem{remark}[theorem]{Remark}
\newtheorem{fact}[theorem]{Fact}

\newtcolorbox{algorithmbox}[2][]{%
	enhanced,colback=white,colframe=black,coltitle=black,
	sharp corners,boxrule=0.4pt,
	fonttitle=\itshape,
	attach boxed title to top left={yshift=-0.3\baselineskip-0.4pt,xshift=2mm},
	boxed title style={tile,size=minimal,left=0.5mm,right=0.5mm,
		colback=white,before upper=\strut},
	title=#2,#1
}

\newcommand{\qdisjointpaths}{\textsc{$q$-disjoint-paths}}

\newcommand{\minproblem}{\textsc{min-problem}}
\newcommand{\maxproblem}{\textsc{max-problem}}
\newcommand{\ancestorsumproblem}{\textsc{ancestor-sum-problem}}
\newcommand{\descendantsumproblem}{\textsc{descendant-sum-problem}}
\newcommand{\rangeproblem}{\textsc{range-problem}}
\newcommand{\sumsubsetproblem}{\textsc{sum-subset-problem}}
\newcommand{\sumtreeproblem}{\textsc{sum-tree-problem}}

\newcommand{\prefixmaxproblem}{\textsc{prefix-maximum-problem}}
\newcommand{\suffixminproblem}{\textsc{suffix-minimum-problem}}

\newcommand{\detectancestor}{\textsc{ancestor-problem}}
\newcommand{\detectdescendant}{\textsc{descendant-problem}}

\newcommand{\contractingALG}{\textsc{contractingALG}}

\newcommand{\selected}{\textrm{Selected}}
\newcommand{\OPT}{\textrm{OPT}}
\newcommand{\COUNT}{\textrm{COUNT}}
\newcommand{\w}{\textrm{w}}

\definecolor{lightred}{rgb}{1,0.6,0.6}

\newtcolorbox{box1}[1][]{colback=gray!20, colframe=black, coltext=black, boxrule=1pt, sharp corners, #1}

\newcommand{\adj}{\mathsf{adj}}
\newcommand{\inc}{\mathsf{inc}}
\newcommand{\vertex}{\mathsf{vertex}}
\newcommand{\edge}{\mathsf{edge}}
\newcommand{\indep}{\mathsf{indep}}

\newcommand{\sj}{j^{*}}

\newcommand{\broadcast}{\textsc{broadcast}}

\newcommand{\markapath}{\textsc{mark-path}}

\newcommand{\reroottree}{\textsc{re-root-tree}}
\usepackage[margin=2.5cm]{geometry}
\usepackage{physics}
\usepackage{thmtools}
\usepackage{amsmath, amsfonts, amssymb}
\usepackage{amsthm}
\usepackage{tikz}
\usepackage{mathdots}
\usepackage{cancel}
\usepackage{color}
\usepackage{url}
\usepackage{hyperref}
\usepackage{cleveref}
\usepackage{siunitx}
\usepackage{array}
\usepackage{multirow}
\usepackage{gensymb}
\usepackage{tabularx}
\usepackage{extarrows}
\usepackage{booktabs}
\usepackage{subcaption}
\usepackage{todonotes}
\usetikzlibrary{fadings}
\usetikzlibrary{patterns}
\usetikzlibrary{shadows.blur}
\usetikzlibrary{shapes}
\usepackage[skins]{tcolorbox}
\usepackage{enumerate}
\usepackage{enumitem}
\usepackage{algorithm}
\usepackage{algpseudocode}

\usepackage{xcolor} 
\usepackage{tcolorbox}

\title{Distributed Treewidth Computation and Courcelle's Theorem in the \CONGEST{} Model}

\usepackage{authblk} 
\author[1,2]{Benjamin Jauregui}
\author[3]{Jason Li} 
\author[4]{Pedro Montealegre}
\author[5]{ Ioan Todinca}

\affil[1]{Universidad de Chile, Santiago, Chile}
\affil[2]{Université Paris Cité, Paris, France}
\affil[3]{Carnegie Mellon University, Pittsburgh, PA, USA}
\affil[4]{Universidad Adolfo Ib\'añez, Santiago, Chile}
\affil[5]{LIFO, Université d'Orléans and INSA Centre-Val de Loire, Orléans, France}

\date{}
\newcommand{\scq}[1]{\mathcal{SQ}(#1)}

\newcommand{\contractingproblem}{\textsc{contracting-problem}}

\begin{document}

\maketitle

\begin{abstract}
    Algorithmic meta-theorems, stating that graph properties expressible in some particular logic can be decided efficiently in graph classes having some specific structural properties, are now standard in sequential graph algorithms. One of the most classic examples is Courcelle's theorem: all properties expressible in Monadic Second-Order logic (MSO) are decidable in linear time in graphs of bounded treewidth.

    We provide here a distributed version of Courcelle's theorem, in the standard \CONGEST{} model for distributed computing: For any MSO formula $\varphi$ and any constant $k$, there is a \CONGEST\ algorithm that, given an input communication network $G$ of treewidth at most $k$ and of diameter $D$, decides if $G$ satisfies property $\varphi$ in $\biglo{D}$ rounds. Simple examples show that the dependency on $D$ is unavoidable. Also, if we drop the assumption of bounded treewidth, deciding MSO properties such as 3-colorability are known to require $\tilde{\Omega}(n^2)$ rounds in the \CONGEST{} model. Our results extend to optimization problems (e.g., computing a maximum size independent set, or a minimum dominating set) and counting (e.g. triangle counting). As usual, the $\tilde{O}$ notation hides polylogarithmic factors in $n$; here it also hides a constant factor depending on $k$ and on the MSO formula $\varphi$.

    We also give a distributed algorithm producing a linear approximation for treewidth: For any $k$, it decides that the treewidth of the input network $G$ is larger than $k$ or computes a tree decomposition of width $\bigo{k}$ and depth $\bigo{\log n}$, in $\biglo{k^{O(k)} D}$ rounds in \CONGEST.

    Our algorithms make use of the low-congestion shortcuts framework introduced by Ghaffari and Haeupler [SODA 2016], and our main technical tool is an $\biglo{k^4 D}$ algorithm for computing $(s,t)$-vertex separators of size at most $k+1$ in graphs of treewidth at most $k$.
\end{abstract}

\thispagestyle{empty}

\newpage
\setcounter{page}{1}
\section{Introduction}

\subsection{Algorithmic Meta-Theorems}

In the area of sequential graph algorithms, one of the most remarkable developments in recent decades is the emergence of \emph{algorithmic meta-theorems}. These theorems are typically of the following form~\cite{GroheK09}: \emph{Families of problems, often defined by logical conditions, can be efficiently solved on certain classes of graphs.} To make such a statement precise, one must specify the class of problems, the relevant logic fragment, the graph classes, and the notion of algorithmic efficiency. The key advantage of this framework is that it allows one to handle large families of problems on broad classes of graphs using a unified algorithmic strategy.

In \emph{first-order logic} (FO), graph properties are expressed as quantified logical formulas with variables ranging over the vertices of the graph. The language includes predicates for adjacency ($\adj$), equality ($=$), and standard boolean connectives ($\land$, $\lor$, $\lnot$). For example, the property ``$G$ is triangle-free'' can be written as:
\[
\varphi = \neg \exists x_1 \exists x_2 \exists x_3 \big( \adj(x_1,x_2) \land \adj(x_2,x_3) \land \adj(x_3,x_1) \big).
\]

Monadic second-order logic (MSO) extends FO by allowing quantification over sets of vertices and edges\footnote{This richer language, sometimes referred to as $\mathrm{MSO}_2$, is more expressive than $\mathrm{MSO}_1$, which only allows quantification over vertices and vertex sets~\cite{CoiurcelleE11}. For example, Hamiltonicity is definable in $\mathrm{MSO}_2$ but not in $\mathrm{MSO}_1$.}. Formally, the graph is viewed as a structure over a domain consisting of vertices and edges, equipped with a binary incidence predicate $\inc(v,e)$, which is true if vertex $v$ is incident to edge $e$; and unary predicates $\vertex(x)$ and $\edge(x)$ to distinguish vertices from edges. 
The adjacency predicate $\adj(u,v)$ can be encoded in $\mathrm{MSO}_2$ as:
\(
\adj(u,v) = \exists e \left( \inc(u,e) \land \inc(v,e) \right).
\)

In MSO, unary predicates represent subsets of the domain. For simplicity, we may write $x \in X$ instead of $X(x)$, and we use the implication symbol $\Rightarrow$ in formulas. For example, the following formula defines when a vertex set $S$ is \emph{independent}:
\[
\indep(S) = \forall x \forall y \left( (x \in S) \land (y \in S) \land \lnot(x = y) \Rightarrow \lnot \adj(x, y) \right),
\]

\noindent
thus one can express 3-colorability as the existence of three disjoint independent sets covering all vertices:
\[
\begin{aligned}
\varphi = \exists S_1 \exists S_2 \exists S_3 \big( & \indep(S_1) \land \indep(S_2) \land \indep(S_3) \land {} \\
& \forall x \left( \vertex(x) \Rightarrow (x \in S_1 \lor x \in S_2 \lor x \in S_3) \right) \big).
\end{aligned}
\]

Many natural graph problems, including 3-colorability, Hamiltonicity, and $H$-minor freeness, are expressible in MSO. One of the most celebrated algorithmic meta-theorems is due to Courcelle~\cite{Courcelle90}, who showed that any property expressible in MSO can be decided in \emph{linear time} on any class of graphs of \emph{bounded treewidth}. The \emph{treewidth} $\tw(G)$ of a graph $G$ informally measures how close $G$ is to a tree. A graph has treewidth at most $k$ if it admits a \emph{tree-decomposition} with bags (subsets of vertices) of size at most $k+1$, such that: (1) each vertex and edge of $G$ is covered by at least one bag, and (2) for each vertex, the set of bags containing it forms a connected subtree. 

By Courcelle's Theorem, properties like 3-colorability or Hamiltonicity can be decided in linear time for graphs of bounded treewidth.  Furthermore, the theorem extends to \emph{optimization and counting variants}. Fix an MSO formula $\varphi(X)$, where $X$ is a set of vertices or edges, and let $\mathcal{G}$ be a class of bounded-treewidth graphs. Then, given a weighted graph $G \in \mathcal{G}$ (with polynomial weights), one can compute in linear time the maximum weight set $S$ such that $G$ satisfies $\varphi(S)$. Consequently, problems such as computing a maximum independent set, minimum dominating set, feedback vertex set, or minimum spanning tree can be solved efficiently on these classes. A similar results holds for counting the solution sets.

These algorithms rely on the input graph being equipped with a tree-decomposition of small width. Therefore, an important preprocessing step is computing such a decomposition. For constant $k$, an optimal-width tree-decomposition can be found in linear time~\cite{Bodlaender96}. A simpler $\mathcal{O}(n \log n)$-time algorithm with approximation factor $4k+3$ and logarithmic depth was given by Reed~\cite{Reed92}. Once a decomposition is available, the evaluation of $\varphi$ (or optimization/counting over $\varphi(X)$) is done by dynamic programming over the tree structure~\cite{Courcelle90,BoPaTo92}.

\subsection{Distributed Computing and Complexity Barriers}

Our aim is to propose a distributed version of Courcelle's theorem. In distributed computation, one must deal with two main difficulties: the \emph{dilation} (or non-locality) and the \emph{congestion} of the input network.

Dilation refers to the fact that, for many problems, nodes that are far apart in the network must exchange information, therefore solving the problem requires $\Omega(D)$ communication rounds, where $D$ is the diameter of the network. More formally, consider the standard \LOCAL\ model for distributed computation~\cite{Linial92}, in which the communication network is represented by a connected undirected graph $G$ with $n$ nodes. The nodes have distinct identifiers represented over $\bigo{\log n}$ bits and exchange messages with their neighbors in synchronous rounds. At each round, each node performs computations based on its identifier, its local input, and the messages received up to that round. Nodes are seen as Turing machines with no constraints in time or space complexity; the computational resources at nodes are not taken into consideration. Distributed algorithms aim to collectively perform a task in a minimum number of rounds. A particular class of problems are \emph{decision} problems on the communication network $G$, where the goal is to decide whether $G$ satisfies some Boolean graph property $\cP$ (e.g., “$G$ is 3-colorable” or “$G$ is Hamiltonian”). In a decision algorithm for property $\cP$, every node must output \emph{accept} or \emph{reject}, such that the property is true on $G$ if and only if all vertices accept.

As observed in~\cite{FFMRT24}, deciding simple FO properties on the input network, such as “$G$ has at most two vertices of degree 3”, requires $\Omega(D)$ rounds in the \LOCAL\ model, even if $G$ is a tree of maximum degree~3.

Congestion arises when the communication network restricts the bandwidth of each communication link. It is well captured by the \CONGEST\ model~\cite{Peleg00}, a variant of the \LOCAL\ model in which each message is limited to $\bigo{\log n}$ bits per round.

While every decision problem on $n$-vertex graphs can be solved in $\mathcal{O}(n^2)$ rounds in the \CONGEST\ model (and in $\mathcal{O}(D)$ rounds in the \LOCAL\ model for graphs of diameter $D$), many problems require far fewer rounds, and some simple ones already exhibit nontrivial lower bounds. Even FO properties are \emph{hard} to decide in the \CONGEST\ model~\cite{Peleg00}, even on graphs of constant diameter. For example, distinguishing whether $G$ has diameter 2 or 3 requires $\tilde{\Omega}(n)$ rounds in \CONGEST~\cite{Censor-HillelPP20}. Another interesting and well-studied example is the minimum spanning tree problem, which requires $\tilde{\Theta}(D + \sqrt{n})$ rounds in \CONGEST~\cite{PelegR00}. Furthermore, properties expressible in MSO, such as $3$-colorability, require $\Omega(n^2 /\log^2 n)$ rounds in \CONGEST~\cite{abboud2021smaller}. Nevertheless, when we restrict ourselves to well-structured graph classes, the general limits of tractability in these models remain unclear.

Recently, there has been growing interest in extending algorithmic meta-theorems from the sequential setting to distributed models. In the context of the $\CONGEST$ model, only a few general results are currently known. One of them is a constant-round algorithm for evaluating MSO properties on graphs of bounded treedepth~\cite{FFMRT24}. Graphs of treedepth $k$ form a subclass of graphs of treewidth $k$, and when restricted to this class, FO logic is as expressive as MSO logic.

Another result shows that FO properties can be decided in $\bigo{D + \log n}$ rounds on graphs of bounded expansion~\cite{FFGMRT25}. This result applies to a broader class of graphs, including planar graphs and all graphs of bounded treewidth, but it is restricted to properties expressible in FO logic only.

On the other hand, in their quest to improve the round complexity of minimum spanning trees and other problems for particular graph classes, Ghaffari and Haeupler~\cite{ghaffari2016distributed-2} introduced the notion of \emph{low-congestion shortcuts of quality $\biglo{D}$}\footnote{Throughout the paper, we use the notation $\tilde{\mathcal{O}}(f(n))$ to hide polylogarithmic factors, i.e., $\tilde{\mathcal{O}}(f(n)) = \mathcal{O}(f(n) \cdot \operatorname{polylog}(n))$.}. Very informally (see Definition~\ref{def:shortcuts} and Subsection~\ref{subsec:shortcuts} for formal statements), in a class of graphs having low-congestion shortcuts of quality $\biglo{D}$, one can solve in parallel ``part-wise aggregation problems'' on disjoint subsets $V_1, V_2, \dots, V_q$ inducing connected subgraphs, in $\biglo{D}$ rounds. This means that each vertex has its own $\bigo{\log n}$-bit input, and each set $V_i$ learns an aggregate function such as $\min$, $\max$, or a more sophisticated associative and commutative operation on $\bigo{\log n}$ bits, over the inputs of the entire set $V_i$. It was shown in~\cite{haeupler2018round} that graphs of constant treewidth admit low-congestion shortcuts of quality $\biglo{D}$, along with tools for solving part-wise aggregation problems within the same round complexity. As we will see in the next sections, these tools are key to overcoming the challenges mentioned above.

\subsection{Our Results}
We provide in this article a~\CONGEST\  version of Courcelle's theorem:

\begin{restatable}{theorem}{mainresult}\label{thm:main}
    Fix a positive integer $k$ and an MSO formula $\varphi$. There is a  \CONGEST{} algorithm that, given an input communication network $G$ with $n$ vertices, of treewidth at most $k$ and of diameter $D$, decides whether $G$ satisfies property $\varphi$ in $\biglo{D}$ rounds. 
\end{restatable}

Due to the afore mentioned lower bounds of $\Omega(D)$, the $\biglo{D}$ complexity is arguably optimal, the only improvable factor being the polylogarithmic dependency on $n$. 

Our result also applies to optimization and counting variants. Given an MSO formula $\varphi(X)$ over a set of vertices or edges, and assuming that vertices of $G$ and edges are endowed with polynomial weights, we can compute the maximum weight set $S$ such that $\varphi(S)$ holds on $S$. We can also count the number of satisfying sets when this number is polynomial in $n$, all within the same running time of $\mathcal{O}(D)$ rounds. Therefore, to cite some possible applications, we can compute in $\biglo{D}$ rounds a minimum spanning tree, or a minimum dominating set, or count the number of triangles of the input graph. We note that the constant hidden in the $\biglo{\cdot}$ notation is a function $f$ depending on $k$ and the formula $\varphi$, and this function can be very high, yet $f(k,\varphi)$ is constant for any fixed $k$ and $\varphi$. Our algorithm works under the promise that the input graph has treewidth at most $k$, but it can easily made \emph{robust}, in the sense that it takes as input an arbitrary $G$, and decides if $G$ satisfies $\varphi$ or correctly outputs that $\tw(G) > k$.

For our algorithm of \Cref{thm:main} we need two ingredients. The first ingredient is to efficiently compute a tree decomposition of width $\bigo{k}$. Throughout this paper, we say that a tree decomposition is computed in a distributed manner if each vertex $v$ in $G$ determines the minimum depth $\rho(v)$ such that it belongs to a bag $B_{\rho(v)}$ of this depth in the decomposition tree, and $v$ learns the identifiers and the subgraphs of $G$ induced by all bags along the path from $B_{\rho(v)}$ to the root bag $B_r$. \Cref{thm:twdistrib} states that a near-optimal tree decomposition can be efficiently computed in the \CONGEST{} model, for fixed $k$. Moreover the decomposition tree will be of depth $\bigo{\log n}$:

\begin{restatable}{theorem}{twdistrib}\label{thm:twdistrib}
    Fix a positive integer $k$. There is a  \CONGEST{} algorithm that, given an input communication network $G$ with $n$ vertices of diameter $D$, computes a tree decomposition of width $\bigo{k}$  and of depth $\bigo{\log n}$ of $G$, or correctly outputs that $\tw(G)>k$, in $\biglo{D}$ rounds.
    The dependency on $k$ hidden in the $\biglo{\cdot}$ is $k^{\bigo{k}}$.
\end{restatable}

In our second ingredient, we simulate the sequential algorithm for decision,  optimization and counting problems expressible en MSO logic, which proceeds by bottom-up dynamic programming on the decomposition tree. Thanks to the logarithmic depth of the decomposition tree $T$, we will have $\bigo{\log n}$ phases, and at each phase $j$ we will update in parallel the dynamic programming tables of all nodes $u$ of depth $j$ of $T$. For the dynamic programming we use an alternative proof of Courcelle's theorem due to Borie, Parker and Tovey~\cite{BoPaTo92} in order to encode partial solutions at each node of the decomposition tree. The encoding only requires constant space for decision problems, and $O(\log n)$ space for optimization and other variants. We use low-congestion shortcuts in order to compute in parallel the required information at all nodes situated at the same depth in the decomposition tree, thanks to some good connectivity properties of the decomposition. 

Hence, most of our work will be devoted to proving \Cref{thm:twdistrib}. The computation of the approximate tree decomposition of logarithmic depth is inspired by a parallel algorithm for treewidth approximation in the PRAM model due to Lagergren~\cite{Lagergren96} (which, in turn, is based on the approach of Reed~\cite{Reed92}), and proceeds by recursively splitting the graph into smaller parts through small, balanced separators.
Informally, we start with an empty root bag and then we recursively split $V$. In general, we treat instances $(C,N(C))$ where $C$ is a connected subgraph, and alternate between finding vertex separators $S$ in $G[C]$ of size $\leq k+1$ that split $C$ into balanced parts, and separators that split $N(C)$ into balanced parts. Then we create a new bag $S \cup N(C)$ attached to the parent bag, and recurse on components $C'$ of $G[C]-S$ and their neighborhoods.

The connection between treewidth and vertex separators is thus central to our approach. In order to carry out this strategy in the distributed setting, one of the main tools, and the most technical component of our work, is an efficient procedure for computing small vertex separators in the \CONGEST\ model. Given two vertices $s, t$ in the input graph $G$ and a constant $q$, the algorithm attempts to find an $(s,t)$-vertex separator of size at most $q$, if it exists, in $\biglo{q^2 \cdot k \cdot D}$ rounds in \CONGEST.

\begin{restatable}{theorem}{congseparator}\label{thm:congseparator}
There is a deterministic \CONGEST\ algorithm that, given $G=(V,E)$ of treewidth at most $k$, two vertices $s,t \in V$ and a parameter $q$, computes in $\biglo{ q^3\cdot k \cdot D}$  rounds $q+1$ vertex-disjoint paths from $s$ to $t$ in $G$, or a set $S$ of at most $q$ vertices such that $S$ is an $(s,t)$ separator set.

\end{restatable}

The algorithm of \Cref{thm:congseparator} is based on the standard reduction of the minimum $(s,t)$-vertex separator problem to a max-flow/min-cut problem in an auxiliary, oriented graph, where each vertex of the initial graph is duplicated. Then we use the Ford-Fulkerson approach to find augmenting paths, and the crux is to find such a path in  $\biglo{D}$ rounds. 

\subsection{Structure of the Paper}

We start the paper with a general overview of our techniques in~\Cref{sec:overview}, emphasizing the novel ideas. Before delving into the technical content, we discuss other related work in~\Cref{s:relatedwork}. Full details begin with the Preliminaries section (\Cref{sec:prelim}), followed by the proof of~\Cref{thm:congseparator} for computing small vertex separators (\Cref{sec:separator}), then~\Cref{thm:twdistrib} on approximating treewidth (\Cref{se:treewidth}), and eventually~\Cref{thm:main} on deciding MSO properties and its generalizations to optimization and counting problems (\Cref{se:MSO}). We conclude with open questions in~\Cref{sec:conclusion}.

\section{Overview of our techniques}\label{sec:overview}

This section is a gentle overview of our techniques, before getting into the technical details.

\subsection{Low congestion shortcuts and part-wise aggregation Problems}

For our distributed algorithms we make an intensive use of \emph{low congestion shortcuts}, introduced by Ghaffari and Haeupler~\cite{ghaffari2016distributed-2}, and their applications to \emph{part-wise aggregation problems}, that we briefly recall here (see~\Cref{subsec:shortcuts} for a complete description).

Let $G=(V,E)$ be a graph of diameter $D$, and consider a partition $\mathcal{P} = \{V_1, \dots, V_k\}$ of its vertex set such that each induced subgraph $G[V_i]$ is connected. Our goal is to efficiently compute some on functions all graphs $G[V_i]$, depending only on information stored at the vertices $V_i$. Typically, we aim to do this in $\tilde{D}$ rounds on all $G[V_i]$ in parallel, which is challenging since the diameter of these induced subgraphs might be much larger than $D$.
A collection of subsets of edges $H_1,\dots, H_k$ of $G$ is called a \emph{$(c,d)$-low-congestion shortcut} for partition $\mathcal{P}$ if: (1) The diameter of $G[V_i]+H_i$ is at most $d$ for each $i \in [k]$, and (2) each edge of $G$ appears in at most $c$ different sets $H_i$. The quality of this low-congestion shortcut is $c+d$. The \emph{shortcut quality $\scq{G}$} of $G$ is the minimum value $q$ such that any partition $\mathcal{P}$ of $V(G)$, admits a low-congestion shortcut of quality $q$.

Haeupler, Hershkowitz, and Wajc~\cite{haeupler2018round} showed that, for any input partition, graphs of diameter $D$ and treewidth at most $k$ have shortcut quality $\bigo{k \cdot D}$, and that such shortcuts can be constructed in $\biglo{k \cdot D}$ rounds in the \CONGEST\ model. Note that the algorithm does not require any knowledge of a tree decomposition of $G$, only the promise that its treewidth is at most~$k$.

Thanks to these tools, several tasks, sometimes referred to as part-wise aggregation problems, can be performed in $\biglo{k \cdot D}$ rounds in \CONGEST{} on graphs of treewidth at most $k$, for any partition $\mathcal{P} = \{V_1, \dots, V_k\}$ of the vertex set of the input network. Among these tasks, we highlight the following operations, all of which are executed independently on each subgraph $G[V_i]$:
\begin{enumerate}
    \item Given to every vertex $v$ an input $x_v$ of $\bigo{\log n}$ bits and a commutative and associative binary operator $\bigoplus$ over sets of size $poly(n)$, all vertices in $V_i$ learn the value $\bigoplus_{v\in V_i} x_v$.
    
    \item For each $G[V_i]$, compute a spanning tree $T_i$  and orient it towards a prescribed root $r_i \in V_i$.

    \item Building on the previous items, each vertex $u \in V_i$ can learn the value $\bigoplus_{v} x_v$ aggregated over all descendants $v$ of $u$ in $T_i$, or over all its ancestors.
\end{enumerate}

Another important application of low-congestion shortcuts is the possibility of identifying, given a subset $S \subseteq V$, the connected components of the subgraph $G[S]$ by simulating a version of Borůvka's algorithm~\cite{ghaffari2016distributed-2}.

\subsection{Computation of disjoint paths and Small Vertex Separators}

As previously explained, the computation of the tree decomposition relies on the ability to find small vertex separators. To approach this task, we begin with a simplified setting: given two vertices $s$ and $t$ in $G$, and a constant $k$, we aim to compute, if it exists, a set of at most $k$ vertices that separates $s$ from~$t$.

The algorithm of \Cref{thm:congseparator} is based on the standard reduction of the minimum $(s,t)$-vertex separator problem to a max-flow/min-cut problem. Roughly speaking, to find an $(s,t)$ separator of size at most $k$, it suffices to show that there are at most $k$ vertex-disjoint $(s,t)$-paths, and then construct a separator by selecting one vertex from each of these paths. 

Obviously, a single $(s,t)$-path can be found in $\biglo{D}$ rounds by running a breadth-first search starting from $s$. To check whether more disjoint paths exist, we aim to implement a distributed version of the Ford--Fulkerson algorithm for maximum flow~\cite{fordfulkerson1956}. This is the most technical part of the paper, introducing several new ideas, and corresponds to the proof of~\Cref{thm:congseparator}, presented in~\Cref{sec:separator}. We now provide an informal overview of the key techniques.

For this purpose we consider as part of the input a set $\mathcal{P}_q$ of vertex-disjoint $(s,t)$-paths, for some $q\leq k$. The quadruple $(G,s,t,\mathcal{P}_q)$ is called a \emph{canonical instance}. Note that the input is provided in a distributed manner: all vertices know $q$ and the identifier of vertices $s$ and $t$, and a vertex on a path $P_j \in \mathcal{P}_q$ knows that it is part of the $j$th path of the instance. 
Using part-wise aggregation techniques we can number the vertices of each path “from left to right”, i.e., following the path from $s$ to $t$, in $\biglo{k\cdot D}$ rounds. Given a canonical instance $(G,s,t,\mathcal{P}_q)$ we need to decide if there exist $q+1$ vertex-disjoint $(s,t)$-paths and mark them, or to compute an $(s,t)$ vertex separator $S$ of size $q$ and communicate it to all vertices. Obviously, the base case $q=1$ can be computed in $\biglo{D}$ rounds by a BFS starting from $s$.

\subsubsection{The residual graph and the bridge graph}

Following the terminology of the Ford--Fulkerson approach, we define the \emph{classical residual graph} $G'_{res}$ associated with the input canonical instance. This is a directed graph obtained by duplicating each vertex~$u$ of $G$ into two vertices, $u_{in}$ and $u_{out}$, and adding two arcs, $(u_{in}, u_{out})$ and $(u_{out}, u_{in})$, for every vertex $u$. For each edge $\{u, v\}$ in $G$, we also add the arcs $(u_{out}, v_{in})$ and $(v_{out}, u_{in})$. 

To account for the current flow paths, we remove from $G'_{res}$ the arc $((u_i)_{out}, (u_{i+1})_{in})$ for each pair of consecutive vertices $u_i, u_{i+1}$ on the same path $P \in \mathcal{P}_q$, where the paths are numbered from $s$  to $t$ (see \Cref{fig:residualgraph}~(a) and (b)). We actually work with a slightly modified version of the classical residual graph, which we simply refer to as the \emph{residual graph} and denote by $G_{res}$. This graph is obtained from $G'_{res}$ by contracting, for each vertex $u$ of $G$ that is not an internal vertex of any path in $\mathcal{P}_q$, the pair $u_{in}, u_{out}$ back into the original vertex $u$ (see \Cref{fig:residualgraph}(c)). An $(s,t)$-directed path in $G_{res}$ is called an \emph{augmenting path}. A key property of $G_{res}$ is that it contains an augmenting path if and only if $G$ has $q+1$ vertex-disjoint $(s,t)$-paths. Therefore, our goal is to find an augmenting path in $G_{res}$.

Another central ingredient in our construction is the \emph{bridge graph}, whose definition is very similar to the one in~\cite{nose}, but which we use in a fundamentally different way to detect augmenting paths. A bridge $B$ of $G$ is defined as a connected component of the subgraph obtained by removing all vertices that lie on the paths of $\mathcal{P}_q$. Observe that each bridge induces a strongly connected component in $G_{res}$. The intuition behind the importance of bridges is that any augmenting path must start in a bridge adjacent to $s$, end in a bridge adjacent to $t$, and, in between, alternate between visiting other bridges and traversing some path in $\mathcal{P}_q$ in the “reverse” direction (that is, following the orientation of the arcs in $G_{res}$). 
 
We define the \emph{bridge graph} $G_B$ (\Cref{def:bridgegraph}), whose vertex set consists of the bridges associated with a canonical instance $(G, s, t, \mathcal{P}_q)$. This graph satisfies several properties that will be instrumental in detecting augmenting paths. The definition of the edge set of $G_B$ is somewhat technical, and we first need to introduce some terminology.

For convenience, given a path $P \in \mathcal{P}_q$ of the canonical instance (that is, a path in the original graph~$G$), we denote by $\vec{P}$ the oriented $(t,s)$-path in $G_{res}$ obtained by duplicating the internal vertices of $P$, and $\vec{\mathcal{P}}_q$ the set of such directed paths. Furthermore, we say that $\vec{P}$ goes \emph{from right to left}. That is, along each such oriented path in $G_{res}$, predecessors appear \emph{to the right} of their successors. In other words, vertices that appear later in the path (those with higher positions) are placed further to the right.

We then denote by $B_1, B_2, \dots$ the bridges induced by the canonical instance $(G, s, t, \mathcal{P}_q)$. Each bridge $B$ is assigned an identifier $\id(B)$ as the maximum identifier among the vertices in $B$.  We say that a bridge $B_i$ \emph{sees} a path $P_j$ if there exists a vertex in $B_i$ adjacent to some vertex of $P_j$. 
 Let us fix $j \in [q]$. We define the set of arcs $D_j$ as the set of all pairs $(B_x, B_y)$ satisfying the following conditions:
\begin{itemize}
    \item[(a)] It is possible to travel from $B_x$ to $B_y$ in $G_{res}$ using some path other than $\vec{P}_j$. 
    \item[(b)] It is possible to travel \emph{back} from $B_y$ to $B_x$ using the path $\vec{P}_j$, but in a more specific way (\Cref{def:bridgegraph})
    
    \item[(c)] We select the bridge $B_x$ that is maximal according to a certain ordering, where ties are broken using bridge identifiers.
\end{itemize}
Note that condition (a) and (b) imply that when $(B_x, B_y)\in D_j$ then the nodes in $B_x \cup B_y$ belong to the same strongly connected component in $G_{res}$ (that contains as well some nodes from the paths in $\mathcal{P}_q$). Note also that due to condition (c), each node has at most one outgoing neighbor in $D_j$, for each $j \in [q]$. 

Finally, we define the edge set of $G_B$ as the union of all the sets $D_j$ over $j \in [q]$. The strict form in which these sets are defined is a carefully designed compromise: it yields structural properties that are useful for identifying augmenting paths, while also making the sets algorithmically tractable in a distributed setting. 

A bridge $B$ is called \emph{$s$-reachable} (resp. \emph{$t$-reachable}) if $s$ (resp. $t$) has a neighbor in $B$ in graph $G$.  A path from a $s$-reachable bridge in $G_B$ to a $t$-reachable bridge is called an \emph{bridge augmenting path}. We show in \Cref{lem:18} that there is an augmenting path in $G_{res}$ if and only if there is a bridge augmenting path in~$G_B$.

An important consideration is that the existing algorithms for computing low-congestion shortcuts (and, consequently, for part-wise aggregation problems) are only applicable to undirected graphs. Therefore, to apply these techniques, we will often need to work with the undirected version of the directed graphs under consideration. For a directed graph $D$, we denote $\overline{D}$ its undirected version. The use of undirected versions of directed graphs introduces  significant challenges.   For instance, given a canonical instance $(G, s, t, \mathcal{P}_q)$, we have that $\overline{G}_{res}$ contains $q$ $(s,t)$-paths corresponding to the paths in $\mathcal{P}_q$, none of which qualify as augmenting paths in the directed graph~$G_{res}$. 

\subsubsection{Representing the bridge graph and detecting an augmenting path}

We firstly compute, using low-congestion shortcuts and part-wise aggregation problems, a distributed representation of the bridge graph $G_B$. In this representation, every vertex in bridge $B$ learns the identifier $\id(B)$ of the bridge (the maximum $\id$ of a vertex in $B$) and, for each $j \in [q]$, the identifier of the parent bridge in $G_B[D_j]$, if it exists.  This is done in $\biglo{k \cdot D}$ rounds (\Cref{lemma:bridgegraph}).

We now outline the general strategy for computing a bridge augmenting path. First, the $s$-reachable bridges are easy to identify. Indeed, any vertex in a bridge $B$ that sees $s$ can broadcast this information to all nodes in $B$ by performing a part-wise aggregation over the bridge. The same holds, of course, for the $t$-reachable bridges.

To identify the remaining nodes that belong to a bridge augmenting path, the idea is to search incrementally within the subgraphs $G_B[D_j]$. We first look for a path in $G_B[D_1]$. If no such path is found, we extend the search to $G_B[D_1 \cup D_2]$, then to $G_B[D_1 \cup D_2 \cup D_3]$, and so on. We stop at the smallest index $j^*$ such that a path exists in $G_B[D_1 \cup \dots \cup D_{j^*}]$, or conclude that no augmenting path exists. Eventually, all vertices are informed of the value $j^*$, or of the fact that no such path exists (\Cref{lemma:detectingapath}). 

For this step, it is convenient to define an additional graph. Fix an index $j \in [q]$, and let $\mathcal{C}$ be a connected component of $\overline{G_B[D_j]}$. We define (\Cref{def:conngr}) the \emph{connectivity graph} $H^j_{\mathcal{C}}$ associated with $\mathcal{C}$ and $j$ as a subgraph of $G_{res}$ as follows. First, for each bridge $B$, we compute a spanning tree $T_B$. Then, for every pair of bridges $B_x, B_y$ in $\mathcal{C}$ such that $(B_x, B_y)$ is an edge of $D_j$, we identify the subset of nodes of $\vec{P}_i$ that connect $B_x$ and $B_y$, where $\vec{P}_i$ is the path referenced in point (a) of the definition of $D_j$. We include these nodes and their edges in $H^j_{\mathcal{C}}$.  Next, for each bridge $B$ with outgoing edges in $D_j$, we choose a vertex $u \in B$ that sees some path $\vec{P}_i$ (as above), and orient the edges of $T_B$ toward $u$, designating it as the root. If $B$ has no outgoing edges in $D_j$, we choose an arbitrary vertex as the root and orient $T_B$ toward it. In other words, $H^j_{\mathcal{C}}$ is a directed tree that includes, as subgraphs, the directed spanning trees of each bridge in $\mathcal{C}$, together with selected subpaths of the paths in $\vec{\mathcal{P}}_q \setminus \{\vec{P}_j\}$.

The connectivity graphs have several useful properties. First (see \Cref{clm:22}), for a fixed $j \in [q]$ and any pair of components $\mathcal{C}_1$ and $\mathcal{C}_2$, the graphs $H^j_{\mathcal{C}_1}$ and $H^j_{\mathcal{C}_2}$ are vertex-disjoint. This implies that we can perform part-wise aggregation problems on them independently. Second, for each $j \in [q]$, if a component $\mathcal{C}$ of $G_B[D_j]$ contains both an $s$-reachable and a $t$-reachable bridge, then $G_{res}$ contains an augmenting path that uses only vertices from $H^j_{\mathcal{C}}$. Third, we show in \Cref{claim:compuofhcj} that all the graphs $H^j_{\mathcal{C}}$, for components $\mathcal{C}$ of $G_B[D_j]$, can be constructed in $\biglo{k \cdot D}$ rounds. Finally, we prove in \Cref{lemma:uvpathingres} that for any pair of vertices $u \in B_x$ and $v \in B_y$ such that $B_x$ and $B_y$ both belong to $H^j_{\mathcal{C}}$, we can compute in $\biglo{D}$ rounds a directed $(u,v)$-path in $G_{res}$.

Thanks to the properties of our connectivity graphs, a bridge augmenting path in $G_B[D_1]$ can be identified in $\biglo{D}$ rounds. To extend this to $G_B[D_1 \cup D_2]$, we define an extended connectivity graph $H^{1,2}_{\mathcal{C}}$ for each component $\mathcal{C}$ of $\overline{G_B[D_1 \cup D_2]}$; this graph retains the useful property that if $\mathcal{C}$ contains both an $s$-reachable and a $t$-reachable bridge, then a bridge augmenting path exists. However, $H^{1,2}_{\mathcal{C}}$ does not guarantee that the corresponding graphs for different components are vertex-disjoint, which obstructs the use of part-wise aggregation techniques in parallel.

The solution is to define a layered graph $G^j$ consisting of $j$ copies of $G_{res}$, where each vertex has one copy per layer and intra-vertex cliques connect the layers. We construct a subgraph ${\mathcal H}^j$ that captures relevant connectivity: bridge trees appear in layer 1, while edges reflecting path-based interactions are added in their corresponding layers. We show that if two bridges belong to the same component of $G_B[D_1 \cup \dots \cup D_j]$, then their copies are connected in ${\mathcal H}^j$ (\Cref{claim:101}). Moreover, any $T$ round algorithm in $G^j$ can be simulated in $\biglo{T}$ rounds in $G$, allowing us to perform all necessary part-wise aggregations in $\biglo{j \cdot k \cdot D}$ rounds. This layered construction underlies our algorithm for detecting augmenting paths, which runs in $\biglo{q^2 \cdot k \cdot D}$ rounds (\Cref{lemma:detectingapath}).

After deciding whether an augmenting path exists in $G_{res}$, we proceed to construct it if it does, or otherwise extract a small separator. When a component of $G_B[D_1 \cup \dots \cup D_{j^*}]$ contains both an $s$-reachable and a $t$-reachable bridge, we construct a directed walk from $s$ to $t$ using the connectivity structure of ${\mathcal H}^{j^*}$. This walk (which could pass several times though the same vertex or arc) is transformed into a valid augmenting path in $G_{res}$ by removing cycles via part-wise aggregation. The resulting path allows us to increment the number of disjoint $(s,t)$-paths from $q$ to $q+1$ (\Cref{lemma:getthepath}).
If no such path exists, we construct an $(s,t)$-separator of size $q$ by selecting one internal vertex per path in the canonical instance. These vertices correspond to the rightmost neighbors of the $s$-reachable bridges on each path, computed in parallel through part-wise aggregation.

To compute a good tree decomposition, we need to apply a generalized version of \Cref{thm:congseparator} in parallel. In this setting, we are given several subgraphs of the input graph, each induced by a connected vertex set and equipped with its own sets \(A_\ell\), \(B_\ell\), and \(X_\ell\). For each such subgraph, the goal is to either find \(k+2\) vertex-disjoint paths from \(A_\ell\) to \(B_\ell\) avoiding \(X_\ell\), or to compute a small separator between \(A_\ell\) and \(B_\ell\). We first show how to handle a single such instance in \Cref{lemma:abseparator}, and then extend this to process all instances in parallel in \Cref{lemma:mabseparator}, while ensuring low congestion even when vertices participate in multiple roles. Full details are given in \Cref{ss:multiplesubsets}.

\subsection{Distributed Construction of a Good Tree Decomposition}\label{sss:distribtw}

Our next goal is to compute a good tree decomposition of the input graph $G$, as in~\Cref{thm:twdistrib}: given a constant $k$, we aim to compute a tree decomposition of $G$ of width $\bigo{k}$ (more specifically $\leq 7k+4$), or to decide that $\tw(G)>k$, in $\biglo{k ^{O(k)}\cdot D}$ rounds in \CONGEST. For reasons that will appear clearly in the next subsection, we also want this decomposition to be of depth $\bigo{\log n}$.

For this purpose we heavily rely on~\Cref{thm:congseparator} and its generalization described in~\Cref{ss:multiplesubsets}, and plug it into the treewidth approximation algorithm of Reed~\cite{Reed92}, also used for an efficient parallel (PRAM) approximation of treewidth by Lagergren~\cite{Lagergren96}. 
Since we mainly implement in \CONGEST{} that approach based on the repeated computation of small separators in disjoint connected subgraphs of $G$, we do not detail it here, please refer to~\Cref{se:treewidth} for a full description. 

Nevertheless we  introduce some notation and describe how the computed tree decomposition is distributedly stored among the vertices of $G$. 

Let $T=(I,F)$ be the decomposition tree produced by the algorithm, we will see it as a rooted tree and call $r$ the root node. We denote by $B_u$ the bag associated to node $u$ of $T$ (recall that bags are vertex subsets of $G$, see also~\Cref{de:treedec}), by $p(u)$ the parent of $u$ in $T$, by $T_u$ the subtree of $T$ rooted at $u$, and by $V_u$ the union of all bags of $T_u$.
We shall see (\Cref{le:twtech}) that our tree decomposition satisfies the following properties:
   \begin{itemize}
        \item At its root $r$, $B_r = \emptyset$ and $V_r = V$.
        \item For each node $u$ of $T$ different from the root, the graph $G[V_u \setminus B_{p(u)}]$ is connected.
        \item For each node $u$ different from the root, there is at least one vertex in $B_u \setminus B_{p(u)}$.
        \item Each vertex $x$ of $G$ knows its depth in the tree $T$, i.e., the minimum depth of a node $u$ such that $x \in B_u$. Moreover, $x$ knows the sequence of bags $B_u, B_{p(u)},\dots, B_r$ on the path from $u$ to the root $r$, and the graphs induced in $G$ by each of these bags.
    \end{itemize}

\subsection{Decision and optimization for MSO properties}

Let us give some insights into the proof of~\Cref{thm:main}, showing that any MSO property $\varphi$ can be decided in $\biglo{D}$ rounds in the \CONGEST{} model, on graphs of treewidth at most $k$. Recall that the $\biglo{\cdot}$ here hides a function depending on $k$ and $\varphi$. Full proofs and extensions to counting and optimization are provided in~\Cref{se:MSO}.

We propose a distributed implementation of Courcelle's theorem~\cite{Courcelle90,BoPaTo92}. Informally, the theorem provides a dynamic programming algorithm on the tree decomposition of width $k$ (or $\bigo{k}$ in our case) to test formula $\varphi$, with tables of constant size computed at every node $u$ of the decomposition, from bottom to top. Let us make it more formal. 

Graphs of treewidth less than $w$ can be represented as \emph{$w$-terminal recursive graphs}, which are graphs equipped with a distinguished set of at most $w$ terminals, corresponding to the root bag of a tree decomposition. These graphs are constructed from small base graphs (with at most $w$ vertices) via a sequence of controlled gluing operations \emph{$\circ_f$}  that identify terminals according to a matrix specification. This recursive structure captures exactly the class of graphs of treewidth $<w$~\cite{Bodlaender98arb}, and any tree decomposition of such width can be translated into a gluing-based representation (see Figure~\ref{fig:TwG} and \Cref{subsec:twterm}). Also note that for any constant $w$ there is a constant number of possible gluing matrices $f$, and a constant number of a base graphs.

The crux of Courcelle's theorem is that, for any MSO property $\varphi$ and any constant $w$, there exists a set $\cC$ of \emph{homomorphism classes} such that we can associate to each $w$-terminal recursive graph $G$ a class $h(G) \in \cC$, with the following properties:
\begin{enumerate}
    \item For any two $w$-terminal recursive graphs $G_1,G_2$, if $h(G_1) = h(G_2)$ then either both graphs satisfy $\varphi$, or none does; the classes corresponding to graphs satisfying the property are called \emph{accepting};
    \item The homomorphism class of a base graph can be computed in constant time, given the graph;
    \item For any glued graph $G = \circ_f(G_1,G_2)$, the class $c=h(G_1)$ depends only on classes $c_1 = h(G_1)$, $C_2 = h(G_2)$ and the gluing matrix $f$; moreover the function $\odot_f$ such that $c = \odot_f(c_1,c_2)$ can be computed in constant time.
\end{enumerate}

To give a quick intuition, think of the property expressing the 3-colorability of $G$. We then take as homomorphism class of a $w$-terminal recursive graph the set of all 3-partitions of its set of terminals that can be extended into a 3-coloring of the whole graph, with one color for each part. Note that a partition of the set $W$ of terminals is actually stored as a partition of the set $\{1,2,\dots,|W|\}$. Since $|W| \leq w$, the set $\cC$ of classes is a constant (depending on $w$), and it is a matter of exercice to see that the above statements are true for this particular property. Observe that the accepting classes are all classes but the empty set. 

To decide MSO properties on graphs of bounded treewidth in the \CONGEST{} model, we rely on computing the homomorphism class $h(G_u)$ of each subgraph $G[V_u]$ defined by the subtree rooted at node $u$ in a tree decomposition $T$. We assume that such a decomposition of width $w = O(k)$ and logarithmic depth has been computed implicitly: each vertex knows the shallowest bag containing it, the bags along the path to the root, and the induced subgraphs of those bags.

The computation proceeds bottom-up in $\bigo{\log n}$ phases, one per depth level. At each phase $j$, we process all nodes $u$ of depth $j$. Each node $u$ has a bag $B_u$ and a base graph $G_u^{base} = G[B_u]$ whose homomorphism class can be computed locally by a vertex  $\ell(u) \in B_u \setminus B_{p(u)}$ called the leader of $u$. For each child $v_i$ of $u$, we construct the $w$-terminal recursive graph $G_u^{=i}$ induced by $V_{v_i} \cup B_u$, and observe that it results from a gluing of $G_{v_i}$ and $G_u^{base}$ along their respective bags. Since $v_i$ knows both bags, it can locally compute the  gluing matrix and apply the operation $\odot$ to obtain $h(G_u^{=i})$.

Crucially, the entire graph $G_u$ can be recovered by successively gluing the graphs $G_u^{=1}, \dots, G_u^{=p}$ using a fixed operation $\odot_{f(B_u,B_u)}$ that preserves terminal structure. This operation is commutative, associative, and admits $h(G_u^{base})$ as neutral element. Therefore, $h(G_u)$ is the aggregation $\bigoplus_{v \in C_u} x_v$, where $C_u = V_u \setminus B_{p(u)}$ and each $x_v$ is either $h(G_u^{=i})$ (if $v$ is a child leader) or $h(G_u^{base})$ otherwise.

This aggregation is implemented via part-wise aggregation over connected sets $C_u$ in $\biglo{D}$ rounds, and the full computation of all $h(G_u)$ completes in $\biglo{D \cdot \log n}$ rounds. Once these classes are known, evaluating an MSO sentence reduces to checking whether the class of the root graph satisfies the formula, completing the decision procedure.

The techniques extend to formulas with free variables $\varphi(X)$, in which we may have several homomorphism classes at a same node $u$ of the tree decomposition. With some more information stored for each class we can deal with optimization, i.e. computing a maximum weighH $S$ satisfying $\varphi(S)$, and counting the number of solutions (\Cref{thm:optcount}). Optimization also requires a top-down phase on the tree decomposition in order to mark an optimal set $S$, for which we need to open a bit the “black box” of low congestion shortcuts and part-wise aggregation problems.

\section{Other Related Work}\label{s:relatedwork}

Regarding tree decompositions in the \CONGEST\ model, Izumi et al. \cite{izumi2022fully} present a randomized algorithm that, given a graph of treewidth $k$ and diameter $D$, constructs a tree decomposition whose width is $\bigo{{k^2} \log n}$. However, this decomposition is not convenient for applying Courcelle’s theorem: the theorem’s running time grows super-exponentially with the decomposition’s width, and since the construction of \cite{izumi2022fully} has a $\log n$ factor in the width of the decomposition, a factor of $n$ would appear in the round complexity.

The interest for algorithmic meta-theorems in distributed algorithm started in the context of \emph{distributed certification}. A \emph{certification scheme} for a boolean graph property $\cP$ consists in a pair \emph{prover} and \emph{verifier}. The prover is a centralized oracle, having full knowledge of the input graph $G$,  computationally unbounded, but non-trustable. The prover assigns a boolean string $c(v)$ to each vertex $v$, called the \emph{certificate} of $v$. Then the verifier is a distributed 1-round algorithm, in which every vertices exchanges its id and certificate with its neighbors, then performs a computation and outputs \emph{accept} or \emph{reject}. The certification scheme for property $\cP$ must ensure two conditions: (1) if property $\cP$ is true on $G$, then there is a set of certificates such that, after the verification on this set, all vertices accept, and (2) if property $\cP$ is false, then for any set of certificates, after verification at least one vertex rejects. Note that certification is trivial if each vertex $v$ receives as certificate the whole graph $G$. Therefore, the aim of certification for property $\cP$ is to provide a certification scheme with small certificates, typically of size $\bigo{\log n}$, or polylogarithmic in $n$. In this context, Feuilloley, Bousquet and Pierron proposed a certification scheme for MSO properties on graphs of bounded treedepth, with certificates of size $\bigo{\log n}$~\cite{FeuilloleyBP22}. Graphs of bounded treedepth are graphs with paths of bounded length, and are also of bounded treewidth. These results were extended to certification of MSO properties on graphs on bounded treewidth, with certificates of size $\bigo{\log^2 n}$~\cite{FMRT24}, and very recently the size of the certificate was reduced to~$\bigo{\log n}$~\cite{CookEM25}. More certification theorems can be found in~\cite{BatersinaC25,FFGMRT25,FraigniaudM0RT23}, for different fragments of logic and different graph classes.
Nevertheless, a certification scheme can be considered as significantly weaker than a decision \CONGEST\ algorithm for the same property, since it relies on a centralized oracle.

To the best of our knowledge, the only algorithmic metatheorems in $\CONGEST$ are a constant-time algorithm for MSO problems on graphs of bounded treedepth~\cite{FFMRT24}, and an $\bigo{D+\log n}$-rounds algorithm for FO, for graphs of bounded expansion~\cite{FFGMRT25}. 

Graphs of treedepth $k$ are also of treewidth $k$, and moreover they are of diameter $D \leq 2^k$. In other terms, our result implies the one of~\cite{FFMRT24}, with the slight drawback that here we have a polylogarithmic factor in $n$ hidden in $\biglo{D}$, while~\cite{FFMRT24} guarantees $\bigo{D + \log n}$ rounds. The treedepth-based algorithm  is also much simpler since in bounded treedepth graphs any depth-first search can be use to construct a “treedepth decomposition", which is a particular case of tree decomposition. Nevertheless, we shall use some ideas of~\cite{FFMRT24} and of~\cite{FMRT24} for the implementation of the dynamic programming.

The result of~\cite{FFGMRT25} uses a totally different approach and is incomparable to ours. Indeed they consider the broader classes of graphs of \emph{bounded expansion}, which include planar graphs and graphs of bounded treewidth, but on the more restricted FO logic. 

In recent years, there has been significant interest in reducing the round complexity of fundamental graph problems, such as Minimum Spanning Tree (MST), DFS trees, maximum flow, and single-source shortest paths, within graph classes that admit optimal low-congestion shortcuts \cite{abd2025distributed, ghaffari2016distributed-2,ghaffari:2017,parter}. Now that the low-congestion shortcut framework has fully deterministic constructions \cite{haeupler2018round}, there is growing attention on developing deterministic algorithms for these problems as well \cite{jauregui2025deterministic}. Although low-congestion shortcuts have been widely applied in the distributed setting, existing applications have focused primarily on specific and individual problems, leaving open the potential for broader and more systematic application as the application shown in this paper.

\section{Preliminaries}\label{sec:prelim}

In this work we consider the message-passing \CONGEST\ model \cite{peleg2000distributed}. In this model, the communication network is represented by a connected undirected graph $G = (V, E)$ with $n = |V|$ nodes. Computation proceeds in synchronous rounds. In each round, every node can perform arbitrary local computation and send a message of size $\mathcal{O}(\log n)$ bits to each of its neighbors. Nodes have unique identifiers of size $\mathcal{O}(\log n)$, and initially, each node knows only its own \id.

Given any integer $n\in\N$, we denote as $[n]$ to the set $\{1,\dots,n\}$. Given any two real values $m,M$, we denote as $[n,M]$ to the real interval $\{x\in\R\colon m\leq x\leq M\}$.


Given an undirected graph $G=(V,E)$, every edge $e\in E$ with extremities $u,v\in V$ will be denoted as $\{u,v\}$. We denote as $N_G(v)$ the set of neighbors of a vertex $v\in V$ and given a subset of vertices $W\subseteq V$, we denote by $N_G(W)$ the set $\cup_{v\in W}N_G(v) \backslash W$. A undirected {\it walk} in $G$ is a sequence of vertices $P=v_1,\dots, v_k$ such that $\{v_i,v_{i+1}\}\in E(G)$ for all $i\in[k-1]$. An undirected {\it path} in $G$ is a walk $P=v_1,\dots, v_k$ such that $v_i\neq v_j$ for all $i\neq j$. 
The {\it length} of $P$ corresponds to the number of vertices and it is denoted as $|P|$.

If $G=(V,E)$ is a directed graph, every arc $e\in E$ from vertex $u$ to vertex $v$ will be denoted as $(u,v)$. Given a vertex $v\in V$, we denote as $N_G^+(v)$ the set of outgoing neighbors of $v$ and as $N_G^-(v)$ the set of ingoing neighbors to $v$, i.e., $N_G^+(v)=\{u\in V\colon (v,u)\in V\}$ and $N_G^-(v)=\{u\in V\colon (u,v)\in V\}$, respectively. Analogously, given a subset of vertices $W\subseteq V$, we denote by $N_G^+(W)$ and $N_G^-(W)$ the sets $\cup_{v\in W}N_G^+(v)\backslash W$ and $\cup_{v\in W}N_G^-(v)\backslash W$, respectively. A directed {\it walk} in $G$ is a sequence of vertices $P=v_1,\dots, v_k$ such that $(v_i,v_{i+1})\in E(G)$ for all $i\in[k-1]$. An directed {\it path} in $G$ is a walk $P=v_1,\dots, v_k$ such that $v_i\neq v_j$ for all $i\neq j$. If $G=(V,E)$ is a directed graph, then we denote by $\overline{G}$ the undirected graph of $G$, this means that $V(\overline{G})=V(G)$ and $E(\overline{G}) = \{\{u,v\}\colon (u,v)\in E(G)\lor (v,u)\in E(G)\}$.

In all the above definitions, we may omit the subindex $G$ if it is clear from the context. 
When dealing with some input graph $G$, we denote by $n$ its number of vertices. We often use the $\biglo{}$ notation, hiding a polylogarithmic factor in $n$.

Given two vertices $s$ and $t$ of $G$, a vertex subset $S$ such that $s$ and $t$ are in different connected components of $G-S$ is called an $(s,t)$ separator set. By Menger's theorem, for any positive integer $q$, there is either an $(s,t)$ separator in $G$, or there exist $q+1$ vertex-disjoint paths from $s$ to $t$. analogously, given two subsets $A,B\subseteq V$, a $(A,B)$-separator set $S\subseteq V$ is a subset such that the vertices of $A$ and $B$ are different components of $G-S$.

\subsection{Tree decomposition and Courcelle's Theorem}

The notions of tree decompositions and treewidth are due to Robertson and Seymour~\cite{RoSe84}:
\begin{definition}[treewidth]\label{de:treedec} 
A \emph{tree decomposition} of a graph $G = (V,E)$ is a pair $(T, B)$  where ${T=(I,F)}$ is a tree, 
and $B=\{B_i, i \in I\}$ is a collection of subsets of vertices of $G$, called \emph{bags}, such that the following conditions hold:
\begin{itemize}
\item For every vertex of $G$, there exists some bag containing this vertex;
\item For every edge $e$ of $G$ there is some bag containing both endpoints of $e$;
\item For every vertex $v \in V$, the set $\{i \in I : v \in B_i\}$ of bags containing $v$ forms a connected subgraph of~$T$. 
\end{itemize}
The \emph{width} of a tree decomposition is the maximum size of a bag, minus one. The \emph{treewidth} of a graph~$G$, denoted by~$\tw(G)$, is the smallest width of a tree decomposition of~$G$.

Throughout this work we will refer to an element of $V$ (a graph node) as {\it vertex} and to the elements of $I$ (a tree node) as {\it nodes}.

\end{definition}

Graphs of treewidth at most $k$ are sparse, in the sense that they have at most $kn$ edges. For any fixed $k$, deciding if an arbitrary graph has treewidth at most $k$ can be computed in $\bigo{n}$ time by an algorithm of Bodlaender~\cite{Bodlaender96}.

The monadic second-order logic was already defined in the introduction. Recall that the graph $G$ is given as a structure on the domain of vertices and edges, together with the incidence binary predicate $\inc(v,e)$ indicating whether vertex $v$ is adjacent to edge $e$, and the unary predicates $\edge$ and $\vertex$ indicating if an element $x$ of the domain is a vertex ($\vertex(x)$ is true) or an edge ($\edge(x)$ is true). An MSO formula uses variables of the domain denoted $x,y,\dots$, and set variables denoted $X,Y...$, corresponding to subsets of the domain. 
An MSO formula is constructed recursively according the following rules from variables and other formulae $\varphi, \psi$\dots:
\[
 x=y \mid x \in X \mid  \vertex(x) \mid \edge(x) \mid \inc(x,e) \mid \varphi\lor\psi \mid  \varphi\land\psi \mid \neg \varphi \mid \exists x \varphi \mid \forall x \varphi \mid \exists X \varphi \mid \forall X \varphi
\]

We are particularly interested in formulae $\varphi$ with no free variables (where all variables are quantified) and formulae $\psi(X)$ with one free set variable, typically corresponding to a vertex or edge subset of the graph.
We write $G \models \varphi$ if formula $\varphi$ is satisfied by graph $G$, and $G \models \psi(S)$ if formula $\psi(X)$ is true on $G$ when assigning set $S$ to variable $X$. 

There exist several proofs of Courcelle's theorem:

\begin{proposition}[Courcelle's theorem~\cite{Courcelle90, BoPaTo92}]\label{prop:Courcelle}
   Fix an integer constant $k \geq 0$, an MSO formula $\varphi$ with no free variables and an MSO formula $\psi(X)$ with one free set variable. Given as input a graph $G$ of treewidth at most $k$, one can: 
   \begin{enumerate}
       \item  decide if $G \models \varphi$ in $\bigo{n}$ time;
       \item compute a maximum weight set $S$ such that $G \models \psi(S)$ in $\bigo{n}$ time, or decide that no such set exists; for this problem, we assume that $G$ is given together with polynomial edges over vertices and edges;
       \item count the number of different sets $S$ such that $G \models \psi(S)$ in $\bigo{n \log(\nbsol(n))}$, where $\nbsol$ is an upper bound on the number of possible solutions in $n$-vertex graphs.
   \end{enumerate}
\end{proposition}

We will give more intuitions on this result, obtained by dynamic programming over the decomposition tree of $G$, in Section~\ref{se:MSO}.

\subsection{Low-Congestion Shortcuts}\label{subsec:shortcuts}

To make this work self-contained for readers unfamiliar with the state-of-the-art in low-congestion shortcuts, we include a brief summary of these notions and of the principal literature results used all throughout the paper.

\begin{definition}[\cite{ghaffari2016distributed-2}, Definition 1]\label{def:shortcuts}
    Given a graph $G=(V,E)$ with diameter $D$ and a partition $\mathcal{P} = \{V_1,...,V_k\}$ of the vertices, such that every induced subgraph $G[V_i]$ is connected, a collection of subset of edges $H_1,...,H_k$ is said to be a $(c,d)$-low-congestion shortcut for the partition $\mathcal{P}$ if
    \begin{enumerate}
        \item[(i)] The diameter of $G[V_i]+ H_i$ is at most $d$, for all $i \in [k]$.
        \item[(ii)] For all edges $e\in E$, $|\{i\in [k]\colon e\in H_i\}|\leq c$.
    \end{enumerate}

    The quality of a given $(c,d)$-low-congestion shortcut $SC = \{H_1,...,H_k\}$ of a graph $G$ it is defined as $q\left(\sc{SC}\right) = c + d$, where $D$ is the diameter of the graph.

    The \emph{quality} $\mathcal{SQ}(G)$ of a graph is defined as the minimum quality between all the low-congestion shortcuts of $G$, i.e., $\scq{G} = \min_{SC\in\mathcal{SC}_G}q\left(\sc{SC}\right)$, where $\mathcal{SC}_G$ is the collection of all low-congestion shortcuts over all partitions $\mathcal P$ of $G$.
\end{definition}

In a follow-up article, the authors of \cite{haeupler2018round} prove the following result.

\begin{proposition}[\cite{haeupler2018round}] Let $G=(V,E)$ be a graph with $\tw(G)~=~k$ and diameter $D$. Then $\mathcal{SQ}(G) = \bigo{k\cdot D}$.
    \label{prop:sqofboundedtrwwidth}
\end{proposition}

Since this seminal work by Ghaffari and Haupler, significant advances have been made in the study and applications of low-congestion shortcuts. In the following, we summarize the most relevant results that will be used in this work.

\begin{proposition}[\cite{haeupler2018round}, simplified]
\label{teo:deterministicshortcut}
There exists a distributed deterministic algorithm in the \CONGEST\ model such that, for any $n$- node graph $G=(V,E)$ with shortcut quality $\scq{G}$, computes a low congestion shortcut of $G$ with quality $\scq{G}$ in $\biglo{\scq{G}}$ rounds.
    
\end{proposition}

\begin{proposition}[\cite{haeupler2018round}]
\label{teo:mstcongest}

There exists a deterministic algorithm in the \CONGEST\ model such that, given any graph $G=(V,E)$ with shortcut quality $\scq{G}$ and a polynomially bounded weight function $\omega\colon E\to \N$ function, computes a MST of $G$ with respect to $\omega$ in $\biglo{\scq{G}}$ rounds. In particular, if $G$ has treewidth $\tw(G) = k$, then an MST is computed in $\biglo{k\cdot D}$ rounds.
\end{proposition}

A direct application of \Cref{teo:mstcongest} allows to compute spanning trees concurrently for a collection of subgraphs.

\begin{lemma}[\cite{jauregui2025deterministic}]\label{teo:kmstcongest}
    There exists a deterministic algorithm in the \CONGEST\ model which, given a graph $G=(V,E)$ with shortcut quality $\scq{G}$, and a family of pairwise-disjoint vertex subsets $\mathcal{V}=\{V_1,...,V_k\}$ of $V$, such that $G[V_i]$ is connected for each $i\in [k]$, computes a collection of trees $T_1,...,T_k\leq G$, such that $T_i$ is a spanning tree of $G[V_i]$ for each $i\in [k]$, in $\biglo{\scq{G}}$ rounds. In particular, if $\tw(G) \leq k$, then this collection of spanning tres can be computed in $\biglo{k\cdot D}$ rounds.
\end{lemma}

{\it Part-wise Aggregation Problems.} Using \Cref{teo:deterministicshortcut} directly, various problems can be solved in $\biglo{D}$ rounds in \CONGEST. However, most of these correspond to variations of broadcast problems, where the objective is to send a message to all other vertices or compute certain local functions that depend on the input from the rest of the vertices.

Subsequently we will denote by $\bigoplus$ a commutative and associative binary operator over sets of size polynomial in $n$.

\begin{definition}[\bf Part-wise aggregation problem]
    Consider a network graph $G = (V, E)$ and suppose that the vertices are partitioned into disjoint subsets $V_1 , V_2 , \dots, V_k$ such that the subgraph induced by each part is connected. In the part-wise aggregation problem, the input is a value $x_v\in \{0,1\}^{\bigo{\log n}}$ for each vertex $v \in V$, and all vertices know the commutative associative function $\bigoplus$. The output is that each vertex $u \in V_i$ learns the aggregate  $\bigoplus_{v\in V_i} x_v$ of the values held by vertices in its part $V_i$, e.g., $\Sigma_{ v\in V_i} x_v$, $\min_{v\in V_i} x_v$, or $\max_{v\in V_i} x_v$. 
\end{definition}

One of the main applications of low-congestion shortcuts, provided in~\cite{haeupler2018round} is the following.

\begin{proposition}[\cite{haeupler2018round}, rephrased]\label{broadcasts}
    There exists a deterministic algorithm in the \CONGEST\ model which, given a specific part-wise aggregation problem and a graph $G=(V,E)$ with shortcut-quality $Q$, solves the part-wise aggregation problem over $G$ in $\biglo{Q}$ rounds.
\end{proposition}

{\it Broadcasts and other problems over disjoint connected subsets of vertices}. An important application of part-wise aggregation functions, besides computing the value of the function itself, is enabling vertices to determine the \id\ of a specific vertex. For example, in $\min$ or $\max$ functions, vertices can identify the \id\ of the vertex with the input $x_v$ that minimizes/maximizes the function. We summarize some results in the following lemma.
\begin{proposition}[\cite{Ghaffari2022},\cite{jauregui2025deterministic}]\label{lemma:somebroadcasts}
    Given a graph $G=(V,E)$ with shortcut quality $\mathcal{SQ}(G)$, a partition $\mathcal{V} = \{V_1,....,V_k\}$ of $V$, $k$ spanning trees $T_1\dots, T_k$ with root $r_i\in V_i$ of each $V_i\in\mathcal{V}$, and a $\log n$-input string $x_v$ for each $v\in V$. There exists deterministic algorithms in the \CONGEST\ model that solves, in parallel, in each $V_i$, the following problems in $\biglo{\mathcal{SQ}(G)}$ rounds.

    \begin{enumerate}[label={\bf \arabic*.}]
        \item \minproblem\ and \maxproblem: all the vertices in $V_i$ learn the \id\ of a vertex $v$ such that $x_v = arg\min_{v\in V_i} x_v$ and $x_v = arg\max_{v\in V_i} x_v$, respectively.
        \item \sumsubsetproblem: All the vertices in $V_i$ learn the value $n_i = \left|V_i\right|$.
        \item \sumtreeproblem: All vertices $v\in V(T_i)$ learn the number of vertices $|V((T_i)_v)|$, where $(T_i)_v$ is the subtree of $T_i$ rooted in $v$.
        \item \rangeproblem: If the input $x_v$ represents real values, and $R_i =[m_i,M_i]_{\Q}\subseteq \Q$ is a range known by the vertices, all the vertices in $V_i$ learn the \id\ of an arbitrary vertex $v\in V_i$ such that $x_v\in R_i$, if any exists, or learn that such vertex does not exist.
        \item \detectancestor\ and \detectdescendant$\colon$ If in each $V_i$ there is a unique vertex $v_i'\in V_i$ known by all vertices in $V_i$, all vertices in $V_i$ can learn if $v_i'$ is ancestor or descendant of them in $T_i$, respectively.

        \item \broadcast$\colon$ In each $V_i$ $k=\bigo{1}$ vertices $v_1^i,\dots,v_k^i$ can broadcast their labels $x_{v_1^i},\dots, x_{v_k^i}$ to all the other vertices in $V_i$.

        \item \markapath$\colon$: Given two vertices $u_i,v_i\in V_i$ known by all the vertices in $V_i$, the unique path between $u_i$ and $v_i$ in $T_i$ can be computed.

        \item \reroottree$\colon$ Given a known vertex $v_i\in V_i$, the tree $T_i$ can be re-rooted in $v_i$, meaning that the parents and depth are updated now with $v_i$ as the root.
    \end{enumerate}

    In particular, given \Cref{prop:sqofboundedtrwwidth}, if the input graph has diameter $D$ and treewidth $k$, all the aforementioned problems can be solved in $\biglo{k\cdot D}$ rounds.
\end{proposition}

\begin{proposition}[\cite{Ghaffari2022}]\label{lemma:hltrees}

Given a graph \( G=(V,E) \) with shortcut-quality $\scq{G}$, a family of pairwise-disjoint vertex subsets $\mathcal{V}=\{V_1,...,V_k\}$ of $V$, such that $G[V_i]$ is connected for each $i\in[q]$, a spanning tree \( T_i \) of each \( G[V_i] \) rooted at \( r_i \in V_i\) for each $i\in[q]$, an associative and commutative binary operator \(\bigoplus\) known by all the vertices, and a $\bigo{\log n}$-bit private input \( x_v \)  in each vertex \( v \in V \), there exist two deterministic algorithms $\mathcal{A}_1$ and $\mathcal{A}_2$ that solve the following problems:

\begin{itemize}
    \item  $\mathcal{A}_1$ solves in $\biglo{\scq{G}}$ rounds the \ancestorsumproblem: Each vertex \( v \in V_i\) learns \( p_v \colon = \bigoplus_{\omega \in \text{anc}_{T_i}(v)} x_\omega \), where anc$_{T_i}(v)$ denotes the set of ancestors of \( v \) with respect to \( T_i \).
    \item $\mathcal{A}_2$ solves in $\biglo{\scq{G}}$ rounds the \descendantsumproblem: Each vertex \( v \) learns \( p_v \colon = \bigoplus_{\omega \in \text{desc}_{T_i}(v)} x_\omega \), where desc$_{T_i}(v)$ denotes the set of descendants of \( v \) with respect to \( T_i \).
\end{itemize}

\label{prop:firstpa}
\end{proposition}

{\it Part-wise aggregation problems in a canonical instance.} One of the main technical contributions of this paper is the computation of small separator sets, through vertex-disjoint paths. Given an undirected graph $G=(V,E)$, two vertices $s,t\in V$ and a collection $\mathcal{P}_q = \{P_1,...,P_q\}$  of $q\in\mathbb{N}$ vertex-disjoint paths between $s$ and $t$, we refer to the tuple as a {\it canonical instance}.

Let $(G,s,t,\mathcal{P}_q)$ be a canonical instance and $P_y\in\mathcal{P}_q$ be a path, the {\it internal vertices} of $P_y$ corresponds to the vertices in $P_y$ different from $s$ and $t$. We refer as a {\it canonical order} of the instance $(G,s,t,\mathcal{P}_q)$ to the following order: In each path $P_y\in\mathcal{P}_q$, the vertices are ordered $v_1 = s, v_2, ...., v_{|P_y|-1}, v_{|P_y|} = t$, where $v_{j-1}v_j\in E\left(P_y\right)$ for each $j\in\left[|P_y|+1\right]$. We may refer to this order of a path $P_y$ as a function $\pi_y\colon V(P_y)\to \left[|P_y|\right]$ with $\pi_y(v_i) = i$.

Part-wise aggregation problems also help to compute key problems efficiently, which we will use repeatedly in the following Sections. The solution to these problems are direct application of previous results but we include them for the sake of completeness.

\begin{lemma}\label{lemma:computingorders}
    There exists a deterministic algorithm in the \CONGEST\ model such that, given any canonical instance $(G,s,t,\mathcal{P}_q)$, computes the canonical order of the instance in $\biglo{\scq{G}}$ rounds.
\end{lemma}
\begin{proof}
    Using $\mathcal V = \{V(P_1)-\{s,t\},\dots, V(P_q)-\{s,t\}\}$ and each set $V_i=V(P_1)-\{s,t\}$ with weight function $\omega_i(e) = 0$ in the edges of the path $P_i$ and $\omega(e) = n^2$
    in the edges of $G[V_i]$ that do not belong to the path $P_i$, we can construct a spanning tree $T_i$ in each $V(P_i)-\{s,t\}$ that uses exactly the edges of the path by using \Cref{teo:kmstcongest}. Then we can use \reroottree\ to re-root each tree $T_i$ in the only neighbor of $s$ in $P_i$. Finally, using \ancestorsumproblem\ with the labels $x(v) = 1$ for each vertex in $V_i$ and addition as the function to be computed, each vertex $v\in V(P_i)-\{s,t\}$ can compute its position in the canonical order, as this position is exactly $\sum_{\omega\in anc_{T_i}(v)}x(\omega) +1$. Finally $s$ sets its position as $1$ in all the paths and $t$ ask to its neighbors $u_j$ in each $P_j\in\mathcal{P}_q$ their position and computes its position $\pi_j(t)$ in each path $P_j$ as $\pi_j(t)=\pi_j(u_j)+1$. All these computations take $\biglo{\scq{G}}$ rounds.
\end{proof}

\begin{lemma}\label{lemma:maxandminprefix}
    There exists a deterministic algorithm in the \CONGEST\ model such that, given a canonical instance $(G,s,t,\mathcal{P}_q)$ along with its canonical order and each vertex $v\in V(\mathcal{P}_q)$ have a $\bigo{\log n}$ bits message $x_v$, then the followings problems can solved in $\biglo{D}$ rounds

    \begin{itemize}
        \item \prefixmaxproblem$\colon$ In each path $P_y\in\mathcal{P}_q$, each the vertices $v_i\in V(P_y)$ learn the value $\displaystyle\max_{v_j\in V(P_y), j\leq i}x_{v_j}$.
        \item \suffixminproblem$\colon$ In each path $P_y\in\mathcal{P}_q$, each the vertices $v_i\in V(P_y)$ learn the value $\displaystyle\min_{v_j\in V(P_y), j\geq i}x_{v_j}$.
    \end{itemize}

\end{lemma}

\begin{proof}
    \prefixmaxproblem\ is solved in the same way that \Cref{lemma:computingorders} but using $\max$ as the function to be computed and the labels $x(v)$ given by the input. \suffixminproblem\ is solved analogously but using \descendantsumproblem.
\end{proof}

\begin{lemma}\label{lemm:detectbridges}
    Given a canonical instance $(G,s,t,\mathcal P_q)$, each connected component $B$ of $G-\cup_{P\in\mathcal P_q}P$ can be detected in $\scq{G}$ rounds.
\end{lemma}

\begin{proof}
    Each vertex knows whether it belongs to a path in $\mathcal{P}_q$ or not. Then, every edge $e$ between two vertices $u$ and $v$ that do not belong to any path in $\mathcal{P}_q$ sets its weight to $\omega(e) = 0$, while any edge $e$ incident to at least one vertex on a path in $\mathcal{P}_q$ sets its weight to $\omega(e) = n^2$ (or any other strictly positive value). Using this setup, the nodes can simulate the distributed Borůvka algorithm from \Cref{teo:mstcongest} until they find a minimum outgoing edge of weight greater than $0$, at which point they stop. After $\biglo{k\cdot D}$ rounds, the algorithm ends, and a spanning tree for each connected component of $G - \bigcup_{P \in \mathcal{P}_q} P$ has been computed and moreover, by construction of the algorithm in \Cref{teo:mstcongest}, all the nodes in the same tree learn the the same $\id$.
\end{proof}

\section{Computation of Disjoint Paths and Separator Sets}\label{sec:separator}

In this section,  we aim to prove the following theorem.

\congseparator*

To obtain the above result, consider first the following problem.  

\begin{algorithmbox}{\qdisjointpaths}

    {\bf Input:} A canonical instance $(G,s,t,\mathcal{P}_q)$.\\ 

    {\bf Output:} A distributed representation of $q+1$ disjoint paths between $s$ and $t$, if exist.  

\end{algorithmbox}

 Our algorithm to solve deterministically \qdisjointpaths\ is inspired by the parallel solution in the PRAM model of the same problem given by Khuller and Schieber~\cite{nose}. Our notion of bridge graph introduced in this section is very similar to theirs, but our construction of augmenting paths, based on successive contractions, is very different of the {\it shortcutting} approach used in \cite{nose}. We develop subroutines in order to obtain our main technical result, stated as follows.

\begin{lemma}\label{theo:disjointpaths}
    There exists a deterministic algorithm in the \CONGEST\ model such that, given a canonical instance $I=(G,s,t,\mathcal{P}_q)$ where $G$ has diameter $D$ and treewidth $k$, solves \qdisjointpaths\ in $I$ in $\biglo{q^2\cdot k\cdot D}$ rounds.
\end{lemma}

Similar to centralized algorithms, given $q$ vertex-disjoint paths between two vertices $s,t\in V$, the main idea to obtain a (distributed representation of a) set of $q+1$ disjoint paths between $s$ and $t$ is to verify if there exists an augmenting path in the residual graph, and if it exists, then to obtain the new collection of $q+1$ disjoint paths. If there is no augmenting path, we will  compute an $s$-$t$ vertex separator of size at most $q$.

The main technical difficulties are about the distributed detection of an augmenting path, given that it is not even clear how to represent the residual graph in a distributed manner or how to simulate computations of the residual graph in the input graph G. 

\subsection{Residual and Bridge Graphs.}

\textbf{Residual Graph}.  Our algorithm is inspired by the Ford-Fulkerson algorithm in centralized models. Therefore, the algorithm tries to find an augmenting path in the so-called {\it residual graph} of a canonical instance. Given a canonical instance $(G,s,t,\mathcal{P}_q)$, the residual graph $G_{res}'$ is build as follows: For each vertex $v\in V$, add two copies $v_{in}$ and $v_{out}$ to the residual graph $G_{res}'$ connected in both directions, i.e., $(v_{in},v_{out}), (v_{out},v_{in})\in E(G_{res}') $. For each edge $\{u,v\}\in E$  of $G$, add the arcs $(v_{out}, u_{in})$ and $(u_{out}, v_{in})$ to the residual graph $G_{res}'$. Finally, for each undirected edge $\{v_j,v_{j+1}\}$ in some path $P_y\in\mathcal{P}_q$, the arc $((v_j)_{out},(v_{j+1})_{in})$ is removed from $G_{res}'$. While this definition corresponds to the classical definition of residual graph, we modify the residual graph $G_{res}'$ into a new graph $G_{res}$ as follows: for each vertex $v\in V$ which not an internal vertex of any of the $q$ vertex-disjoint paths, we contract the vertices $\{v_{in},v_{out}\}$ into a single vertex $v$. It is this new graph $G_{res}$ that we will call the residual graph and will be used throughout the rest of the paper. For each path $P_j\in\mathcal P_q$, we denote as $\overrightarrow{P_j}$ the directed version of $P_j$ in $G_{res}$. See \Cref{fig:residualgraph} for a graphical example of these definitions.

\begin{figure}[h]
    \centering
    \scalebox{0.7}{\input{img/newresidualgraph}}
    \caption{(a) A canonical instance $(G,s,t,\mathcal{P}_1)$ with only one path $P_1={sv_1v_2t}$. (b) the classical residual graph $G_{res}'$ of that instance. (c) The simplified residual graph $G_{res}$ to be used throughout this work.}
    \label{fig:residualgraph}
\end{figure}

We refer to any $(s,t)$-directed path in $G_{res}$ as an $(s,t)$-augmenting path, or simply augmenting path. The following fact is a standard argument for constructing vertex-disjoint $s,t$-paths based on the Ford-Fulkerson approach~\cite{fordfulkerson1956} by using the original definition of residual graphs.

\begin{proposition}\label{pr:AugPath}
    Given a canonical instance $(G,s,t,\mathcal{P}_q)$ and its respective residual graph $G_{res}'$, there exists $q+1$ disjoint $s,t$-paths in $G$ if and only if there is a directed $(s_{in},t_{out})$-augmenting path in $G_{res}'$. 
\end{proposition}

We prove that in our modified and simplified residual graph $G_{res}$ the same holds.

\begin{corollary}\label{cor:AugPath}
There is an $(s_{in},t_{out})$-augmenting path in $G_{res}'$ iff there is an $(s,t)$-path in $G_{res}$. The later will also be called augmenting path.
\end{corollary}
\begin{proof}
Observe that the contraction to create $G_{res}$ from $G_{res}'$ does not create new, simple $(s, t)$-directed paths, since given any simple $(s,t)$-directed path $P$ in $G_{res}$, for each vertex $v\in P$ not on one of the $r$ vertex-disjoint paths, replace the occurrence of $v$ with $v_{in},v_{out}$ in that order; the resulting path is a $(s, t)$-directed path $P$ in $G_{res}'$. Clearly, since we only contract vertices, we do not destroy any $(s, t)$-directed paths. Therefore, there is a $(s,t)$-directed path in $G_{res}$ iff there is one in $G_{res}'$.
\end{proof}

In fact, since there is always a $(t, s)$-directed path in $G_{res}$ (through any path $P\in\mathcal{P}_q$ if $q\geq 1$, trivially for $q=0$), we can translate this statement in terms of strong connectivity.

\begin{claim}\label{cor:SC}
There is an augmenting path  in $G$ iff $s$ and $t$ are strongly connected in $G_{res}$.
\end{claim}

Recall that given a canonical instance $(G,s,t,\mathcal{P}_q)$, we assume that each vertex in a path $P_y\in\mathcal{P}_q$ has a known position from $1$ in vertex $s$ to $|P_y|$ in vertex $t$. We can naturally extend this order to the directed path in $G_{res}$ from $t$ to $s$ in this way: Suppose $P_y\in \mathcal P_q$ is an $(s,t)$-path in $G$ with  $h$ vertices, then in $G_{res}$ this path will convert in a directed path $\overrightarrow{P_y}=t\omega_{2(h-1)-1}\omega_{2(h-1)-2}\dots\omega_{3}\omega_{2}s$ of $2h-2$ vertices from $t$ to $s$. Every interval vertex $v_i$ of $P_y$ now has two vertices $\omega_{2i-2}$ and $\omega_{2i-1}$ in $\overrightarrow{P_y}$ corresponding to the vertices $(v_i)_{out}$ and $(v_i)_{in}$, respectively, and $s$ has position $1$ and $t$ has position $2|P_y|-1$. We assume that every directed path $\overrightarrow{P_y}$ in $G_{res}$ induced by a path $P_y\in \mathcal{P}_q$ has this order. See \Cref{fig:orderespath}.

\begin{figure}[h!]
    \centering
    \scalebox{0.7}{\input{img/newnewenumeration}}
    \caption{Above a canonical instance $(G,s,t,\mathcal{P}_1)$ with two bridges $B_1$ and $B_2$. Below the corresponding representation in the residual graph $G_{res}$, where the path $P_1$ of length $6$ now is a directed path of $10$ vertices in $G_{res}$. Each internal vertex $v_i$ has now two vertices $\omega_{2i}$ and $\omega_{2i+1}$, with $i\in[4]$}.
    \label{fig:orderespath}
\end{figure}

\begin{fact}\label{fact:simofgres}
    Consider a canonical instance $(G,s,t,\mathcal P_q)$ of $G$, and let $\overline{G_{res}}$ denote the undirected version of $G_{res}$ (i.e., directed edges are turned into undirected ones and multiple edges are removed). Every $T$-round \CONGEST\ algorithm in $G_{res}$ or $\overline{G_{res}}$ can be simulated in $\bigo{T}$-rounds in $G$.
\end{fact}

This fact is just an easy observation, since every vertex of $G$ simulates its one or two copies in $\overline{G_{res}}$.
By $\Cref{fact:simofgres}$, every distributed algorithm in $\overline{G_{res}}$ or $G_{res}$ will be described directly in this graph, omitting how the input graph $G$ simulates the algorithm. Also observe that the diameter and treewidth of $\overline{G_{res}}$ and $G$ differ by at most a factor of 2, hence we can speak of algorithms running in $\biglo{k \cdot D}$ rounds in $\overline{G_{res}}$ thinking of these parameters in graph $G$.

Now, in order to compute distributed and deterministically the new collection of $q+1$ disjoint paths, instead of deciding directly the residual graph $G_{res}$ the existence of an $(s,t)$-augmenting path, we decide the existence of this augmenting path in the so-called {\it bridge graph} of a canonical instance.\\

\textbf{Bridge Graph.} Given a canonical instance $(G,s,t,\mathcal{P}_q)$, we call a {\it bridge} to every connected component $B$ of $G-\cup_{j\in [q]}P_j$. We may say that $B$ is a bridge of the instance $(G,s,t,\mathcal{P})$. Throughout this work, we are going to refer to the bridges of a canonical instance with subindex $B_1,B_2, \dots, B_f$ where $f$ is the number of bridges in that instance. The \id\ of a bridge $B_i$ is defined as the maximum \id\ of a vertex in $B_i$ and it is denoted by $\id(B_i)$. Since each bridge $B_i$ of $(G,s,t,\mathcal{P}_q)$ has the same set of vertices in $G_{res}$, and between two neighbors $u,v\in B_i$ we add two arcs $(u,v),(v,u)\in E(G_{res})$, we may abuse notation and refer also as $B_i$ to the connected components of $G_{res}$ without $s,t$ and the vertices $v_{in},v_{out}$ of each vertex $v$ in a path of $\mathcal P_q$. 

Given a bridge $B_i$ and a path $P_j\in\mathcal{P}_q$, $\ell_i^j$ is defined as the lowest position of an ingoing neighbor of $B_i$ on the path $\overrightarrow{P_j}$ in $G_{res}$, and analogously $r_i^j$ is defined as the highest position of an outgoing neighbor of $B_i$ in $\overrightarrow{P_j}$ in $G_{res}$. We set these values as $+\infty$ and $-\infty$, respectively, if such neighbors don't exist. For example, in the canonical instance of \Cref{fig:orderespath}, for the bridge $B_1$ we have that $\ell_1^1 =2$ and $r_1^1 = 7$. For the bridge $B_2$ we have that $\ell_2^1 = 10$ and $r_2^1 = 10$ (the position of $t$). By abusing notation, we may refer to the vertices in $P_j$ with positions $\ell_i^j$ and $r_j^i$, respectively, as $ \ell_i^j$ and $ r_j^i$. In the case of \Cref{fig:orderespath}, $\omega_2$ corresponds to the internal vertex with position $\ell_1^1$ and $\omega_7$ corresponds to the vertex with position $r_1^1$.

Given a canonical instance $(G,s,t,\mathcal{P}_q)$, the directed {\it bridge graph} $G_B$ is defined as follows. See \Cref{fig:bridgegraph} for an illustration.

\begin{definition}
\label{def:bridgegraph}

Given a canonical instance $(G,s,t,\mathcal{P}_q)$, the {\it bridge graph} $G_B$ is defined as follows: 

\begin{enumerate}
    \item The set of vertices $V(G_B)$ of $G_B$ is the set $\{\beta_i:B_i$ is a bridge of $(G,s,t,\mathcal{P}_q)\}$.

    \item For every $j\in[q]$ and two bridges $B_i,B_x$, add an arc $(\beta_i,\beta_x)$ to the set $D_j$ if:
  \begin{itemize}

  \item[(a)] For some path $\overrightarrow{P_y}$, $ l_x^y\le r_i^y$. Intuitively, this means that we can reach $B_x$ from $B_i$ in $G_{res}$ by traveling leftward from $r_i^y$ to $l_x^y$ along path $P_y$. 
  \item[(b)] We have $r_x^j>r_i^j$. Intuitively, this means that we make ``progress'' along path $\overrightarrow{P_j}$, in that we can now reach a vertex in $\overrightarrow{P_j}$ further to the right.
  \item[(c)] There is no $\beta_z$ such that $\beta_z$ satisfies the above two conditions, and either $r_z^j>r_x^j$, or $r_z^j=r_x^j$ and $\id(B_z)>\id(B_x)$. In other words, ties are broken by the \id\ of the bridges.
  \end{itemize}
    \item The set of directed edges $E(G_B)$ of $G_B$ corresponds to $\cup_{j\in[q]} D_j$.
\end{enumerate}

We say that a vertex $\beta_i$ is \emph{ $s$-reachable} if there is a vertex in $B_i$ with $s$ as an in-neighbor.
Likewise, we say that a vertex $\beta_i$ is \emph{$t$-reachable} if there is a vertex in $B_i$ with $t$ as an out-neighbor. Finally, given any subset $X\subseteq \cup_{j\in[q]}D_j$, we say that a bridge $B_i$ is $(s,X)$-reachable (analogously,$B_i$ is $(t,X)$-reachable) if there exists an $s$-reachable bridge $B_s$ (analogously, $t$-reachable bridge $B_t$)  such that there exists a $(\beta_s,\beta_i)$-directed path (analogously, $(\beta_i,\beta_t)$-directed path) in $G_B[X]$.

\end{definition}

\begin{figure}
    \centering
    \scalebox{0.7}{\input{img/newbridgegraph}}
    \caption{On the top a canonical instance $(G,s,t,\mathcal P_2)$ with two paths $P_1=s,v_1,v_2,v_3,v_4,t$ and $P_2 = s,u_1,u_2,t$ and two bridges $B_1$ and $B_2$. In the middle the residual graph $G_{res}$ of the instance and on the bottom the bridge graph $G_B$ of the instance.}
    \label{fig:bridgegraph}
\end{figure}

This definition of the {\it bridge graph} is slightly different from the one in~\cite{nose}, in order to make it more amenable to distributed computing. Nevertheless, we have the following Lemma.

\begin{lemma}\label{lem:18} 
There is an $(s,t)$-augmenting path in $G_{res}$ iff there exists $s$-reachable bridge $B_i$ and $t$-reachable bridge $B_x$ such that there exists a $(\beta_i,\beta_x)$-directed  path in $G_B$.
\end{lemma}

\begin{proof}
    
For the \textit{if} direction, suppose there is a $(\beta_i,\beta_x)$-directed path in $G_B$ where $B_i$ is $s$-reachable and $B_x$ is $t$-reachable; we will transform this path into an $(s, t)$-augmenting path in $G_{res}$. First, replace each arc $(\beta_j,\beta_{j'})$ with a directed path in $G_{res}$ from a vertex in $B_j$ to a vertex in $B_{j'}$ as follows. By definition of $G_B$, there is $y\in[q]$ satisfying $r_j^y\ge l_{j'}^y$. Let $\omega_j\in B_j$ be the in-neighbor of $r_i^y$ and $\alpha_{j'}\in B_{j'}$ the out-neighbor of $\ell_{j'}^y$. We replace arc $(\beta_j,\beta_{j'})$ with the path $\omega_j\to r_j^y\overset{\overrightarrow{P_y}}{\to}\ell_{j'}^y\to\alpha_{j'}$, where $\overset{\overrightarrow{P_y}}{\to}$ denotes a directed path in $\overrightarrow{P_y}$. Finally, in each $\beta_j$ participating in the path from $\beta_s$ to $\beta_t$, add a directed path from $\alpha_j$ to $\omega_j$ of $G_B[B_j]$ (this directed path always exists since each $B_j$ is strongly connected in $G_{res}$). The union of each path forms an $(s,t)$-augmenting walk, but removing the cycle of this walk we obtain an $(s,t)$-augmenting path.

For the \textit{only if} direction, suppose there is a $(s,t)$-directed path $P$ in $G_{res}$; without loss of generality, assume that $P$ is simple. A simple path in $G_{res}$ from $s$ to $t$ has to go through vertices in some bridges $B_i$ and vertices in some of the $q$ disjoint paths. If the path has a vertex in a bridge $B_i$, before going to a vertex in another bridge, it has to go from $B_i$ with a leftward path along some path $P_j\in\mathcal{P}_q$. Let $\beta_{x_1},\dots,\beta_{x_\ell}$ be all the bridges that $P$ goes through from $s$ to $t$; 
by definition, $B_{x_1}$ is $s$-reachable and $B_{x_\ell}$ is $t$-reachable. We now construct a walk from $\beta_{x_1}$ to possibly a different $t$-reachable vertex in $G_B$.

For $i\in[\ell-1]$, let $y_i$ be such that the path $P$ travels along $\overrightarrow{P_{y_i}}$ from $\beta_{x_i}$ to $\beta_{x_{i+1}}$, and let $y_\ell$  be an arbitrary integer in $[r]$. First, set $x_{1}'\gets x_1$.
Then, one by one, for $i$ from $2$ to $\ell$, we will replace the vertex $\beta_{x_i}$ with a vertex $\beta_{x_i'}$ such that (i) either $\beta_{x_{i-1}'}=\beta_{x_i'}$ or the arc $(\beta_{x_{i-1}'},\beta_{x_{i}'})$ exists in $G_B$, and (ii) we have $r_{x_i'}^{y_i}\ge r_{x_i}^{y_i}$. Note that condition (ii) is satisfied by definition for $i=1$.

Fix an $i\in[2,\ell-1]$; we assume that the invariant is satisfied at $i$. Since $r_{x_i'}^{y_i}\ge r_{x'}^{y_i}$, we can travel from $\beta_{x_i'}$ to $\beta_{x_{i+1}}$ along path $\overrightarrow{P_{y_i}}$. Consider the arc $(x_i,x')\in D_{y_{i+1}}$; we first assume that it exists. By condition 2(c) of the construction of the bridge graph with $\beta_{x_{i+1}}$ as $\beta_z$, we have $r^{y_{i+1}}_{x'} \ge r^{y_{i+1}}_{x_{i+1}}$; otherwise, we contradict condition 2(c). Therefore, setting $\beta_{x_{i+1}'}\gets\beta_{x'}$ maintains the two properties for index $i+1$. 

Now suppose arc $(x_i,x')$ does not exist in $D_{y_{i+1}}$. Then, we must have $r_{x'}^{y_{i+1}}\ge r_{x_{i+1}}^{y_{i+1}}$; otherwise, $\beta_{x_{i+1}}$ satisfies conditions 2(a) and 2(b) of the bridge graph, so an arc in $D_{y_{i+1}}$ must exist. Therefore, setting $\beta_{x'_{i+1}}\gets\beta_{x'_i}$ maintains the two properties for index $i+1$.

Since $B_{x_\ell}$ is $t$-reachable, we have $r_{x_\ell}^{y}=\left|\overrightarrow{P_y}\right|$ for all $y\in[q]$, i.e., the rightmost out-neighbor of vertex $\beta_x$ is $t$ on each path $P_y$. Since $r_{x_\ell'}^{y_\ell}\ge r_{x_\ell}^{y_\ell}=\left|\overrightarrow{P_{y_\ell}}\right|$, vertex $\beta_{x'_\ell}$ is also $t$-reachable. We remove duplicates from the sequence $\beta_{x_1'},\beta_{x_2'},\dots,\beta_{x_{\ell-1}'}$, obtaining our desired path in $G_B$ from $s$-reachable $\beta_{x_1}$ to some $t$-reachable vertex.
\end{proof}

Our next goal is to determine whether there is an $(s, t)$-augmenting path in the residual graph $G_{res}$. Given \Cref{lem:18}, this search will be mainly implemented through the bridge graph $G_B$. The reason for this is that directed reachability is a difficult problem in general, and therefore, we need to exploit the special structure of $G_B$.

\begin{remark}
    From now on, we abuse notation, sometimes referring to $D_j$ as the directed graph whose arcs are precisely $D_j$, i.e., $G_B[D_j]$. Observe that for each $j\in[q]$, every vertex has out-degree at most $1$ in $D_j$.  Also, the directed graph $D_j$ is acyclic, since an arc $(\beta_i,\beta_x)$ implies that $r_x^j>r_i^j$. It follows that $D_j$ is composed of rooted trees, where the arcs point from away from the leaves towards the root. 
\end{remark}

\subsection{Search of Augmenting Paths in Bridge and Residual Graphs}\label{subsec:searchofap}

Given the properties exhibited of the bridge graph, the framework to compute (if exists) an augmenting path in the residual graph (through the bridge $G_B$) of any canonical instance is the following:
\begin{enumerate}
\item Compute a distributed representation of the bridge graph $G_B$ in $G$ (\Cref{lemma:bridgegraph}), 
\item\label{i:2} Use this representation to decide if there exists an augmenting path or not (\Cref{lemma:detectingapath}), 
\item\label{i:3} If (\ref{i:2}) decides that the exists an augmenting path, mark this path in $G_{res}$ (\Cref{prop:gettheaugmentingpath}), and
\item If an augmenting path was marked in (\ref{i:3}), compute and mark a new set of $q+1$ disjoint paths in $G$ (\Cref{lemma:getthepath}).
\end{enumerate}

\subsubsection{Distributed Representation of the Bridge Graph}

The first question to be addressed is about how to distributively construct and represent a bridge graph.

\begin{lemma}\label{lemma:bridgegraph}
    Given a graph canonical instance $I=(G,s,t,\mathcal{P}_q)$ such that $G$ has shortcut-quality $\scq{G}$, there exists a deterministic algorithm in the \CONGEST\ model such that, after $\biglo{\scq{G}}$ rounds, each vertex $v$ in a bridge $B_i$ (with respect to the instance $I$) learns the following.

    \begin{itemize}
        \item The \id\ of its component $\id(B_i)$.
        \item For each $j\in [q]$, the $\id(B_x)$ of the (unique) parent $B_x$ of $B_i$ in $D_j$ (if any exists).
    \end{itemize}

    In particular, if the graph has diameter $D$ and $\tw(G) \leq k$, then the algorithm works in $\biglo{k\cdot D}$ rounds.
\end{lemma}

\begin{proof}
    Each bridge $B_i$ of the residual graph can be identified in $\biglo{\scq{G}}$ rounds by running \Cref{lemm:detectbridges}. Then, through \maxproblem\ in each disjoint bridge, every vertex in a bridge $B_i$ learns the maximum \id\ between all the vertices of $B_i$. This value is exactly $\id(B_i)$. As the bridges are disjoint, it can be computed in parallel in each bridge in $\biglo{\scq{G}}$.

    Afterward, by applying $q$ times \minproblem\ and \maxproblem\ of \Cref{lemma:somebroadcasts}, in each bridge $B_x$ the vertices can learn the values $r_x^j$ and $\ell_x^j$ with respect $\overrightarrow{P_j}$ for each path $P_j\in\mathcal{P}_q$. Then, these values are sent directly from the vertices in $B_x$ to their neighbors along the corresponding path $\overrightarrow{P_j}$.

    Then the vertices inside each path $\overrightarrow{P_y}$ can compute $q$ different \prefixmaxproblem (in $\overline{G_{res}}$): In the $j\in [q]$ computation of \prefixmaxproblem, each vertex $v$ of $\overrightarrow{P_y}$ have as label the value $(r_i^j, \id(B_i))$ (if $v$ were informed previously that it is a vertex $r_i^j$ for some bridge $B_i$), and then the vertices of $P_y$ learn the maximum $r_i^j$, breaking ties according to $\id(B_i)$. This can be done in $\biglo{q \cdot \scq{G}}$ deterministic rounds, given \Cref{lemma:maxandminprefix}.

    At this point each vertex $v\in \overrightarrow{P_y}$ learns the pair $(r_i^j,\id(B_i))$, for each $j\in [q]$, such that $r_i^j$ is maximum between all the components $B_i$ such that $\pi_y\left(\ell_i\right)\leq \pi_y(v)$, by definition of the \prefixmaxproblem\ solved. Then, each vertex $v\in \overrightarrow{P_y}$ shares this information with its neighbors in each bridge $B_i$. We note that $q\cdot \log n$ is $\bigo{\log n}$; therefore, this information can be shared in a single round of communication. Now, in each bridge $B_i$, for each $j\in [q]$, through a \maxproblem\ of \Cref{lemma:somebroadcasts}, all the vertices in $B_i$ can learn the maximum value $\left(r_x^j,\id(B_x)\right)$ (respect lexicographic order), between all the values given by the neighbors of $B_i$ in the paths $P_1,..., P_q$.

    Let $\left(r_x^j,\id(B_x)\right)$ be the $j$th tuple obtained in a bridge $B_i$. This value should be sent by a vertex $v$ in a path $\overrightarrow{P_y}$ as a result of a \prefixmaxproblem. Therefore, it follows that $\ell_x^y\leq r_x^y\leq \pi(v) \leq r_i^y$. Then, condition $1.$ of \Cref{def:bridgegraph} is satisfied. Conditions $2.$ and $3.$ are satisfied by construction of $r_x^j$.

    Therefore, $(r_x^j,\id(B_x))$ corresponds to the unique neighbor (more precisely, out-neighbor) of $B_i$ in~$D_j$. Finally, if the treewidth of $G$ is $k$, then directly by \Cref{prop:sqofboundedtrwwidth} we have that $\scq{G}=\biglo{k\cdot D}$.
\end{proof}

Even \Cref{lemma:bridgegraph} does not give a fully distributed representation of the bridge graph, as only one endpoint of every edge learns the edge in the bridge graph. Nevertheless this will be enough to detect an augmenting path from $s$ to $t$ in $G_{res}$. We call a {\it distributed representation of the bridge graph} to the information learned by the vertices in $G$ in \Cref{lemma:bridgegraph}.

\paragraph{\bf Connectivity Subgraphs $H_{\mathcal{C}}^j$ and layered graph $G^j$} \label{subsec:hcjandgj}

It will be necessary for the following steps to move from properties in the bridge graph $G_B$ to computations in the residual graph $G_{res}$ and vice-versa. Specifically, the bridge graph will firstly be used to decide if there exists an $(s,t)$-augmenting path in $G_{res}$ (\Cref{lemma:detectingapath}). If such an augmenting path exists then we need to compute it explicitly (\Cref{prop:gettheaugmentingpath}). In order to compute the $(s,t)$-augmenting path, we introduce the notion of {\it connectivity subgraphs} $H_{\mathcal{C}}^j$ of $G_{res}$, for each $j\in [q]$ and each connected component $\mathcal{C}$ of $G_B[D_j]$. The usefulness of these subgraphs will appear clearly in the proofs of \Cref{lemma:detectingapath} and \Cref{prop:gettheaugmentingpath}. 
For the moment, we state the definition, properties, and distributed algorithms that can be developed in these graphs.

\begin{definition}[Connectivity Graph]\label{def:conngr}
Given a canonical instance $(G,s,t,\mathcal{P}_q)$ and the corresponding residual graph $G_{res}$ and bridge graph $G_B$ of this instance, for each $j\in[q]$ and each connected component $\mathcal C$ of $G_B[D_j]$, the connectivity graph $H_{\mathcal C}^j$ is defined as follows:
     
     \begin{itemize}
         \item Add to $H_C^{j}$ a spanning tree of each bridge $B_i\in \mathcal{C}$.
         \item For each arc $(\beta_i,\beta_x)\in D_{j}$, let $y_{i,x}$ be the minimum value\footnote{I.e., the minimum $y$ such that path $P_y$ satisfies condition $(a)$ of \Cref{def:bridgegraph} for the edge $(\beta_i,\beta_x)$.} $y\in [q]$ for which $l_x^{y_{i,x}}\le r_i^{y_{i,x}}$.  Add to $H^{j}_\mathcal{C}$ all the vertices $v_h\in \overrightarrow{P_{y_{i,x}}}$ such that $l_x^y\le \pi_y(v_h) = h\le r_i^y$. Add also the arc $(v_{i+1},v_{i})$ for all $i\in \{\ell_x^{y_{i,x}},\dots, r_i^{y_{i,x}}-1 \}$.
         \item For each edge $(\beta_i,\beta_x)\in D_{j}$, let $\omega_i\in V(B_i)$ be an in-going neighbor of $r_i^{y_{i,x}}$ and $\omega_x\in V(B_x)$ be an out-going neighbor of $\ell_x^{y_{i,x}}$ with minimum \id\ in $B_i$ and $B_x$, respectively. then also add to $H^{j}_\mathcal{C}$ the arcs $(\omega_i,r_i^{y_{i,x}})$ and $(\ell_x^{y_{i,x}}, \omega_x)$.
     \end{itemize} 
\end{definition}

     See \Cref{fig:hcj} for a graphical example. The following properties of hold.

     \begin{figure}[H]
         \centering
         \scalebox{0.6}{\input{img/hcj}}
         \caption{The subgraph $G_B[D_1]$ of the instance of \Cref{fig:bridgegraph} and the corresponding $H_{C}^1$ and $\overline{H_{C}^1}$ for the connected component marked with green.}
         \label{fig:hcj}
     \end{figure}

     \begin{lemma}\label{clm:22}
Given any canonical instance $(G,s,t,\mathcal{P}_q)$, $j\in [q]$ and $\mathcal{C},\mathcal{C}'$ two connected components of $\overline{G_B[D_j]}$, the following properties hold.
\begin{enumerate}  
\item[(1)] The subgraphs $H^j_{\mathcal{C}}$ and $H^j_{\mathcal{C}'}$ are vertex disjoint. 
\item[(2)] The undirected graph $\overline{H_{\mathcal C}^{j}}$ is an undirected tree.

\item[(3)] Given $\beta_i,\beta_x$ in the same connected component $\mathcal C$ of $G_B[D_j]$ such that $\beta_x$ is ancestor of $\beta_i$ in the directed tree of component $\mathcal C$, then given any pair of vertices $v\in B_i$ and $u\in B_x$, there exists an $(v,u)$-directed path in $G_{res}$ that corresponds exactly to orient the edges of a undirected path between $u$ and $v$ in $\overline{H_{\mathcal C}^j}$.

\end{enumerate}
\end{lemma}
\begin{proof}
To prove (1), first consider the following auxiliary claim.

\begin{claim}\label{prop:disjointpj}
    Let $(G,s,t,\mathcal{P}_q)$ be a canonical instance and $G_B$ the corresponding bridge graph of this instance. Consider any $j\in [q]$ and two edges $(\beta_i,\beta_x),(\beta_{i'},\beta_{x'})\in D_j$, such that both edges satisfies condition $(1)$ of \Cref{def:bridgegraph} for the same path $P_y$. Let $P[i,x]$ and $P[i',x']$ be the segments of $\overrightarrow{P_y}$ satisfying condition $(1)$ for $(\beta_i,\beta_x)$ and $(\beta_{i'},\beta_{x'})$, respectively.
    
    If $P[i,x]\cap P[i',x'] \neq \emptyset$, then $\beta_x = \beta_{x'}$.
\end{claim}

\begin{proof}[Proof of \Cref{prop:disjointpj}]
    Suppose by contradiction that $P[i,x]\cap P[i',x'] \neq \emptyset$ but $\beta_x\neq\beta_{x'}$. We assume w.l.o.g. that $\id(B_x)>\id(B_{x'})$.

    Recall that by definition, as $(\beta_i,\beta_x),(\beta_{i'},\beta_{x'})\in D_j$, the following conditions hold: (1) $\ell_x^y\leq r_i^y$, (2) $\ell_{x'}^y\leq r_{i'}^y$, (3) $r_x^j > r_i^j$ and (4) $r_{x'}^j > r_{i'}^j$.  We proceed through cases, and in every case, we obtain a contradiction.

    {\it Case 1.} First, suppose that $r_{x}^j\neq r_{x'}^j$. As $P[i,x]\cap P[i',x'] \neq \emptyset$, then $\min(r_i^y, r_{i'}^y) > \max(\ell_x^y, \ell_{x'}^y)$. If $r_{x'}^j > r_x^j$, then as $\ell_{x'}^y < r_i^y$, by condition $3.$ of \Cref{def:bridgegraph}, the parent of $\beta_i$ in $D_j$ should be $\beta_{x'}$ and not  $\beta_x$. Analogously, if $r_{x}^j > r_{x'}^j$, then as $\ell_{x}^y < r_{i'}^y$, by condition $3.$ of \Cref{def:bridgegraph}, the parent of $\beta_{i'}$ in $D_j$ should be $\beta_x$, and not $\beta_{x'}$. In both cases, we obtain a contradiction.

    {\it Case 2.} Second, if $r_{x'}^j = r_x^j$, then again we have two cases. If $\id(\beta_x)>\id(\beta_{x'})$, by the same argument as above it should hold that $(\beta_{i'},\beta_x)\in D_j$. If $\id(\beta_{x'})>\id(\beta_x)$, by the same argument it should hold that $(\beta_i,\beta_{x'})\in D_j$. Thus all cases lead to a contradiction.  
\end{proof}

Let $\mathcal C$ and $\mathcal C'$ be two connected components of $G_B[D_j]$. By definition the vertices of $H^j_{\mathcal{C}}$ and $H^j_{\mathcal{C}'}$ correspond to the vertices in the bridges of each connected component plus vertices in some paths of $\mathcal{P}_q$, but
Claim~\ref{prop:disjointpj} implies the following statement: for any two arcs $(\beta_i,\beta_x),(\beta_{i'},\beta_{x'})$ in $D_j$, either the $ (r_i^{y_{i,x}}, l_x^{y_{i,x}})$-directed on $\overrightarrow{P_{y_{i,x}}}$ and the $( r_{i'}^{y_{i,x}}, l_{x'}^{y_{i,x}})$-directed path on $\overrightarrow{P_{y_{i,x}}}$ are disjoint, or $x=x'$. If $x=x'$, then the arcs $(\beta_i,\beta_x),(\beta_{i'},\beta_{x'})$ belong to the same connected component in $G_B[D_1]$. Therefore, if $(\beta_i,\beta_x),(\beta_{i'},\beta_{x'})$ belong to different connected components, then the $ (r_i^{y_{i,x}}, l_x^{y_{i,x}})$-directed path  and $(r_{i'}^{y_{i,x}} , l_{x'}^{y_{i,x}})$-directed path do not intersect. This directly implies that the graphs $H^j_{\mathcal{C}}$ and $H^j_{\mathcal{C}'}$ are disjoint.

For property (2), fix a subgraph $\overline{H_{\mathcal C}^j}$. For each $\beta_i\in\mathcal C$, the corresponding bridge $B_i$ induces a tree in $\overline{H_{\mathcal C}^j}$ by construction. Therefore, the subgraph $\overline{H_{\mathcal C}^j}$ is acyclic iff the graph obtained by contracting each bridge $B_i$ in $\overline{H_{\mathcal C}^j}$ is acyclic. Suppose we take $\overline{H_{\mathcal C}^j}$ and contract each $B_i$ into a vertex $\beta_i$, so that we obtain a new graph, called $\Tilde H_{\mathcal C}^j$. To prove that $\Tilde H_{\mathcal C}^j$ is acyclic, we start with $\mathcal{C}$, a directed tree in $G_B[D_j]$, and transform it into $H$ while preserving the acyclicity of $\mathcal{C}$. For each $\beta_x\in \mathcal C$ with positive in-degree in $D_j$, let $\beta_{x_1},\dots,\beta_{x_\ell}$ be the in-neighbors of $\beta_i$. In $\overline{H_{\mathcal C}^j}$, the algorithm adds the union of undirected version of the $(r_{x_h}^{y_{x,x_h}}, l_{x_h}^{y_{x,x_h}})$-directed paths to $H_{\mathcal C}^j$, for all $h\in [h]$. In addition, for each $\beta_{x_i}$, the algorithm adds an edge connecting $B_{x_i}$ to $r_{x_i}^{y_{i,x}}$. The set of edges added is a tree connecting the vertices $\beta_x,\beta_{x_1},\dots,\beta_{x_\ell}$ of $\Tilde H_{\mathcal C}^j$, call this tree $T_x$. Moreover, by \Cref{prop:disjointpj}, the set of all edges added is acyclic. For each $\beta_x$ in $G_B[D_j]$, we delete the arc $(\beta_x,\beta_{x_i})$ in $\mathcal{C}$ for all $i\in[k]$ and add $T_x$. Since we always delete a tree and add back a tree, the graph remains acyclic. At the end, we have added exactly the edges in $\Tilde H_{\mathcal C}^j$, proving that $\overline{H_{\mathcal C}^j}$ is acyclic. Lastly, $\overline{H_{\mathcal C}^j}$ is connected by construction.

For $(3)$, let $\beta_x$ be an ancestor of $\beta_i$ in $G_B[D_j]$ and $\mathcal C$ the connected component of $G_B[D_j]$ that contains both $\beta_i$ and $\beta_x$. Let $P=\beta_1,\dots,\beta_r$ be a$(\beta_i,\beta_x)$-directed path in $G_B[D_j]$ with $\beta_1=\beta_i$, $\beta_r=\beta_x$ and $r=|P|$. For each $(\beta_k,\beta_{k+1})\in E(P)$, let $\alpha_k\in B_k$ be the neighbor of $r_k^{y_{k,k+1}}$ and $\epsilon_{k+1}\in B_{k+1}$ be the neighbor of $\ell_{k+1}^{y_{k,k+1}}$. Given two vertices $u\in B_x$ and $v\in B_i$, it is direct by construction of  $H_{\mathcal{C}}^j$, it contains the path $\overline{P}(v,u)$ between $v$ and $u$ that goes from $v$ to $r_1^{y_{1,2}}$, then from $r_1^{y_{1,2}}$ to $\ell_2^{y_{1,2}}$, and for all $k\in \{2,\dots, r-1\}$ goes inside $B_k$ from $\epsilon_k$ to $\alpha_{k}$, and from $r_k^{y_{k,k+1}}$ to $\ell_{k+1}^{y_{k,k+1}}$, and finally from $\epsilon_{r}$ to $u$. Since each bridge is strongly connected in each path $P_j\in\mathcal P_q$ the path $\overline{P}(v,u)$ goes from left to right, then the oriented version of this path from $v$ to $u$ is a $(v,u)$-directed path in $G_{res}$.
\end{proof}
Using \Cref{clm:22} and part-wise aggregation problems, the following problems can be solved efficiently.

\begin{lemma}\label{claim:compuofhcj}
    Given a canonical instance $(G,s,t,\mathcal{P}_q)$ $j\in[q]$, a spanning tree of $\overline{H^{j}_{\mathcal{C}}}$ for every connected component $\mathcal C$ of $G_B[D_j]$ can be computed in $\biglo{\scq{G}}$ rounds.
\end{lemma}

\begin{proof}[Proof of \Cref{claim:compuofhcj}]

First, all the bridges in parallel can compute a spanning tree of the bridge in $\biglo{\scq{G}}$ rounds (\Cref{teo:kmstcongest}). All the edges of the respective spanning tree are marked. Afterwards, let $B_i$ be any bridge. Recall that in the algorithm of \Cref{lemma:bridgegraph}, if $(\beta_i,\beta_x)\in D_j$ and $\overrightarrow{P_y}$ is such that $P_y\in \mathcal{P}_q$ and $\ell_x^y\leq r_i^y$, then all the vertices $v\in \overrightarrow{P_y}$ that are neighbors of the vertex $r_i^y$ in $\overrightarrow{P_y}$ learn the following values: $\id(B_x), \ell_x^y$ and $\ell_x^j$. Therefore, through a \minproblem\ of \Cref{lemma:somebroadcasts} over $\mathcal V= \{V(B_i)\colon B_i\text{ is a bridge}\}$, in each bridge $B_i$ such that $(\beta_i,\beta_x)\in D_j$, the vertices can learn the value $y_{i,x}$.

Then, each vertex $v$ in some bridge $B_i$ such that $(\beta_i,\beta_x)\in D_j$ and it is neighbor of the vertex $r_i^{y_{i,x}}$ can send to this vertex the information $x(r_i^{y_{i,x}}) = (\ell_x^{y_{i,x}}, r_x^{y_{i,x}})$ and therefore through \suffixminproblem\ and \prefixmaxproblem\ in each path $P_y$, each vertex $v\in \overrightarrow{P_y}$ can learn if there exists bridges $B_i,B_x$ such that  $\ell_x^{y_{i,x}}\leq \pi_y(v)\leq r_x^{y_{i,x}}$. If these bridges $B_i$ and $B_x$ exist, then vertex $v$ marks itself. Following the same idea, it is also possible to mark all the edges of the path $\overrightarrow{P_y}$ between the vertices with a position in $(\ell_x^{y_{i,x}}, r_x^{y_{i,x}})$. By \Cref{lemma:maxandminprefix} all these computations can be performed in $\biglo{\scq{G}}$ rounds, and thanks to \Cref{prop:disjointpj} (2) we know that every edge in a path $\overrightarrow{P_y}$ belongs to a unique $\overline{H_{\mathcal{C}}^j}$ because these subgraphs are vertex disjoint.

Finally, by using \Cref{teo:kmstcongest} in the subgraph induced by the marked vertices and edges, a spanning tree of each $\overline{H_{\mathcal C}^{j}}$ can be computed.
    
\end{proof}

\begin{lemma}\label{lemma:uvpathingres}
    Given a canonical instance $(G,s,t,\mathcal{P}_q)$, $j\in [q]$ and two bridges $B_i$ and $B_x$ such that $\beta_i $ and $\beta_x$ belong to the same connected component of $G_B[D_j]$, then for any pair of vertices $v\in B_i$ and $u\in B_x$, there exists a $(v,u)$-directed path in $G_{res}$ and it can be computed in $\biglo{\scq{G}}$ rounds.
\end{lemma}

\begin{proof}[Proof of \Cref{lemma:uvpathingres}]

Given a canonical instance $(G,s,t,\mathcal P_q)$, two bridges $B_i$ and $B_x$ and a value $j\in [q]$ such that $\beta_i,\beta_x$ are in the same connected component of $G_B[D_j]$, let $v\in B_i$ and $u\in B_x$ be two vertices. 

First in $\biglo{\scq{Q}}$ rounds a representation of the bridge graph $G_B$ can be computed as described in \Cref{lemma:bridgegraph}. Then, by \Cref{claim:compuofhcj} a spanning tree of each undirected graph $\overline{H_{\mathcal C}^j}$ can be computed for each connectivity subgraph $H_{\mathcal C}^j$. Through a \broadcast\ protocol in each spanning tree (since they are disjoint) we can detect the subgraph $\overline{H_{\mathcal C_0}^j}$ of the connected component $\mathcal C_0$ of $G_B[D_j]$ such that $\beta_i,\beta_x\in \mathcal C$. In point, each bridge $B_r$ knows if it is part of the connected component $C_0$ or not (since the vertices of each bridge $B_r$ learn if they are part of $\overline{H_{\mathcal C_0}^j}$ or not). Also, the connected component $\mathcal C_0$ is a directed tree in $G_B[D_j]$ and therefore the root $\beta_r$ of this tree can be detected in $\biglo{\scq{G}}$ rounds, since the root is the only bridge $B_r$ with vertices $V(B_r)\subseteq V$ in $\overline{H_{\mathcal C_0}^j}$ and with out-degree $0$ in $G_B[D_j]$ (i.e. there exists no $\beta_f$ such that $(\beta_r,\beta_f)\in D_j$).

Let $v_0\in B_r$ be the vertex with minimum \id\ in $B_r$ (and therefore it can be marked in $\biglo{\scq{G}}$ rounds using \minproblem\ just in $B_r$). By \Cref{clm:22} (3), there exists a directed path in $H_{\mathcal C_0}^j$ from $v$ to $v_0$, and this directed path induces the unique undirected path in $\overline{H_{\mathcal C_0}^j}$ between $v$ and $v_0$. This path in $\overline{H_{\mathcal C_0}^j}$ can be marked in $\biglo{\scq{G}}$ rounds using \markapath\ protocol in $\overline{H_{\mathcal C_0}^j}$. (\Cref{lemma:somebroadcasts}). This path induces a $(v,v_0)$-directed path in $G_{res}$. Finally, as $B_r$ is the root of a directed tree $G_B[D_j]$, then $r_r^j\geq r_x^j$ and we always have that $r_x^j> \ell_x^j$. Therefore $r_r^j\geq \ell_x^j$ and then the $(r_r^j,\ell_x^j)$-directed path through $\overrightarrow{P_j}$ in $G_{res}$. Therefore, we can go from $v_0$ to $r_r^j$ (with a directed path in the strongly connected component $B_r$), then from $r_r^j$ to $\ell_x^j$ in $P_j$, and then from $\ell_x^j$ to $u$ in $B_x$. These last three paths can also be computed in ${B_i}$, ${P_j}$ and ${B_x}$, respectively, in $\biglo{\scq{G}}$ rounds using \markapath.
\end{proof}

The intuition for the connectivity subgraphs $H_{\mathcal C}^j$ is that they allow us to efficiently compute a $(u,v)$-directed path for any pair of vertices $u\in B_i$ and $v\in B_x$ such that $\beta_i$ and $\beta_x$ are in the same connected component of $G_B[D_j]$ (\Cref{lemma:uvpathingres}). Now, by \Cref{lem:18}, in order to decide if there exists a $(s,t)$-augmenting path in $G_{res}$, we need to decide if there exists a $(\beta_t,\beta_t)$-directed path in $G_B$ such that $B_s$ is $s$-reachable and $B_t$ is $t$-reachable. This  $(\beta_s,\beta_t)$-directed path in $G_B$ may pass through edges in different sets $D_1,\dots,D_q$, and moreover, to later compute the augmenting path, we need to translate this $(\beta_s,\beta_t)$-directed path in $G_B$ into a $(v_s,u_t)$-directed path for vertices $v_s\in B_s$ and $B_t$. Since $B_s$ and $B_t$ may not be in the same connectivity subgraph $H_{\mathcal C}^j$ for any $j\in [q]$, \Cref{lemma:uvpathingres} is not enough to compute our desired $(v_s,u_t)$-directed path. To solve this problem, we define the layered graph $G^j$ and specific subgraphs of it that will be helpful to solve this problem.

\begin{definition}\label{def:gj}
    Given a canonical instance $(G,s,t,\mathcal P_q)$ and a value $j\in [q]$, the layered graph $G^j$ is defined as follows: Create $j$ copies of the graph $\overline{G_{res}}$, i.e., each vertex $v\in V(G_{res})$ is replaced with $j$ copies $v^1,\dots,v^j$ in $G^j$ every edge $\{u,v\}\in E(G)$, is replaced in $G^j$ with $j$ copies $\{u^i,v^i\}$ for all $i\in[j]$; eventually, for each $v\in  V(G_{res})$ add edges to $G^j$ to turn its copies $v^1,\dots,v^j$ into a clique. See \Cref{fig:clique} for a graphical example.
\end{definition}

The subgraph induced by the vertices $\{v^i\}_{v\in V(G)}$ of a fixed $i\in [j]$ is called the {\it layer $i$} of $G^j$.  A vertex $v^h$ in the layer $h$ is $G^j$ is said to be the $h$-copy of vertex $v\in V(G)$. Analogously, given a subgraph $H$ of $\overline{G_{res}}$, we refer as the $h$-copy of $H$ to the subgraph in layer $h$ of $G^j$ corresponding to the subgraph $H$ but with vertices labeled as $v^h$ instead of $v$, for all $v\in V(H)$.

\begin{definition}\label{def:hj}
    Given a canonical instance $(G,s,t,\mathcal P_q)$ and a value $j\in [q]$, the subgraph $\mathcal H^j\subseteq G^j$ is constructed as follows:
    
    \begin{itemize}
        \item For each bridge $B_x$ of $G_{res}$, add a $1$-copy of a spanning tree of $B_x$ to $\mathcal H^{j}$, i.e., a copy of the spanning tree in layer $1$.
        \item For each $i\in [j]$, and each connected component $\mathcal C$ of $G_B[D_i]$, add a copy of $\overline{H_{\mathcal C}^i}$ in layer $i$, and delete of this $i$-copy all the edges $\{u,v\}\in E\left(\overline{H_{\mathcal C}^i}\right)$ such that $u$ and $v$ belong to the same bridge.
        \item Connect all the copies of a vertex $v$ in $\mathcal H^j$, such that $v$ belongs to some path $P_y\in \mathcal P_q$, through a clique in $\mathcal H^j$.
    \end{itemize}

\end{definition}

\begin{figure}[H]
    \centering
    \scalebox{0.5}{\input{img/g3}}
    \caption{To the left an example of a graph $\overline{G_{res}}$ and to the right the graph $G^3$. The cliques corresponding to the $3$ copies of each vertex are marked with red. Each layer is marked with dashed boxes.}
    \label{fig:clique}
\end{figure}

The following technical properties holds for the graph $G^j$ and the subgraph $\mathcal H^j$. These properties will be use for \Cref{lemma:detectingapath} and \Cref{lemma:getthepath}.

\begin{lemma}\label{claim:101}
    Given any canonical instance $(G,s,t,\mathcal P_q)$ such that $G$ has diameter $D$ and treewidth $k$, then for each $j\in[q]$ the following properties hold.
    
    \begin{enumerate}
        \item If  $D$ is the diameter of $G$ and $k$ is the treewidth of $G$, then $diam( G^j) = \bigo{diam(G)}$ and $\tw( G^j) = j\cdot \tw(G)$.  Therefore $\scq{G^j} = \bigo{j\cdot k\cdot D}$.
        \item Every $T$-rounds algorithm in $G^j$ can be simulated in $\bigo{T}$ rounds in $G$.
        \item For each pair of bridges $B_i,B_x$ such that the $1$-copy of $B_i$ and the $1$-copy of $B_x$ belongs to the same connected component $\mathcal C$ of $\mathcal H^j$, there exists an $(u,v)$-directed path in $G_{res}$ for all $u\in B_i$ and $v\in B_x$.
        \item For each pair of bridges $B_i,B_x$ such that there exists a $(\beta_i,\beta_x)$-directed path in $G_B[D_1\cup\dots\cup D_j]$, then the $1$-copy of $B_i$ and a $1$-copy of $B_x$ belong to the same connected component $\mathcal C$ of $\mathcal H^{j}$.
        
    \end{enumerate}
\end{lemma}

\begin{proof}[Proof of \Cref{claim:101}]

    It is direct by construction that for all $u,v\in V$ and $i_1,i_2\in [j]$, $dist_{\mathcal H^j}(u_{i_1}, v_{i_2}) = dist_G(u,v) +1$. Then, to prove that $\tw(G^j) = j\cdot \tw(G)$ we build a tree composition of $G^j$ with bags of size at most $j\cdot \tw(G)$ as follows. 

    Since $G$ has treewidth $k$, there exists a tree decomposition $\mathcal T$ of $G$ with maximum bag size $k+1$. Define $\mathcal T'$ as the following tree decomposition of $G^j$ with maximum bag size $(k+1)j$, which is sufficient to prove the claim. $\mathcal T'$ has the same bags that $\mathcal T$ but we replace all occurrences of $v$ in a bag of $\mathcal T$ with the vertices $v^1,\dots,v^j$ in $\mathcal T'$. Clearly, the maximum bag size is now at most $(k+1)j$. We now claim that the new decomposition $\mathcal T'$ is a treewidth decomposition of $\mathcal H^j$. By construction of $\mathcal T'$ from a tree decomposition of $G$, clearly $\mathcal T'$ is a tree decomposition of $\mathcal H^j$.

   For condition $(2)$, observe that by (1), $dist_{G^j}(u_{i_1}, v_{i_2}) = \bigo{dist_G(u,v)}$ for all $u,v\in V$ and $i_1,i_2\in [j]$, then each $m$-round algorithm over $\mathcal H^j$ can be simulated in $\bigo{m}$ rounds in $G$ since every vertex $v \in V$ can simulate its $j$ corresponding copies $v^1, \dots, v^j$ in the graph $G^j$, and two neighbors $u,v$ of $G$ have at most $j$ copies that are neighbor between them, therefore in each round $u$ and $v$ need to share at most $\bigo{j\cdot \log n}$, and $j$ is a constant that do not depend on $n$. 
   
    For $(3)$, let $B_i$ and $B_x$ be two bridges such that the $1$-copy of $B_i$ and the $1$-copy of $B_x$ belongs to the same connected component $\mathcal C$ of $\mathcal H^j$ and two vertices $u\in B_i,v\in B_x$. By definition of $\mathcal H^j$ there exists an undirected path $P(u^{1}, v^{1})$ between $u^{1}$ and $v^{1}$ (the copies of $u$ and $v$ in layer $1$) in $\mathcal H^j$. If $u$ and $v$ belongs to the same connectivity graph $H_{\mathcal C'}^{j'}$ for some connected component $C'$ of $G_B[D_{j'}]$ with $j'\leq j$, then there exists a $(u,v)$-directed path in $G_{res}$ by \Cref{lemma:uvpathingres}. In the rest of the proof then assume that $u,v$ are not in the same connectivity graph $H_{\mathcal C'}^{j'}$ for any connected component $C'$ of $G_B[D_{j'}]$ with $j'\leq j$. As the path  $P(u^{1}, v^{1})$ exists in $\mathcal H^j$, then this path needs to go through vertices in different layers of $\mathcal H^j$. Let $P_1,\dots,P_r$ be the subpaths of $P(u^{1}, v^{1})$, such that each $P_h$ is entirely contained in some layer $\ell_h\in[j]$ and $P_h$ correspond to the $\ell_h$-copy of a subpath $P$ of some disjoint path $P_y\in \mathcal P_q$. Therefore, between two consecutive subpaths $P_h$ and $P_{h+1}$ the path $P(u^{1}, v^{1})$ moves exclusively through a $1$-copy of some bridge $B_x$. Let $B_h$ and $B_{h+1}$ be two consecutive bridges in the path $P(u^{1}, v^{1})$, meaning that the path goes from $B_h$ to $P_h$ and then to $B_{h+1}$. By definition of $\mathcal H^j$, this means that exists a $\alpha_h\in[j]$ such that $(\beta_h,\beta_{h+1})\in D_{\alpha_h}$ or $(\beta_h,\beta_{h+1})\in D_{\alpha_h}$, since this path $P_h$ correspond to the undirected path of a $(r_i^y,\ell_x^y)$-directed path or a $(r_x^y,\ell_i^y)$-directed path (see \Cref{def:conngr}). In any case, we know that there exists a $(u_h,v_{h+1})$-directed path in $G_{res}$ from any vertex $u_h\in B_h$ and any vertex $v_{h+1}\in B_{h+1}$ by \Cref{lemma:uvpathingres}, since $B_h$ and $B_{h+1}$ lie in the same connected component of $G_B[D_{\alpha_h}]$. Let $\overrightarrow{P}(u_h,v_{h+1})$ be such $(u_h,v_{h+1})$-directed path in $G_{res}$. By definition of the paths $P_1,\dots,P_r$, the vertices $u_h$ and $v_{h}$ lie in the same bridge $B_h$, and therefore there exists too a $(v_h,u_{h})$-directed path $\overrightarrow{P}(v_h,u_{h})$ in $G_{res}$, since the bridges are strongly connected. Then, if we join all these paths we obtain a $(u_1,v_r)$-directed walk in $G_{res}$. As $u_1\in B_i$ and $v_{r+1}\in B_x$, then we can extend these walk to obtain a $(u,v)$-directed walk from any vertex $u\in B_i$ and $v\in B_x$. By deleting cycles of this walk we obtain a $(u,v)$-directed path in $G_{res}$ for any pair of vertices $u\in B_i$ and $v\in B_x$.

    Finally, for $(4)$ consider a $(\beta_i,\beta_x)$-directed path $P$ in $G_B[D_1\cup\dots\cup D_j]$ and enumerate the path $P=\beta_1,\dots,\beta_r$ from $\beta_1=\beta_i$ to $\beta_r=\beta_x$ where $r=|P|$. Then,  each arc $(\beta_h,\beta_{h+1})$ in $P$ belongs to a subset $D_{j_h}$ for some $j_h\in[q]$. Consider the $1$-copy of each bridge $B_z$ such that $\beta_z\in P$. To conclude we prove that all these $1$-copies belong to the same connected component of $\mathcal H^j$. For each arc $(\beta_h,\beta_{h+1})$ in $P$, the vertices $\beta_h$ and $\beta_{h+1}$ belong to the same connected component $\mathcal C$ of $G_B[D_{j_h}]$ and therefore the bridges $B_h$ and $B_{h+1}$ belong to the same connectivity subgraph $H_{\mathcal C}^{j_h}$. By definition of $\mathcal H^j$ this means that the $1$-copies of $B_h$ and $B_{h+1}$ belongs the same connected component of $\mathcal H^j$, since in layer $j_h$ there is a copy of the path in $H_{\mathcal C}^{j_h}$ connecting $B_h$ and $B_{h+1}$. This proves then that the $1$-copies of $B_i$ and $B_x$ are in the same connected component of $\mathcal H^j$.  
\end{proof}

\subsubsection{Contracting the Bridge Graph}

The classical strategy to find an augmenting path involves searching for an augmenting path in the residual graph $G_{res}$. By \Cref{lem:18}, the search now is reduced to identifying an $s$-reachable bridge $B_i$ and a $t$-reachable bridge $B_x$ such that there exists a $(\beta_s,\beta_t)$-directed path in $G_B$.

We solve this problem with a {\it contraction-base} algorithm. Since each vertex in the bridge graph corresponds to a connected component of $G - \cup_{j=1}^q P_j$, the contraction algorithm minimizes the number of communication rounds required. Simulating $G_B$ directly can be very costly in terms of rounds, as each round in $G_B$ must be simulated in $G$. Each vertex in $G_B$ corresponds to a subgraph of $G$ that may contain up to $\Theta(n)$ vertices and have a diameter of $\Theta(n)$. Therefore, even if we use low-congestion shortcuts, the communication inside a bridge could take up to $\Theta(n)$ rounds.

While our goal is to implement an algorithm that decides if there exists a $(s,t)$-directed path in $G_{res}$, by the properties (3) and (4) of \Cref{claim:101}, it is equivalent to solve the following problem: 

\begin{algorithmbox}{\contractingproblem}

    {\bf Input:} A canonical instance $(G,s,t,\mathcal{P}_q)$, along with the distributed representation of the bridge graph.\\

    {\bf Output:} Each vertex of $G$ learns if there exists an $(s,t)$-augmenting path in $G_{res}$, and if exists, the vertices also learn the same {\bf minimum} value $j^*\in [q]$ such that there exists an $s$-reachable bridge $B_s$ and a an $t$-reachable bridge $B_t$ such that there exists a $(\beta_s,\beta_t)$-directed path in $G_B[D_1\cup\dots\cup D_{j^*}]$.
    
\end{algorithmbox}

Although our aim in this Section is simply to decide the existence of an $(s,t)$-directed path in $G_{res}$, without any extra rounds beyond those needed for the decision, we can compute a critical index $j^*\in[q]$ such that there is an $s$-reachable bridge $B_s$ and a $t$-reachable bridge $B_t$ for which a $(\beta_s,\beta_t)$-directed path lies in $G_B[D_1\cup\cdots\cup D_{j^*}]$.  This index will be useful to build the augmenting path in the next Section.

Now we describe the algorithm that solves \contractingproblem, to later show in \Cref{lemma:detectingapath} how it can be implemented in the \CONGEST\ model and in \Cref{lemma:correctcontracting} its correctness.
\bigskip

\centerline{
{\bf Algorithm} \contractingALG
}
\smallskip
At the beginning of the algorithm, each vertex $v$ in a connected component $B_i$ of $G- \cup_{i=1}^q P_i$ sets its label $x(v)$ as

\begin{equation*}
    x\left(v\right) = \begin{cases}
        \id(s) & B_i\text{ is s-reachable $\land$ $B_i$ is not t-reachable} \\
        \id(t) & B_i\text{ is not s-reachable $\land$ $B_i$ is t-reachable}\\
        \id(s)\#\id(t) & B_i\text{ is s-reachable $\land$ $B_i$ is t-reachable} \\
        \bot &  B_i\text{ is not s-reachable $\land$ $B_i$ is not t-reachable}
    \end{cases}
\end{equation*}

We fix the order $\preceq$ such that $\bot \preceq \id(s)\preceq\id(t)\preceq\id(s)\#\id(t)$, known to all vertices. The algorithm will proceed through $q+1$ phases such that, 
\begin{itemize}
    
    \item At the end of each phase $j$, the vertices $v$ in a bridge $B_i$, such that the $1$-copy of $B_i$ in $\mathcal H^j$ (see \Cref{fig:hcj}) is in a connected component $\mathcal{C}$ of $\mathcal H^j$ will update its label $x(v)$, as $x(v) = \max_\preceq\{x(\omega)\colon \omega\in V(\mathcal C)\}$. 
    \item If at phase $j\in[q]$ there exists a bridge $B_i$, such that at the end of phase $j$, the vertices of $B_i$ updates its label to $\id(s)\#\id(t)$, then the algorithm indicates that an $(s,t)$-augmenting path, returns $j$ and stops.
    \item If the algorithm reaches phase $q+1$, then it stops and informs that there is no augmenting path.
\end{itemize}

\begin{lemma}\label{lemma:detectingapath}
    Given a canonical instance $(G,s, t, \mathcal{P}_q)$ such that $G$ has diameter $D$ and treewidth $k$, \contractingALG\ can be implemented in the \CONGEST\ model in $\biglo{q^2\cdot k\cdot D}$ rounds.
\end{lemma}

\begin{proof}
    
At the beginning of the algorithm, through two \broadcast\ in each bridge $B_i$, the vertices $v\in B_i$ can learn if $B_i$ is $s$-reachable and/or $t$-reachable, and therefore set its initial label $x(v)$. By \Cref{lemma:somebroadcasts} this can be done in $\biglo{k\cdot D}$ rounds. Now we describe how each phase can be implemented. \\

{\bf Description of phase $j$.} Given $j\in[q]$, the algorithm compute the all connectivity subgraphs $H_{\mathcal C}^j$ using \Cref{claim:compuofhcj} in $\biglo{j\cdot k\cdot D}$ rounds.

Also, in previous phases $j'<j$ also were created the connectivity bridges $\overline{H_{\mathcal{C'}}^{j'}}$ for each connected component $\mathcal C'$ of $G_B[D_{j'}]$, and then the vertices in $G$ in this point can simulate computations in the subgraph $\mathcal H^j$ (as defined in Section \ref{subsec:hcjandgj}) in $\biglo{j\cdot \tw(G)\cdot D}$ rounds according to Claim \ref{claim:101}). For any $j\in [q]$, each vertex $v$ in a connected component $\mathcal C_{\mathcal H^j}$ of $\mathcal H^j$ updates its input $x(v)$ as 

\begin{equation*}
    x(v) = 
        \max_\preceq\{x(\omega)\colon\omega\in V(\mathcal C_{\mathcal H^j})\} 
\end{equation*}

Then, each vertex $v$ in a bridge $B_i$ update its label as the label obtained by its $1$-copy in $\mathcal H^j$ (recall that each vertex in a bridge have a $1$-copy in $\mathcal H^j$).This updating step, by Claim \ref{claim:101}, can be done in $\biglo{j \cdot \tw(G)\cdot D}$ rounds (in $G$) using \maxproblem\ of \Cref{lemma:somebroadcasts} in each disjoint undirected component $\mathcal C_{\mathcal H^j}$ of $\mathcal H^j$. 

Finally, using \broadcast\ of \Cref{lemma:somebroadcasts}, it can be learned by all the vertices if there exists a bridge $B_x$ such that the label of the vertices in $B_x$ have updated its label to $\id(s)\#\id(t)$, in which case the vertices learn $j$. After the $q$ phases, all the vertices learn the minimum value $\sj$ such that the vertices in a bridge have $\id(s)\#\id(t)$ as label at the end of phase $\sj$ As every phase takes $\biglo{q\cdot \scq{G}}$ rounds and there are $q$ phases, the total round complexity is $\biglo{q^2\cdot k \cdot D}$ rounds. 

\end{proof}

Now we prove the correctness of \contractingALG.

\begin{lemma}\label{lemma:correctcontracting}
    At the end of \contractingALG, there exists a bridge $B_i$ such all all its vertices $v\in B_i$ have label $x(v) = \id(s)\#\id(t)$ if and only if there exists an $(s,t)$-augmenting path in $G_{res}$ passing through vertices in $B_i$. 
\end{lemma}

\begin{proof}
For the {\it only if} condition, consider first the following claim:

\begin{claim}\label{claim:102}
    Let $B_i$ be a bridge such that the vertices $v$ of $B_i$ in phase $h\in[q]$ of \contractingALG\ have label $x(v) = \id(s)$ (analogously, $x(v) = \id(t)$). Then, there exists an $(s,v)$-directed path (analogously, $(v,t)$-directed path) in $G_{res}$ for all $v\in B_i$.
\end{claim}

\begin{proof}[Proof of Claim \ref{claim:102}]
    We prove the Claim when the vertices have label $\id(s)$, as the proof for $\id(t)$ is analogous. 
    
    By induction in the phases $j\in[q]$. For $j=1$, if at the end of phase $1$ the vertices of a bridge $B_i$ have $\id(s)$, then these vertices had this $\id$ at the start of phase $1$, in which case $B_i$ is $s$-reachable, or there exists other bridge $B_x$ such that the vertices of $B_x$ had the label $\id(s)$ at the start of phase $1$ and the vertices of $B_x$ and $B_i$ were in the same connected component of $\mathcal H^j$. As the vertices of $B_x$ had the label $\id(s)$ at the start of phase $1$, then $B_x$ is $s$-reachable, and the connected components of $\mathcal H^1$ corresponds exactly to the connectivity subgraphs $\overline{H_\mathcal C^1}$. Therefore, $B_i$ and $B_x$ are in the same connected component of $G_B[D_1]$, in which case there exists an $(u,v)$-directed path in $G_{res}$ for all vertices $u\in B_x$ and $v\in B_i$ by \Cref{lemma:uvpathingres}, proving the base case.

    Now, suppose that the Claim is true for all the bridges such that their vertices have label $\id(s)$ at the end of phase $j'<j$. If a bridge has vertices with labels $\id(s)$ at the end of phase $j+1$, then suppose that the vertices didn't have this label at the start of phase $j+1$ (if not, we conclude by induction hypothesis). In this case, there exists a bridge $B_x$ such that its $1$-cop belongs to the same connected component $\mathcal C$ of $\mathcal H^j$ that the $1$-copy of $B_i$, and the vertices of $B_x$ had label at the start of phase $j+1$. In this case, there exists a directed path from $s$ to a vertex in $B_x$ by the induction hypothesis, and there exists a path from $B_x$ to $B_i$ by \Cref{claim:101}. Therefore, we conclude that there exists a directed path from $s$ to a vertex in $B_i$.
\end{proof}

We prove then that if the vertices of a bridge $B_i$ have label $x(v) = \id(s)\#\id(t)$ at the end of phase $h\in[q]$, then there exists an $s$-reachable bridge $B_s$ and a $t$-reachable bridge $B_t$ such that there exist a $(v_s,v_t)$-directed path in $G_{res}$ with $v_s\in B_s$ and $v_t\in B_t$. We prove this by induction in $h\in [q]$. If $h=1$ and at the end of phase $1$ the vertices of $B_i$ have label $x(v) = \id(s)\#\id(t)$, then either the vertices had this label at the start of phase $1$, in which case $B_i$ is $s$-reachable and $t$-reachable, in which case we conclude since each bridge $B_i$ is strongly connected in $B_{res}$, or there exists two bridges $B_s,B_t$ in the same connected component of $G_B[D_1]$ that are $s$-reachable and $t$-reachable, respectively, in which case we conclude by \Cref{lemma:uvpathingres}. 

Now, assume the proposition holds for every $t\leq h$. If the vertices of $B_i$ have label $x(v) = \id(s)\#\id(t)$ at the end of phase $h+1$, then either the vertices already had this label at the start of phase $h+1$, in which case they receive this label in a phase $t\leq h$ and we conclude by induction hypothesis, or they compute this label in phase $h+1$, in which case there exist two bridge $B_s, B_t$ with their $1$-copy in the same connected component $\mathcal C$ of $\mathcal H^{h+1}$ that the $1$-copy of $B_i$, such that the vertices of $B_s$ had the label $\id(s)$ and the vertices of $B_t$ had labels $\id(t)$ a the beginning of phase $h+1$, since this is the only way a vertex can compute label $\id(s)\#\id(t)$ at the end of the algorithm. By Claim \ref{claim:102}, there exists a $(s,\omega_s)$-directed path and a $(\omega_t,t)$-directed path in $G_{res}$ for all vertices $\omega_s\in B_s$ and $\omega_t\in B_t$. By Claim \ref{claim:101} (3), there exists also a $(\omega_s,\omega_t)$-directed path in $G_{res}$. Finally, by joining these three directed paths and deleting cycles, we obtain a $(s,t)$-augmenting path in $G_{res}$.
    \newcommand{\dist}{\textup{dist}}

For the \textit{if} condition, given a bridge $B_i$, consider all directed simple paths in $G_B$ that start from $\beta_i$ and end at some $t$-reachable bridge $\beta_t$. Define $\dist(\beta_i)$ to be the minimum possible number $h\in[q]$ such that there exists a path in $G_B[D_1\cup\dots\cup D_h]$ from $\beta_i$ to a vertex $\beta_t$ such that $B_t$ is $t$-reachable. In particular, $\dist(\beta_i)=\infty$ iff there is no such path, and $\dist(\beta_i)=0$ iff $B_i$ is $t$-reachable.  We prove the following claim:

\begin{claim}\label{claim:100}
    All bridges $\beta_i$ with $\dist(\beta_i)=d<\infty$ satisfy that at the end phase $d$ of \contractingALG, the vertices $v\in B_i$ have label $\id(t)$ or $\id(s)\#\id(t)$.
\end{claim}

\begin{proof}[Proof of \Cref{claim:100}]
    We prove the Claim by induction on $d\ge0$. The base case $d=0$ is trivial; we now consider the case $d=1$. Suppose a bridge $\beta_i$ satisfies $\dist(\beta_i)=1$. Then, there exists a $(\beta_i,\beta_t)$-directed path in $G_B[D_1]$ with $\beta_t$ such that $B_t$ is $t$-reachable, so on iteration $1$ of the algorithm, $\beta_i$ and $\beta_t$ are in the same connected component of $G_B[D_1]$ and therefore the vertices of $B_i$ update their labels to $\id(t)$, since the connected components of $\mathcal H^1$ corresponds exactly to the subgraphs $\overline{H_{\mathcal C}^1}$ for each connected component $\mathcal C$ of $G_B[D_1]$. At the of the algorithm, this label can only change to $\id(s)\#\id(t)$ by the order $\preceq$.

Suppose \(\dist(\beta_i)=d+1\) for some \(d\ge1\).  Let \(P\) be a directed path in
\(
G_B\bigl[\bigcup_{j=1}^{d+1}D_j\bigr]
\)
from \(\beta_i\) to a \(t\)-reachable vertex \(\beta_t\) attaining \(\dist(\beta_i)=d+1\).  Enumerate its vertices as
\(
P = \beta_1,\dots,\beta_r,
\)
with \(\beta_1 = \beta_i\) and \(\beta_r = \beta_t\), so that \(r = |P|\).  Recall that each arc \((\beta_h,\beta_{h+1})\) lies in one of the subsets \(D_1,\dots,D_{d+1}\).  Define
\[
h^* \;=\;\min\Bigl\{\,h\in\{1,\dots,r\}\;\Bigm|\;\text{the subpath from }\beta_h\text{ to }\beta_t\text{ is contained in }G_B[D_1\cup\cdots\cup D_d]\Bigr\},
\]
i.e.\ the first index for which \(\beta_h\) is \((t,\bigcup_{j=1}^d D_j)\)-reachable.  By construction \(\dist(\beta_{h^*})=d\), so by the induction hypothesis at the end of phase \(d\) every vertex of \(B_{h^*}\) have the label \(\id(t)\) or \(\id(s)\#\id(t)\).

Finally, since there is a $(\beta_i,\beta_{h^*})$-directed path in $G_B[D_1\cup\dots\cup D_{d+1}]$, by (4) of\Cref{claim:101}, there exist copy of \(B_i\) at layer \(h_1\) and the copy of \(B_{h^*}\) at layer \(h_2\) that lie in the same connected component of \(\mathcal H^{d+1}\).  Therefore the label \(\id(t)\) or \(\id(s)\#\id(t)\) propagates to the copy of \(B_i\) at layer \(h_1\).  Since each \(v\in B_i\) chooses its final label as the maximum (with respect to \(\preceq\)) over its \(d+1\) copies, it follows that every \(v\in B_i\) is ultimately labeled \(\id(t)\) or \(\id(s)\#\id(t)\), as required. 
\end{proof} 

To conclude, if there exists an $s$-reachable bridge $B_i$ and a $t$-reachable bridge $B_t$ such that there exists a direct path from $\beta_i$ to $\beta_t$ in $G_B$, by Claim~\ref{claim:100} respect $\beta_t$, in phase $j'=dist(\beta_s)$ the vertices of $B_s$ have label either $\id(t)$ or $\id(s)\#\id(t)$. Since at the beginning of the algorithm the vertices in $B_s$ had label $\id(s)$ (because $B_s$ is $s$-reachable), then in phase $j'$ the vertices of $B_s$ have label $\id(s)\#\id(t)$.
\end{proof}

A direct Corollary implied in the proof of \Cref{lemma:correctcontracting} is the following.

\begin{corollary}\label{cor:bsandbt}
    Given a canonical instance $(G,s,t,\mathcal{P}_q)$, let $j^*$ be the value returned by \contractingALG. Then there exists a $s$-reachable bridge $B_s$ and a $t$-reachable bridge $B_t$ such that there exists a $(\beta_s,\beta_t)$-directed path in $G_B[D_1\cup\dots\cup D_{j^*}]$ and the $1$-copies of $B_s$ and $B_t$ that lie in the same connected component $\mathcal C$ of $\mathcal H^{j^*}$.
\end{corollary}

\subsubsection{Recovering the paths from the Bridge Graph}\label{sec:obtainthepath}

Given \Cref{lemma:detectingapath}, we are able to recover an $(s,t)$-augmenting path.

\begin{lemma}\label{prop:gettheaugmentingpath}
    Given a canonical instance \( (G, s,t, \mathcal{P}_q) \) such that $G$ has diameter $D$ and treewidth $k$, suppose that \( \sj \in [q] \) is the value returned by \Cref{lemma:detectingapath}. Then it is possible to compute an augmenting path $P_{aug}$ from $s$ to $t$ in $G_{res}$ in \( \biglo{q\cdot k \cdot D} \) rounds.

\end{lemma}

\begin{proof}
    
Let $\sj\in [q]$ by the number returned by \contractingALG\ of \Cref{lemma:detectingapath}. Then, the algorithm will continue to obtain a distributed representation of an augmenting path in $G_{res}$. Recall that by \Cref{cor:bsandbt}, if \contractingALG\ returns $j^*\in [q]$ then there exist two bridges $B_s$ and $B_t$, such that: 

\begin{enumerate}
    \item $B_s$ is $s$-reachable.
    \item $B_t$ is $t$-reachable.
    \item  There exists a $(\beta_s,\beta_t)$-directed path in $G_B[D_1\cup\dots \cup D_{\sj}]$.
\end{enumerate} 

Moreover, the $\id$ of the bridges \( B_s \) and \( B_t \) can be learned by all the vertices via \broadcast\ through the bridges in $G$, and by $(3)$ and $(4)$ of \Cref{claim:101}, there exists a $(u,v)$-directed path in $G_{res}$ for all the vertices $u\in B_s$ and $v\in B_t$. The algorithm to compute the $(s,t)$-augmenting path is in fact the distributed implementation of the proof of $(3)$ of \Cref{claim:101}. The algorithm will compute a $(u_s,v_t)$-directed path in $G_{res}$ such that $u_s\in B_s$ is a neighbor of $s$ and $v_t\in B_t$ is neighbor of $t$. Then is trivial to add $s$ and $t$ to this $(u_s,v_t)$-directed path and obtain a $(s,t)$-augmenting path. 

To compute the  compute a $(u_s,v_t)$-directed path in $G_{res}$ we proceed as follows. First, through two \broadcast\ in $\overline{G_{res}}$ all the vertices can learn the \id\ of two vertices $u_s\in B_s$ and $v_t\in B_t$, such that $u_s$ is out-neighbor of $s$ and $v_t$ is in-neighbor of $t$ in $G_{res}$. Then, trough $\sj$ calls of \Cref{claim:compuofhcj}, the vertices can compute all the connectivity subgraphs $H_{\mathcal C'}^{j'}$ for all $j'\leq \sj$ and each connected component $\mathcal C'$ of $G_B[D_{j'}]$. This step takes in total $\biglo{\sj\cdot k\cdot D}$ rounds. At this point the vertices of $G$ can simulate $\mathcal{H}^{\sj}$. Therefore, using \Cref{teo:mstcongest} the vertices in $G$ can compute an spanning tree $\mathcal T^{\sj}$ of $\mathcal{H}^{\sj}$ in $\biglo{\sj\cdot k\cdot D}$ rounds ((2) of \Cref{claim:101}). Then, by using \markapath\ in the spanning tree $\mathcal T^{\sj}$ it is possible to compute an arbitrary undirected path $\overline{P}(v^1_s, v^1_t)$ between the copies of $u_s$ and $v_t$ in later $1$ in $\mathcal{T}^{\sj}$. The following claim holds:

\begin{claim}\label{claim:10001}
    There exists a $\delta\in \mathbb N$ and $2\delta+1$ vertices $\beta_1,\dots,\beta_{2\delta+1}$ in $\mathcal H^j$ such that the undirected path $\overline{P}(u_s^1,v_t^1)$ in $\mathcal H^{\sj}$ can be decomposed in subpaths $P_1,\dots, P_\delta, Q_1,\dots,Q_{\delta +1}$ such that 

    \begin{itemize}
        \item Each $Q_i$ is a $1$-copy of an undirected path in some bridge $B_{x_i}$.
        \item Each path $P_i$ is a $\ell_i$-copy of some subpath $P$ of a path $\overrightarrow{P_{j_i}}\in\mathcal{P}_q$ of the canonical instance
        
    \end{itemize}

    And $\overline{P}(u_s,v_t)=Q_1\beta_1R_1\beta_2Q_2\dots R_\delta \beta_{2\delta+1}Q_{\delta +1}$. Moreover, the vertices $\{\beta_i\}_{i\in [2\delta+1]}$ and the subpaths $R_1,\dots, R_\delta, Q_1,\dots,Q_{\delta +1}$  can be detected in $\biglo{\sj \cdot k\cdot D}$ rounds.
\end{claim}

\begin{proof}[Proof of \Cref{claim:10001}]

Enumerate the path $\overline{P}(u^1_s,v^1_t) = \alpha_1\dots \alpha_r$ where $\alpha_1=u^1_s$, $\alpha_r = v^1_t$ and $r=\left|\overline{P}(u^1_s,v^1_t)\right|$. The path $\overline{P}(u^1_s,v^1_t)$ start in vertex $v^1_s$ which belongs to the $1$-copy of the bridge $B_s$. Define $\alpha_{i_1^*}$ as the maximum vertex\footnote{Respect the numbering $\alpha_1,\dots,\alpha_r$} $\alpha_i$ such that the subpath from $\alpha_1$ to $\alpha_i$ is contained in the $1$-copy of $B_s$, since $\alpha_1$ is in the $1$ copy of $B_s$ and $\alpha_r$ is in the $1$-copy of a different bridge $B_t$, then $1\leq i^*\leq r$. Define $Q_1 = \alpha_1,\dots ,\alpha_{i_1^*}$. By definition of $\alpha_{i_1^*}$, $\alpha_{i_1^*+1}$ is not in the $1$-copy of $B_s$, and since the bridges are disconnected (and this property is maintained in $\mathcal H^{\sj}$), then $\alpha_{i_1^*+1}$ is a $\ell_1$-copy (in layer $\ell_1$) of the same vertex in $G_{res}$ such that its $1$-copy is $\alpha_{i_1^*}$. Define then $\beta_1  =\alpha_{i_1^*+1}$. Then, the copies of the bridges in layers different of $1$ are only connected to copies of the paths of $\mathcal P_q$, and therefore $\alpha_{i_1^*+2}$ has to be a $\ell_1$-copy of a vertex $v\in \overrightarrow{P_{j_1}}$ for some $P_{j_1}\in \mathcal P_q$. Define $\alpha_{h_1^*}$ as the maximum vertex after $\alpha_{i_1^*+1}$ such that the subpath from $\alpha_{i_1^*+2}$ to $\alpha_{h_1^*}$ is fully contained in the $\ell_1$ layer of $\mathcal{H}^{\sj}$ and the subpath $\alpha_{i_1^*+2},\dots,\alpha_{h_1^*}$ is a $\ell_1$-copy of a subpath of $\overrightarrow{P_{j_1}}$, since the path $\overline{P}(u_s^1,v_t^1)$ ends in the $1$-copy of the bridge $B_t$, then $h_1^*\leq r$. Define $R_1 = \alpha_{i_1^*+2},\dots,\alpha_{h_1^*}$. By definition of $\mathcal H^{\sj}$ (\Cref{def:hj}) and the connectivity subgraphs (\Cref{def:conngr}), the vertex $\alpha_{h_1^*+1}$ has to be a $\ell_1$-copy of a vertex in some bridge $B_x$, since we don't have edges in $\mathcal{H}^{\sj}$ between copies of vertices in different disjoint subpaths of $\mathcal P_q$. Therefore we can set $\beta_2 = \alpha_{h_1^*+1}$, and then by construction of $\mathcal H^j$, $\alpha_{h_1^*+2}$ is either the $1$-copy of the same vertex in $B_x$, or the $\ell_1$-copy of a vertex in a path $P_{j_2}$. In the former case, we continue computing $Q_2$ as explained before. In the latter case we set $Q_2=\emptyset$, $\beta_3 = \beta_2$ and continue computing $P_2$ in the same way as explained before .Therefore we can define analogously $Q_2,\beta_3, P_2,\beta_4,Q_3,..$ until we arrive to some subpath $Q_{\delta+1}$ such that $\alpha_{i_{\delta+1}^*}= v_t^1$. This proves the existence of the decomposition.

To compute this subpaths, using \Cref{lemma:computingorders} in $\overline{P}(u_s^1,v_t^1)$ the numbering of $\overline{P}(u_s^1,v_t^1)=\alpha_1,\dots,\alpha_r$ can be computed in $\biglo{\sj\cdot k\cdot D}$ rounds. Then, each vertex $\alpha_i$ can verify if its neighbors in $\overline{P}(u_s^1,v_t^1)$ are a $1$-copy of a vertex in a bridge or a disjoint path of $G_{res}$. If the vertex $\alpha_i$ is a $1$-copy of a vertex in a bridge and the neighbor $\alpha_{i-1}$ (analogously, $\alpha_{i+1})$ is the $\ell_{i-1}$-copy of the vertex, then the edge $\{\alpha_i,\alpha_{i-1}\}$ is removed (analogously, the edge $\{\alpha_i,\alpha_{i+1}\}$. Analogously, we can remove all the edges between vertices in different sets as described above. Then, using only the remaining edges of $\overline{P}(u_s^1,v_t^1)$, using \Cref{teo:kmstcongest} in $\mathcal H^j$ we can compute a spanning tree of each connected component of the remaining forest. Each connected component corresponds to either a subpath $Q_i$, a subpath $P_i$, or a unique vertex $\{\beta_i\}$. This last step can also be simulated in $\biglo{\sj\cdot k\cdot D}$ rounds.
\end{proof}

Following \Cref{claim:10001}, in $\biglo{\sj\cdot k\cdot D}$ rounds the decomposition $R_1,\dots,R_\delta,Q_1,\dots, Q_{\delta+1}$ and the vertices $\beta_1,\dots,\beta_{2\delta +1}$ can be computed.

By construction, the vertices $\beta_i$ are just the connection between paths $R_i$ and $Q_i$ or between $Q_i$ and $R_{i+1}$. Therefore, Now we prove that a $(u_s^1,v_t^1)$-directed {\it walk} can be computed from the decomposition $R_1$,$\dots$, $R_\delta$, $Q_1$, $\dots$, $ Q_{\delta+1}$ of $\overline P(u_s^1,v_t^1)$.

\begin{claim}\label{claim:10002}
    Given the undirected path $\overline P(u_s^1,v_t^1)$ and its decomposition $R_1,\dots,R_\delta,Q_1,\dots, Q_{\delta+1}$ of \Cref{claim:10001}, a $(u_s,v_t)$-directed walk $P(u_s,v_t)$ in $G_{res}$ can be computed in $\biglo{\sj\cdot k\cdot D}$ rounds. This means that all the vertices in $P(u_s,v_t)$ learn their predecessor and successor.\footnote{If the walk passes twice or more through the same vertex $v$, the vertex learns all it's predecessor ans successor}
\end{claim}

\begin{proof}[Proof of \Cref{claim:10002}]
    
First, for each $j\in[\delta +1]$, the subpath $Q_j$ is completely contained in some ($1$-copy of a) bridge $B_x$, and as the bridges are strongly connected in $G_{res}$, we can convert $Q_j$ in a directed path of $G_{res}$ by simply converting any edge $\{\alpha_b,\alpha_{b+1}\}$ into the arc $(\alpha_b, \alpha_{b+1})$ for all $b\in\{h_{j-1}^*+1,\dots, i_h^*\}$, and replace the $1$-copy of each vertex by the original vertex in $B_x$ to obtain a $(\alpha_{h_{j-1}^*+1}, \alpha_{i_{j}^*})$-directed path $\overrightarrow{P}(\alpha_{h_{j-1}^*+1}, \alpha_{i_{j}^*})$ in $G_{res}$. This can be done locally in $\bigo{1}$ rounds.

We cannot do the same and simply orient the edges of the subpaths $R_{z}$ of the decomposition, because even we know that $R_z$ is a $1$-copy of a subpath of $\overrightarrow{P_{j_z}}$ with $P_{j_z}\in\mathcal P_q$, we don't know if the path $R_z=\alpha_{i_z^*+1},\dots,\alpha_{h_z^*}$ goes from left to right, or from right to left through $\overrightarrow{P_{j_z}}$. If the path goes from right to left, then we can simply orient the edges as before and this directed path is in $G_{res}$, but if the path goes from left to right, the directed subpath is not in $G_{res}$, since in $G_{res}$ we only have arcs from right to left in each path of $\mathcal P_q$.

This problem is solved as follows. For each $z\in[\delta]$, the undirected path $R_z$ is in layer $\ell_z$ and this undirected path $R_z$ connects a vertex in (a $1$-copy of) a bridge $B_{x_z}$ (the last vertex of $Q_z$, such that $Q_z$ is contained in bridge $B_{x_z}$) to a vertex in ( a $1$-copy of) a bridge $B_{x_z'}$ (the first vertex of $Q_{z+1}$, such that $Q_{z+1}$ is contained in bridge $B_{x_z'}$). By construction of $\mathcal H^{\sj}$, this means that either $(\beta_{x_z},\beta_{x_z'})\in D_{\ell_z}$ or $(\beta_{x_z'},\beta_{x_z})\in D_{\ell_z}$, since the only paths connecting a bridge $B_{x_z}$ with a bridge $B_{x_z'}$ through the layer $\ell_z$ in $\mathcal H^{\sj}$ is a subpath of a path in $\mathcal{P}_q$ satisfying condition 2.1. of \Cref{def:bridgegraph}, and this subpath is added in layer $\ell_z$ of $\mathcal H^{\sj}$. 

\begin{claim}\label{claim:10003}
        For each $z\in [\delta]$, such that $Q_z$ is contained in bridge $B_{x_z}$ and $Q_{z+1}$ is contained in bridge $B_{x_z'}$, the nodes of the subpath $R_z$ can learn if $(\beta_x,\beta_{x'})\in D_{\ell_z}$ or $(\beta_x,\beta_{x'})\in D_{\ell_z}$. This procedure takes $\biglo{\sj \cdot k\cdot D}$ rounds. 
    \end{claim}

    \begin{proof}[Proof of \Cref{claim:10003}]

    For each $z\in [\delta]$, the endpoint of $R_z$ can ask to their neighbors in $Q_z$ and $Q_{z+1}$\footnote{Formally, they ask to their neighbors $\beta_z$ as these vertices can learn the \id\ of the bridges of their respective $1$-copies} the \id\ of its unique parent in $D_{\ell_z}$. Through two \broadcast\ in each $R_z$ the vertices can learn these \id\ and decide if $(\beta_{x_z},\beta_{x_z'})\in D_{\ell_z}$ or $(\beta_{x_z'},\beta_{x_z})\in D_{\ell_z}$. The \broadcast\ can be implemented in $\biglo{\sj\cdot k\cdot D}$ rounds in $G^{\sj}$, and therefore in the same rounds in $G$ (\Cref{claim:101}).      
    \end{proof}

    By \Cref{claim:10003}, in $\biglo{\sj \cdot k\cdot D}$ the vertices in each subpath $R_z$ can learn if $(\beta_{x_z},\beta_{x_z'})\in D_{\ell_z}$ or $(\beta_{x_z},\beta_{x_z'})\in D_{\ell_z}$. Assume that $(\beta_{x_z},\beta_{x_z'})\in D_{\ell_z}$ (the argument if $(\beta_{x_z},\beta_{x_z'})\in D_{\ell_z}$ is analogous and therefore omitted). In this case the subpath \(R_z\) corresponds to an \(\ell_z\)-copy of either the undirected path
\(\alpha_{i^*_z+1} = r^y_x,\dots,\alpha_{h^*_z} = \ell^y_{x'}\) or the undirected path \(\alpha_{i^*_z+1} = \ell^y_{x'},\dots,\alpha_{h^*_z} = r^y_x\) for some path $\overrightarrow{P_y}$ with $P_y\in\mathcal P_q$ satisfying $y=y^{x,x'}$ of \Cref{def:conngr} for $j=\ell_z$. Each vertex in $R_z$ can decide locally in $\bigo{1}$ rounds if the path $R_z$ corresponds to \(\alpha_{i^*_z+1} = r^y_x,\dots,\alpha_{h^*_z} = \ell^y_{x'}\) or to path \(\alpha_{i^*_z+1} = \ell^y_{x'},\dots,\alpha_{h^*_z} = r^y_x\), for example by checking if its successor in $R_z$ have greater or lower position in $\overrightarrow{P_y}$.

If the path $R_z$ is such that $R_z$ is the path $ \alpha_{i^*_z+1} = r^y_{x_z},\dots,\alpha_{h^*_z} = \ell^y_{x_z'}$, then we can simply orient the edges into arcs in $G_{res}$ and replace the $1$-copy of each vertex by the original vertex in $\overrightarrow{P_y}$. This is then a $(r^y_x,\ell^y_{x'})$-directed path $\overrightarrow{P}(r^y_x,\ell^y_{x'})$ in $G_{res}$ (moving from right to left through $\overrightarrow{P_y}$).

Finally, suppose $R_z$ is such that $R_z $ is the path $ \alpha_{i^*_z+1} = \ell^y_{x'},\dots,\alpha_{h^*_z} = r^y_x$, In this case, by condition 2.2. of \Cref{def:bridgegraph}, we know that $r_{x_z}^{\ell_z}\geq \ell_{x_z'}^{\ell_z}$, and therefore, if $\alpha_{i_{z}^*}$ is the last vertex of $Q_z$ and $\alpha_{h_{z}^*+1}$ is the first vertex of $Q_{z+1}$, we mark the arcs 
$(\alpha_{i_{z}^*},r_x^{\ell_z})$ and $(\ell_{x'}^{\ell_z},\alpha_{h_{z}^*+1})$ in $G_{res}$. Here we assume that $\alpha_{i_{z}^*}$ is neighbor of $r_{x_z}^{\ell_z}$ and 
$\alpha_{h_{z}^*+1}$ is neighbor of $\ell_{x_z'}^{\ell_z}$, since if this is not the case, by already established argument is easy to change the directed paths $\overrightarrow{Q}(\alpha_{h_{z}^*-1},\alpha_{i_{z}^*})$ such that the endpoint is a neighbor $r_{x_z}^{\ell_z}$ and the same for the directed path already computed in $Q_{z+1}$.

To simplify notation let us define as $A\subseteq [\delta]$ all the index such that $R_z$ is the path $ \alpha_{i^*_z+1} = \ell^y_{x'},\dots,\alpha_{h^*_z} = r^y_x$ and $(\beta_{x_z},\beta_{x_z'})\in D_{j_{\ell_z}}$. To complete the $(u^s,v^t)$-directed walk in $G_{res}$, we want to mark all the $(r_{x_z}^{\ell_z},\ell_{x_z'}^{\ell_z})$-directed paths $\overrightarrow{P}\left(r_{x_z}^{\ell_z},\ell_{x_z'}^{\ell_z}\right)$ in $G_{res}$ for all $z\in A$. We do this in $\sj$ phases: In each phase $\eta\in [\sj]$,  we aim to mark all the vertices $\omega\in \overrightarrow{P_\eta}$ with $P_\eta\in\mathcal P_q$ such that exists a $z\in A$ that satisfies that $r_{x_z}^{\ell_z}\geq \pi_\eta(\omega)\geq \ell_{x'_z}^{\ell_z}$ and $\ell_z = \eta$. We prove that in each phase we can do this in $\biglo{k\cdot D}$ rounds.

\begin{claim}\label{claim:1000006}
    For each $\eta\in [\sj]$, all the vertices $\omega\in \overrightarrow{P_\eta}$ such that exists a $z\in A$ such that $r_{x_z}^{\ell_z}\geq \pi_\eta(\omega)\geq \ell_{x'_z}^{\ell_z}$ and $\ell_z = \eta$ can be marked in $\biglo{k\cdot D}$ rounds
\end{claim}

\begin{proof}[Proof of \Cref{claim:1000006}]

Let $\rho$ be the number of values $z\in A$ such that $\ell_z = \eta$. For each such index we have a pair of vertices $(\ell_{x'_z}^{\eta},r_{x_z}^{\eta})$ in $R_z$ marked. To simplify notation, we refer to these pairs just as $(\ell_1,r_1),\dots,(\ell_\rho,r_\rho)$. Then the problem is to mark all the vertices $\omega\in \overrightarrow{P_\eta}$ such that exists a $i\in [\rho]$ such that $\pi_\eta(\omega)\in (\ell_i,r_i)$. Define for each $\omega\in P_\eta$ the values $\psi_\omega$ and $\kappa_\omega$ as the number of values $i\in [\rho]$ such that $\ell_i\leq \pi_\eta(\omega)$ and $r_i\leq \pi_\eta(\omega)$, respectively. It is direct to see the following fact

\begin{fact}
    For each vertex $\omega\in P_\eta$, there exists a $i\in [\rho]$ such that $\ell_i\leq\pi_\eta(\omega)\leq r_i$ iff $\psi_\omega> \kappa_\omega$
\end{fact}

Therefore, through two \sumsubsetproblem\ in the path $\overrightarrow{P_\eta}$ in $\overline{G_{res}}$ the vertices can compute the value $\psi_\omega$ and $\kappa_\omega$ and decide if there exists an $i\in [\rho]$ such that $\ell_i\leq \pi_\eta(\omega)\leq r_i$, in which case the vertex mark itself. Two implementations of \sumsubsetproblem\ can be done in $\biglo{k\cdot D}$ rounds in $G$ by \Cref{lemma:somebroadcasts} and \Cref{fact:simofgres}.    
\end{proof}

By \Cref{claim:1000006}, in $\biglo{k\cdot D}$ rounds (in $\overline{G_{res}}$) each vertex $\omega$ in each path $\overrightarrow{P_\eta}$ with $P_\eta\in \mathcal P_q$  can learn if exists a $z\in A$ such that $\ell_z=\eta$, the $(r_{x_z}^\eta, \ell_{x'_z}^\eta)$-directed path passes through $\omega$. Finally, mark all the arcs in $G_{res}$ between two marked vertices in $P_\eta$. Let's say that the set of arcs marked in each path $\overline{P_\eta}$ in this step is $\mathcal E_\eta$.

Finally, we obtain a $(u_s, v_t)$-directed path in $G_{res}$ by joining the following directed paths in $G_{res}$: (1) The $(\alpha_{h_{j-1}^*+1}, \alpha_{i_{j}^*})$-directed path $\overrightarrow{P}(\alpha_{h_{j-1}^*+1}, \alpha_{i_{j}^*})$ for all $Q_j$ with $j\in [\delta +1]$, (2) the $(r^y_x,\ell^y_{x'})$-directed path $\overrightarrow{P}(r^y_x,\ell^y_{x'})$ for all $P_z$ such that $P_z$ is the path $ \alpha_{i^*_z+1} = r^y_{x_z},\dots,\alpha_{h^*_z} = \ell^y_{x_z'}$ and $(\beta_{x_z},\beta_{x_z'})\in D_{\ell_z}$, (3) The arcs the arcs 
$(\alpha_{i_{z}^*},r_x^{\ell_z})$ and $(\ell_{x'}^{\ell_z},\alpha_{h_{z}^*+1})$ for all  $R_z$ is such that $P_z $ is the path $ \alpha_{i^*_z+1} = \ell^y_{x'},\dots,\alpha_{h^*_z} = r^y_x$ and $(\beta_{x_z},\beta_{x_z'})\in D_{\ell_z}$, and (4) The arcs in $\mathcal E_\eta$ for all $\eta\in [\sj]$. 

In this construction, $(1), (2)$ and $(3)$ correspond to simply orient the undirected path $\overline{P}(u_s^1,v_t^1)$ and the edges of $(4)$ correspond to replace the undirected paths $R_z$ of the decomposition whose orientation is not in $G_{res}$ by a directed path in $G_{res}$.
\end{proof}

    Given \Cref{claim:10002} we can compute a $(u_s,v_t)$-directed walk in $\biglo{\sj\cdot k\cdot D}$ rounds. Then by adding the arcs $(s,u_s)$ and $(v_t,t)$ we obtain a $(s,t)$-directed walk $P(s,t)$ in $G_{res}$. Now we explain how we delete cycles in $P(s,t)$ to obtain a $(s,t)$-augmenting path in $G_{res}$. Notice that by construction of $P(s,t)$, the walk never passes twice by a vertex in a bridge, so if $P(s,t)$ has cycles, is because it passes twice or more times by a same vertex in a disjoint path of $\mathcal P_q$.

    Before to explain the distributed implementation, we describe the main idea: Let $v$ be a vertex in a path $\overrightarrow{P_y}$ with $P_y\in \mathcal{P}_q$ that appears more than one time in $P(s,t)$, intuitively, if $v$ appears $c$ times in $P(s,t)$, and each appearance $i\in [c]$ has position $v_i$, then between any consecutive appearance $v_i,v_{i+1}$ we have a cycle in $P(s,t)$, and therefore the vertices in this cycle can be deleted from $P(s,t)$. For every vertex $v$ in some path of $\mathcal{P}_q\cap P(s,t)$, define $x(v)$ as the maximum position of $v$ in an appearance of $v$ in $P(s,t)$ and $y(v)$ as the minimum position of $v$ in an appearance of $v$ in $P(s,t)$. Now, let's consider a subpath $P'$ of $P_y$ such that this subpath is a subgraph of $P(s,t)$ and some vertices of $P'$ appear more than once in $P(s,t)$. In this subpath, consider the vertices $v_1,v_2\in V(P')$ such that $v_1$ has the minimum label $y(v)$ and $v_2$ has the maximum label $x(v)$ between all the vertices $v\in V(P')$. By construction $v_1$, the predecessor of $v_1$ in $P(s,t)$ has to be some vertex $\omega_1$ in some bridge $B_i$ and the successor of $v_2$ in $P(s,t)$ has to be some vertex $\omega_2$ in a bridge $B_x$. Then, all the vertices with positions lower than $v_2$ or greater than $v_1$ in $P'$ mark themselves as outside the augmenting walk $P(s,t)$, as well all the vertices in a bridge $B_i$ connected to some of these deleted vertices. This deletion of vertices deletes all the directed cycles of $P(s,t)$ and then we obtain the desired $(s,t)$-augmenting path. See \Cref{fig:duplicated} for a graphical example of this deletion.

  \begin{figure}[H]
            \centering
            \scalebox{0.5}{\input{img/newfig5}}
            \caption{An $(s,t)$-augmenting walk $P_{aug}\colon s\to B_s\to \omega_4\to\omega_3\to\omega_2\to\omega_1\to B_x\to \omega_5\to\omega_4\to\omega_3\to\omega_2\to B_t\to t$ in a canonical instance $G(s,t,\{P_1\})$ with only one path from $s$ to $t$. In $P_1$, as $\omega_4$ is the first vertex in the augmenting path and $\omega_2$ is the last one, then $\omega_1$ and $\omega_5$ mark themselves out of the augmenting walk. As the neighbors of $B_x$ are marked out of the path, the vertices in $B_x$ also mark them out of the path, Finally obtaining to the right the $(s,t)$-augmenting path.}
            \label{fig:duplicated}
        \end{figure}

    To implement the idea explained above, we need to compute an numbering of the $(s,t)$-directed walk $P(s,t)$. While \Cref{lemma:computingorders} gives an algorithm to compute an ordering in a (undirected) path, it is not trivial to implement in the (undirected) walk $\overline{P(s,t)}$. We prove that a partial ordering in this case can be computed. To avoid heavy notation, let $\mathcal L$ be the collection of all the vertices $r_{x_z}^{\ell_z}$ and $\ell_{x'_z}^{\ell_z}$ of $P(s,t)$ for some $P_z$ such that $P_z $ is the path $ \alpha_{i^*_z+1} = \ell^y_{x'},\dots,\alpha_{h^*_z} = r^y_x$ and $(\beta_{x_z},\beta_{x_z'})\in D_{\ell_z}$.

    \begin{claim}\label{claim:100005}
        In $\biglo{\sj\cdot k \cdot D}$ rounds, all the vertices of $\mathcal L$ can learn the position of all its appearances in $P(s,t)$.
    \end{claim}

    \begin{proof}[Proof of \Cref{claim:100005}]
        Consider the undirected path $\overline{P}(u_s^1,v_t^1)$ in $\mathcal H^{\sj}$ computed at the beginning of the algorithm. Recall that $P(s,t)$ is the oriented version of $\overline{P}(u_s^1,v_t^1)$ (using the original vertices in $G_{res}$ instead of the copies of $\mathcal H^{\sj}$) except in the subpaths $P_z$ is such that $P_z $ is the path $ \alpha_{i^*_z+1} = \ell^y_{x'},\dots,\alpha_{h^*_z} = r^y_x$ and $(\beta_{x_z},\beta_{x_z'})\in D_{\ell_z}$. In this case we 'replace' $P_z$ by the $(r_{x_z}^{\ell_z},\ell_{x_z'}^{\ell_z})$-directed path in $G_{res}$. Let $\mathcal D$ be the collection of all the subpaths $P_z$ is such that $P_z $ is the path $ \alpha_{i^*_z+1} = \ell^y_{x'},\dots,\alpha_{h^*_z} = r^y_x$ and $(\beta_{x_z},\beta_{x_z'})\in D_{\ell_z}$. For each vertex $\omega\in \overline{P}(u_s^1,v_t^1)$, define the following labels $x(\omega)$ 

        \begin{itemize}
            \item If $\omega$ is the last vertex of a subset $Q_z$, then $x(v) = 1$ if $P_{z}\notin \mathcal D$ and $x(v) = 1 + (r^{\ell_z}_{x_z}-\ell^{\ell_z}_{x'}+1) $ if $P_{z}\in \mathcal D$. In the last case, $x(v) = 1 + (r^{\ell_z}_{x_z}-\ell^{\ell_z}_{x_z'}+1) $ counts the vertex $\omega$ plus all the vertices in the $(r^{\ell_z}_{x_z},\ell^{\ell_z}_{x_z'})$-directed path added to $P(s,t)$ instead of $P_z$.
            \item If $\omega$ is in some path $P_z\in\mathcal D$, then $x(\omega)=0$. These vertices are not included in $P(s,t)$ and therefore we don't count them.

            \item In any other case, $x(\omega)=1$.
        \end{itemize}

        Notice that by previous computations, all the vertices in $\overline{P}(u_s^1,v_t^1)$ can compute their label locally without extra communication. Then, through \sumsubsetproblem\ in $\overline{P}(u_s^1,v_t^1)$ (considering it as a tree rooted in $v_t^1$), each vertex $\omega$ can compute the sum of labels from $u_s^1$ to $\omega$ in $\biglo{\sj\cdot k\cdot D}$ rounds. For all the vertices that do not belong to a path in $\mathcal D$, this sum corresponds exactly to their position in $P(s,t)$. Since all the vertices that do not belong to a path in $\mathcal D$ appear once in $P(s,t)$, then this position is unique. In this point, all the vertices of $P(s,t)$ that are neighbors of a vertex in $\mathcal L$ can inform to this vertex their position in $P(s,t)$, since this position is the position of $\omega$ plus or minus one (depending if the neighbor in $\mathcal L$ is a predecessor or successor of $\omega$ in $P(s,t)$). Then all the vertices in $\mathcal L$ learn the position of all of its appearances in $P(s,t)$
    \end{proof}
    Then, by \Cref{claim:100005}, in $\biglo{\sj\cdot k\cdot D}$ rounds, each vertex in $v\in\mathcal L$ can compute their label $y(v)$ and $x(v)$ as described above, and the rest of the vertices in each path of $\mathcal{P}_q$ sets its labels as $y(v) = 0$ and $x(v) = n^2$\footnote{Or any value greater than the length of $P(s,t)$.}. Distributively, this deletion of cycles in $P(s,t)$ can proceed as follows: Each subpath $R$ of $P_j\cap P(s,t)$ can be detected in $\biglo{k\cdot D}$ analogously as in \Cref{lemm:detectbridges} (by using only the edges that participate in $P(s,t)$ in each path $P_j$). Then, through a \maxproblem\ of the labels $y(v)$ and a \minproblem\ of the labels $x(v)$ in $\overline{G_{res}}$, the vertices $v$ in each subpath $R$ of $P_j\cap P(s,t)$ can learn the maximum and minimum value $\Tilde{y}(v)$ and the minimum $\Tilde{x}(v)$ in $R$, along with the position $\id$ and the position of the nodes with such positions in $P_j$. If a vertex $v\in R$ is in $P(s,t)$ but its position in $P_j$ is lower than the position of the node with label $\Tilde{y}(v)$ or greater then the position of the node in $P_j$ with label $\Tilde{x}(v)$, mark itself out of the walk $P(s,t)$. By \Cref{lemma:maxandminprefix}, this can be done in $\biglo{\scq{G}}$. Finally each vertex marked out of the walk that has a neighbor in $P(s,t)$ in some bridge $B_i$ inform to this neighbor that it is out the path and through \broadcast\ all the vertices in $B_i$ also mark themselves out of the path. The remaining vertices still marked in $P(s,t)$ form our augmenting path that can be detected in $\biglo{k\cdot D}$ rounds by using \Cref{teo:mstcongest} using only the vertices and edges still marked in $P(s,t)$. 

    Every computation described in the algorithm takes at most $\biglo{j^*\cdot k \cdot D}$ rounds and $j^*\leq q$, obtaining a total round complexity of $\biglo{q\cdot k\cdot D}$ rounds.

\end{proof}

    \subsubsection{Computing $q+1$ Disjoint Paths and Proof of \Cref{thm:congseparator}}

    Given \Cref{prop:gettheaugmentingpath}, it is possible to recover the $q+1$ disjoint paths.

    \begin{lemma}\label{lemma:getthepath}
        Given a canonical instance $(G,s,t,\mathcal{P}_q)$ along with the $(s,t)$-augmenting path $P_{aug}$ of $G_{res}$ given by \Cref{prop:gettheaugmentingpath}, a new collection of $q+1$ vertex disjoint  $(s,t)$-paths in $G$ can be computed in $\bigo{1}$ rounds.
    \end{lemma}
\begin{proof}
    
Once the augmenting path $P_{aug}$ in $G_{res}$ has been computed in \Cref{prop:gettheaugmentingpath}, it is direct to transform this directed path in $G_{res}$ into a $(s,t)$-undirected path in $G$ by contracting all the vertices $v_{in}, v_{out}$ of $P_{aug}$ and considering the undirected edges in $G$ of each arc of $G_{res}$. This can be done locally without communication by each vertex in $G$. See \Cref{fig:bridgegraph} for a graphical example.

\begin{figure}
    \centering
    \scalebox{0.5}{\input{img/paug}}
   \caption{(a) red arcs represents the augmenting path $P_{aug}$ in $G_{res}$. (b) red edges represents the augmenting path in $G$.}
    \label{fig:paug}
\end{figure}

For the rest of the proof, we abuse notation and refer to $P_{aug}$ as the $(s,t)$-undirected path in $G$. At this point we have $q+1$ paths distributively computed in $G$: The $q$ disjoint paths of $\mathcal{P}_q$ and $P_{aug}$.  To obtain $q+1$ disjoint $(s,t)$-paths of $G$, the vertices in these $q+1$ paths will either update their predecessors and successors, or they will mark themselves out of the paths. This procedure is done as follows: Each vertex in one of the input paths $P_y\in\mathcal{P}_q$ that does not participate in $P_{aug}$ will set its predecessor and successor as the same vertices that are its predecessor and successor in $P_y$. Analogously the vertices in $P_{aug}$ that do not belong to any disjoint path of $\mathcal P_q$ (i.e., the vertices of $P_{aug}$ in the bridges) also set as their predecessor and successor the corresponding predecessor and successor in $P_{aug}$.

Now, let $W\subseteq V$ be all the vertices of  graph $G$ that belong to a path $P_y$ for some $y\in[q]$ and are also part of the augmenting path $P_{aug}$. A vertex $v\in W$ is an extreme vertex if $v$ has a neighbor in $P_{aug}$ (either its predecessor or successor) that does not belong to a path of $\mathcal{P}_q$. Notice that vertex can decide locally if they belong to $W$ and/or are an extreme vertex. Every vertex in $W$ that is not an extreme vertex will mark itself outside the paths. Every vertex $v\in W$ that is an extreme vertex and belongs to the path $P_y\in\mathcal P_q$ will set its predecessor and successor as follows: If $v$ is an extreme vertex because in predecessor (analogously, its successor) in $P_{aug}$ is not on a path of $\mathcal P_q$, it will set its predecessor as its predecessor in $P_{aug}$ (analogously, $P_y$) and its successor as its successor in $P_y$ (analogously, in $P_{aug}$).

After these modifications, by using \Cref{teo:kmstcongest} in the subgraphs induced by the vertexs still inside the path and edges between predecessor and successor in these paths, a new collection of $q+1$ disjoint paths can be detected. See \Cref{fig:newq1paths} for a graphical example.
\end{proof}

\begin{figure}[H]
    \centering
    \scalebox{0.5}{\input{img/newpaths}}
    \caption{To the left a canonical instance with two disjoint paths $P_1=s,v_2,\dots,v_5,t$, $P_2=s,u_2,u_3,u_4,t$ and a augmenting path $P_{aug}$ marked with red. To the right the $3$ $(s,t)$-disjoint path marked with red edges.}
    \label{fig:newq1paths}
\end{figure}

We are now able to conclude with the proofs of \Cref{theo:disjointpaths}, and eventually the main result of this section, \Cref{thm:congseparator}.

\begin{proof}[Proof of \Cref{theo:disjointpaths}.]

Given a canonical instance $(G,s,t,\mathcal{P}_q)$, we first apply the algorithm described in \Cref{lemma:bridgegraph} to construct a distributed representation of the associated bridge graph $G_B$. This step can be completed in $\biglo{\scq{G}}$ rounds.

Next, using \Cref{lemma:detectingapath}, we determine whether an augmenting path exists within $\biglo{q^2\cdot k\cdot D}$ rounds. If no augmenting path is found, all vertices become aware of this, and the algorithm terminates.

If an augmenting path does exist, we invoke \Cref{lemma:getthepath} to obtain $q+1$ disjoint paths, achieving this in $\biglo{q \cdot k \cdot D}$ rounds. Moreover, recalling \Cref{prop:sqofboundedtrwwidth}, if graph $G$ has treewidth $k$ and diameter $D$, the total round complexity of our algorithm is $\biglo{q^2 \cdot k \cdot D}$.
\end{proof}

We proved that if an augmenting path exists, we detect it and compute a new set of $q+1$ vertex-disjoint $(s,t)$-paths. In order to prove~\Cref{thm:congseparator}, it remains to deal with the case when no augmenting path exists.

\begin{proof}[Proof of \Cref{thm:congseparator}] 
Given a canonical instance $(G,s,t,\mathcal{P}_q)$, we can compute the first path by simply building a BFS of $G$ and marking an $(s,t)$-path in the BFS. For the rest of the disjoint paths, by using \Cref{theo:disjointpaths}, a new collection of $2$ disjoint paths can be computed, if it exists. By applying \Cref{theo:disjointpaths} at most $q$ times, a new collection of $q+1$ disjoint paths, if exists, can be computed. If this $q+1$ disjoint paths don't exist, then this is informed by \Cref{theo:disjointpaths} and this information is conveyed to the vertices at the end of \contractingALG. This process takes at most $\biglo{q^3\cdot k \cdot D}$ rounds, since we use at most $\bigo{q}$ times the algorithm of \Cref{theo:disjointpaths}. 

If $q+1$ disjoint paths don't exit, this was informed by \contractingALG\ at the end of one of the at most $q$ calls to \Cref{theo:disjointpaths}. In that application of \contractingALG, the vertices have label $\id(s)$, $\id(t)$ or none (but there is no vertex with label $\id(s)\#\id(t)$), and all the nodes in the same bridge have the same label. Let $\mathcal B_s$ denote all bridges $B_i$ such that the vertices of $B_i$ received $\id(s)$ as a label at the end of \contractingALG, when it was decided that there is no $q+1$ disjoint paths. We know from \Cref{lemma:correctcontracting} that these bridges correspond exactly to the bridges $B_i$ such that there exists a $(\beta_s,\beta_i)$-directed path in $G_B$, and these vertices $v$ of the bridges $B_i$ are the vertices of $G_{res}$ such that there exists an $(s,v)$-directed path in $G_{res}$.

Now, for each path $P_j\in\mathcal{P}_q$, let $w'_j$ be the rightmost $r_j^i$ over all bridges $B_i\in\mathcal B_s$. Necessarily,$w_j' = s$ or $w_j'=v_{in}$ for some internal node $v$ on the vertex-disjoint path in $G_{res}$ corresponding to $P_j$. In the former case let $w_j\in V(G)$ be the vertex on $P_j$ adjacent to $s$, and in the latter case let $\omega_j$ be the vertex $v$ (i.e. $\omega_j$ is the vertex $v\in V(G)$ such that $\omega_j' = v_{in}$). 

    \begin{claim}\label{claim:99}
        The set $\{\omega_1,\dots,\omega_q\}$ is an $(s,t)$ separator set.
        \end{claim}
    \begin{proof}[Proof of \Cref{claim:99}]
       Suppose for contradiction that there is a simple $s$-$t$ path $P$ in $G-\{w_1,\dots, w_r\}$. Consider the subsequence $v^1,\dots, v^k$ of vertices in $P$ that are also in $\bigcup_jP_j$. Consider the numbering as the order of appearance of the vertices in $P$ from $s$ to $t$.

At some point, we must have two vertices $v,v'\in\bigcup_jP_j$ consecutive on this subsequence (i.e. $v = v^i$ and $v' = v^{i+1}$ for some $i\in [k]$) such that for the paths $P_j,P_{j'}$ containing $v$ and $v'$ respectively, we have $v<w_j$ and $v'>w_{j'}$. The vertices $v,v'$ cannot be adjacent in $P$ (since the path can only connect through bridges), so there must be vertices inside a single bridge $B_i$ in between the occurrences of $v$ and $v'$ on $P$. This bridge $B_i$ satisfies $l_i^j<w_j$ and $r_i^{j'}>w_{j'}$. Since $w_j=r_{i'}^j$ for some $B_{i'}\in\mathcal  B_s$, vertex $w_j$ is reachable from $s$. Therefore, vertex $l_i^j$ is also reachable from $s$, and so is $B_i$, which means that $B_{i'}$ and $B_i$ are strongly connected in $G_{res}$. In particular, $B_i\in\mathcal  B_s$, so $w_{j'}\ge r_i^{j'}$ by definition of $w_{j'}$, contradicting the assumption that $r_i^{j'}>w_{j'}$.
    \end{proof}
        The computation of these vertices $\omega_y$ can be done in $\biglo{\scq{G}}$ rounds by using the \maxproblem\ of \Cref{lemma:somebroadcasts} in each disjoint path $P_j$ (using only the internal vertices of each $P_j$), where each vertex $v\in P_j$ uses as label $x(v) = r_j^i$ if $v$ is neighbor of a vertex in a bridge $B_i\in \mathcal B_s$ and $x(v) = 0 $ otherwise. Therefore, after $\biglo{\scq{G}}$ rounds an $(s,t)$ separator set has been computed. Combining \Cref{theo:disjointpaths} and \Cref{claim:99} the proof of \Cref{thm:congseparator} is direct. 
\end{proof}

\subsection{Disjoint Paths and Separator Sets in Multiple Subsets}
\label{ss:multiplesubsets}
The application of \Cref{thm:congseparator} will primarily focus on computing separator sets. However, unlike \Cref{thm:congseparator}, the separator sets we are interested in will not separate just two vertices $s,t$, but rather two subsets of vertices $A,B \subseteq V$. Furthermore, we aim to compute separators between an arbitrary number of pairs of subsets $\{A_1,B_1\}, \dots,\{A_i, B_i\}$ simultaneously, for any $i \in [n]$.

To achieve this, we will generalize the algorithm presented in \Cref{sec:obtainthepath} to address the specific problem discussed in the next section, which involves computing a tree decomposition of a graph $G$. 

\begin{lemma}[Generalization of \Cref{thm:congseparator}]
    \label{lemma:abseparator}
    There exists a deterministic algorithm in the \CONGEST\ model, such that given a graph $G$ with diameter $D$ and treewidth at most $k$, two sets $U, X\subseteq V$, two disjoint sets $A,B\subseteq N[U]$ with no edge between them, such that each vertex $v\in V$ knows if belongs to these sets, and a parameter $q\in\N_+$, it compute in $\biglo{q^3\cdot k\cdot D}$ rounds either:
    \begin{itemize}
        \item A set of $q+1$ disjoint paths between $A$ and $B$, such that all the inner vertices belongs to $U \setminus X$, or
        \item An $(A,B)$ separator set $S\subseteq U \setminus X$ in graph $G[U \cup A \cup B \setminus X]$, of size at most $q$.
    \end{itemize}
\end{lemma}

\begin{proof}
    To generalize the algorithm presented for \qdisjointpaths, first we run the algorithm of \Cref{theo:disjointpaths} in $$G'=G[U \cup A \cup B \setminus X].$$ In $\bigo{1}$ rounds, vertices can determine which edges and neighbors belong to this subgraph and execute the algorithm considering only those vertices. As all the computations of \Cref{theo:disjointpaths} are either part-wise aggregation problems or local computations. Running in a subgraph is not a problem, because every computation of \Cref{theo:disjointpaths} will be through part-wise aggregation problems in vertex disjoint subsets of $G$, and therefore the same round complexity holds.

    Now, consider the graph $G'$ plus two (fake) vertices $s$ and $t$, such that $s$ is connected to all the vertices in $A$ and $t$ is connected to all the vertices in $B$. If we refer to $G_{sim}$ to the graph $G'$ plus these vertices $s$ and $t$, then the following holds. By definition of $G_{sim}$, there exist $q+1$ disjoint paths between $A$ and $B$ with internal vertices in $U \setminus X$ iff there exists $q+1$ disjoint paths between $s$ and $t$ in $G_{sim}$, and therefore if the algorithm of \qdisjointpaths\ in instance $G_{sim}$ gives $q+1$ disjoint paths between $s$ and $t$, then the internal vertices of these $q+1$ paths correspond exactly to $q+1$ paths between $A$ and $B$ with internal vertices in $U \setminus X$, and if the algorithm of \qdisjointpaths\ in instance $G_{sim}$ gives a $(s,t)$ separator set $S$ , then this set $S$ is a $(A,,B)$-separator set.

    Therefore, to conclude, it remains to show how to simulate the algorithm of \qdisjointpaths\ in $G_{sim}$  within $\biglo{q^2\cdot k \cdot D}$ rounds.

    To this end, we observe that the algorithm for \qdisjointpaths\ uses the vertices $s$ and $t$ in only two parts:
(i) to compute the order $\pi_j$ of each path, and
(ii) to detect the bridges that are $s$-reachable and $t$-reachable. Informally, vertices of $A$ will simulate $s$ and those of $B$ will simulate $t$.

    To simulate (i), it is a straightforward modification to ensure that all internal vertices obtain the correct order without relying on vertices $s$ and $t$.
    To simulate (ii), we note that by construction of $G_{sim}$, since $s$ and $t$ are connected only to vertices in $A$ and $B$, respectively, then given a canonical instance $(G_{\text{sim}}, s, t, \mathcal{P}_q)$, a bridge $B_x$ in $G_{\text{sim}} - \cup_{j \in [q]} P_j$ is $s$-reachable if and only if $B_x$ contains a vertex from $A$. Similarly, a bridge $B_x$ is $t$-reachable if and only if it contains a vertex from $B$. Hence, $s$-reachable and $t$-reachable bridges can be detected using only the vertices in $A$ and $B$, respectively.

    Both (i) and (ii) can be simulated without any additional round complexity, and the rest of the algorithm is executed entirely on $G_{\text{sim}} - \{s,t\}$. Therefore, the graph $G'$ can simulate \Cref{thm:congseparator} on $G_{\text{sim}}$ in $\biglo{q^3 \cdot k\cdot D}$ rounds.
\end{proof}

In fact, \Cref{lemma:abseparator} can be implemented by running in multiple subsets in parallel.

\begin{lemma}
    \label{lemma:mabseparator}
    There exists a deterministic algorithm in the \CONGEST\ model, such that given a graph $G$ with diameter $D$ and treewidth at most $k$, a parameter $q\in\N$, and $i\in[n]$ 4-tuples $\{(U_j,X_j,A_j,B_j)\}_{j\in[i]}$ such that the sets $U_j$ are connected and pairwise disjoint, and for each $j\in[i]$,  $U_j\subseteq V$, $A_j,B_j, X_j\subseteq N[U]$, and sets $A_j$ and $B_j$ are disjoined and no edge connects them, the algorithm computes for each tuple $(U_j,X_j,A_j,B_j)$ either
    \begin{itemize}
        \item A set of $q+1$ disjoint paths between $A_j$ and $B_j$, such that all the inner vertices belong to $U_j \setminus X_j$, or
        \item An $(A_j,B_j)$ separator set $S\subseteq U_j \setminus X_j$ in graph $G[U_j \cup A_j \cup B_j  \setminus X_j]$ with size at most $q$.
    \end{itemize}

    The total round complexity is $\biglo{q^3\cdot k \cdot D}$.
\end{lemma}

\begin{proof}
    To prove the Lemma it is enough to prove that we can run the algorithm of \Cref{lemma:abseparator} in parallel in each tuple $(U_j, X_j, A_j, B_j)$.

    In each instance $(U_j,X_j,A_j,B_j)$, the algorithm of \Cref{lemma:abseparator} runs in $G'_j = G[U_j \cup A_j \cup B_j  \setminus X_j]$. Then, in each instance, the algorithm is run in vertices of $U_j$, $A_j$, and $B_j$. As the sets $U_j$ are pairwise disjoint, then each vertex in $U_i$ can run the algorithm in parallel with the rest of the instances. Although the sets $A_j, B_j$ are disjoint between them for each fixed $j\in[i]$, there may be different $j_1, j_2 \in [i]$ such that the sets $A_{j_1} \cup B_{j_1}$ and $A_{j_1} \cup B_{j_1}$ intersect.

    Therefore, a vertex $v \in A_j$ may participate in up to $\Theta(n)$ parallel invocations of the algorithm from \Cref{lemma:abseparator}, since it may belong to up to $n$ sets $A_f$ or $B_f$ from $\{A_f, B_f\}_{f\in[n]}$.  Now, in each instance $(U_f, X_f, A_f, B_f)$, the algorithm is executed over $G'_f$, in which vertex $v$ is only adjacent to vertices in $U_f$. Since the sets $U_f$ are disjoint from each other, vertex $v$ can run the algorithm from \Cref{lemma:abseparator} in parallel across all tuples in which it participates. In each round, $v$ can receive messages from its neighbors in each $U_f$, compute the required information for each tuple, and send the corresponding message to each neighbor in some $U_f$.  Because these sets $U_f$ are pairwise disjoint, $v$ uses disjoint subsets of incident edges to communicate with different instances. Therefore no congestion issues arise, thus the complexity is the same as in \Cref{lemma:abseparator}. 
\end{proof}

\section{Distributed construction of a tree decomposition}\label{se:treewidth}

Let us restate Theorem~\ref{thm:twdistrib} with more details. 

\begin{theorem}\label{th:treedec}
    There is an $\biglo{D}$-rounds \CONGEST{} algorithm that takes as input a constant $k$ and correctly outputs one of the following:
    \begin{itemize}
        \item The algorithm accepts and produces a tree-decomposition of the input network, of width at most $7k+4$ and of depth $O(\log n)$, or
        \item The algorithm rejects, in which case $\tw(G) > k$.
    \end{itemize}
\end{theorem}

We precise some technical features of the tree decomposition produced by the algorithm, that will be useful for solving decision and optimization problems in Section~\ref{se:MSO}, and detail how the output is distributed among the nodes. 

The algorithm produces a rooted tree decomposition $(T=(I,F),\{B_u\}_{u \in I})$. Denote by $r$ the root of the tree,  by $T_u$ the subtree of $T$ rooted at $u$, and by $V_u$ the union of all bags $B_v$, for all vertices $v$ of $T_u$. Let $p(u)$ be the parent of $u$ in the tree. 

\begin{lemma}\label{le:twtech}
    The tree decomposition $(T=(I,F),\{B_u\}{u \in I})$ produced by the algorithm satisfies the following properties:
    \begin{itemize}
        \item At its root $r$, $B_r = \emptyset$ and $V_r = V$.
        \item For each node $u$ of $T$ different from the root, the graph $G[V_u \setminus B_{p(u)}]$ is connected.
        \item For each node $u$ different from the root, there is at least one vertex in $B_u \setminus B_{p(u)}$.
        \item Each vertex $x$ of the graph knows its depth in the tree $T$, i.e., the minimum depth of a node $u$ such that $x \in B_u$. Moreover, $x$ knows the sequence of bags $B_u, B_{p(u)},\dots, B_r$ on the path from $u$ to the root $r$, and the graphs induced in $G$ by each of these bags.
    \end{itemize}
\end{lemma}

The algorithm proceeds in $\bigo{\log n}$ phases, and we shall see that it satisfies both Theorem~\ref{th:treedec} and the technicalities of Lemma~\ref{le:twtech}. At each phase $i$, it treats nodes at depth $2i$ in the decomposition tree, and runs over instances $(U,X)$, where $U \subseteq V$ is a connected set, $X = N(U)$ and $X \cup U$ corresponds to set $V_u$ of some node $u$ of the decomposition. The border set $X$ corresponds to the intersection of $V_u$ with the bag of the parent node. At depth $i=0$, we start with instance $(V,\emptyset)$. 

Then we alternate an even and an odd step at each phase. Even steps split $U$ into smaller components, through a separator $S$ of size at most $k+1$, ensuring that each component of $G[U] - S$ has size at most $10 |U|/12$. This will ensure that the algorithm ends after a logarithmic number of phases. At such an even step, we create a new bag $X \cup S$, and then we need to recursively consider instances $(U',X')$ over all components $U'$ of $G[U] - S$ and sets $X' = N(U')$. Note that $X'$ can be larger that $X$, by an additive term of $k+1$. This is why we need odd steps, that will find a separator $S'$ of $G[U' \cup X']$, splitting $X'$ into roughly balanced parts through some separator $S'$. Then we create a bag $X' \cup S'$ and recurs on components $U''$ of $G[U' \cup X'] - S'$ and their neighborhoods $X''$. This will insure that in the next recursion, the new border set $X''$ will be small enough.  After each phase, the new instances $(U'',X'')$ will satisfy that $|U''| \leq 10 |U|/12$ and $|X''|\leq 6k+4$. At phase $i$ we add new nodes to the decomposition tree, of depth $2i+1$ and $2i+2$, with bags containing at most $7k+5$ vertices.

We heavily use balanced separators of graphs of treewidth $k$, so let us define them. Given a graph $G=(V,E)$, a constant $\alpha\in(0,1)$ and a set $W\subseteq V$, a set $S\subseteq V$ is called a $(W,\alpha)$-balanced separator of $G$ there is a partition $(A,B)$ of $V \setminus S$ such that $S$ separates $A$ and $B$ in $G$, and $|A\cap W|, |B\cap W|\leq\alpha |S|$. The following fact is known, see e.g.~\cite{Reed92}.

 \begin{proposition}\label{prop:sepbt}
     Let $G=(V,E)$ be a graph of treewidth $k$, then for every $X\subseteq V$ there exists a $(X,2/3)$-balanced separator of $G$ of size $k+1$.
 \end{proposition}

\subsection{Even Steps}\label{subsec:even}

We are given a pair $(U,X)$ where $U$ is a connected subset of $G$ and $X = N(U)$ is of size at most $6k+4$. The purpose is to find $10/12$-balanced separator $S$ of $U$, of size at most $k+1$ (if such a set does not exist, we can directly conclude that $\tw(G)>k$ by Proposition~\ref{prop:sepbt}).

\begin{lemma}\label{lemma:sepofU}
    Let $G$ be a graph of treewidth at most $k$ and diameter $D$.
    There exists an $\biglo{k^{O(k)}D}$-rounds deterministic algorithm in the \CONGEST\ model such that, given a connected subset $U$ of $G$, computes a $10/12$-balanced separator $S$ of $U$, of size at most $k+1$. 

    Moreover, the same computation can be preformed within the same number of rounds in parallel, on disjoint connected sets $U_1, \dots, U_{\ell}$.
\end{lemma}

The whole subsection is devoted to the proof of this lemma. We focus on the description of a unique instance $U$. As we shall see, everything is based on part-wise aggregation functions on an arbitrary spanning tree of $G[U]$ and on computing separators of size $O(k)$ as in Theorem~\ref{thm:congseparator} and its parallel version~\ref{lemma:mabseparator}, so the parallelization on multiple subsets is immediate.
For obtaining the separators we use a technique of~\cite{Lagergren96, Reed92}. It consists in taking an arbitrary spanning tree $A$ of $G[U]$ and then computing a $b$-splitter of $A$, for $b = |U|/(12k+12)$, defined as follows:

\begin{definition}[$b$-splitter]
    Given a spanning rooted tree $A$ of $G[U]$, a $b$-splitter is a set $R \subseteq A$ such that $|R| \leq n/b$ and every component of $A - R$ has less than $b$ vertices. 
\end{definition}

Such splitters are easy to find, and since we set $b=n/(12k+12)$ they are of size at most $12k+12$. Moreover, they will prove very useful to find a small balanced separator of $U$. We first describe how to find the splitter. As before we denote by $A_v$ the subtree of $A$ rooted at vertex $v$. We will denote by $child_A(u)$ the children of $u$ in the tree $A$. 

\begin{proposition}[Theorem 9.2 in \cite{Lagergren96}]
The following set is a $b$-splitter:

    $$\displaystyle R=\{u\in U\colon 1 +  \Sigma_{v\in child_A(u)}(|A_v| \mod b) > b\}$$   
\end{proposition}

In particular, this makes $b$-splitters easy to compute, as follows.
    Using the \sumsubsetproblem\ problem of \Cref{lemma:somebroadcasts}, each node $u\in U$ can learn the value $|A_u|$ and by using the same protocols of \Cref{lemma:somebroadcasts} each node can compute the value $1 +  \Sigma_{v\in child_A(u)}(|A_v| \mod b)$, verify if this value is greater than $b$ and mark itself as part of $R$. All these computations can be done in $\biglo{\scq{G}}$ rounds.

The efficient computation of $b$-splitters allows to obtain a deterministic distributed algorithm to compute a separator set in any $U\subseteq V$.

In order to use this splitter to find a balanced separator for $U$ we need more notations. The weight $\omega_v$ of a vertex $v \in R$ corresponds to the components of $A_v-R$ “seeing” $v$:

    $$\omega_v = |A_v| -\sum_{u\in desc_A(v)\cap R}|A_u|,$$

where $desc_A(v)$ is the set descendants of $v$ in $A$. For a set $W \subseteq R$, we denote 

$$\displaystyle \omega(W) =  \sum_{v \in W} \omega_v.$$

We then use the following crucial statement, observing that if some small separator $Z$ of $G[U]$ does not intersect $R$, then the sizes of the components of $G[U]-Z$ roughly correspond to their weights.:

\begin{proposition}
[\cite{Lagergren96, Reed92}]\label{prop:balancedsplitter} Consider a set $U\subseteq V$, a spanning tree $A$ of $G[U]$ and a corresponding $b$-splitter $R$.
Let  $Z\subseteq U$ be a subset such that $|Z|\leq k$, $Z\cap R = \emptyset$, and  $Z$ separates $G[U]$ into two parts $W_1$ and $W_2$. Then for any such part $W_j$,

    $$||W_j| - \omega(W_j\cap R)|\leq k \cdot b.$$
\end{proposition}

We are now ready to prove~\Cref{lemma:sepofU}.
\begin{proof}[Proof of~\Cref{lemma:sepofU}]
By Proposition~\ref{prop:sepbt}, $U$ admits a $2/3$-balanced separator $Z$ of size at most $k+1$. If, by chance, this separator does not intersect $R$, this gives us a way to construct another separator $S$ of size at most $k+1$, that is only $10/12$-balanced. Recall that the $b$-splitter $R$ can be computed in $\biglo{k\cdot D}$ rounds, and the same holds for the weights $\omega_v$ of the vertices $v \in R$. Let $W_1,W_2$ be two parts of $G[U]$ separated by $S$, of size at most $2|U|/3$. Let now $R_1 = W_1 \cap R$ and $R_2 = W_2 \cap R$. By Proposition~\ref{prop:balancedsplitter}, $$\omega(R_1) \leq |W_1| + (k+1)b \leq 2|U|/3 + (k+1)|U|/(12k+12) \leq 9|U|/12,$$ and the same holds for $\omega(R_2)$.

We claim that any separator $S$ of size at most $k+1$, separating $W_1$ from $W_2$ and containing $R - (R_1 \cup R_2)$, is a $12/10$ balanced separator of $U$ (and such a separator exists, e.g. by taking $S=Z$). Indeed for any such $S$, each component $W$ of $G[U]-S$ satisfies that $W \cap R$ is contained into $R_1$ or into $R_2$. Assume w.l.o.g. that $W \cap R \subseteq R_1$. Then by Proposition~\ref{prop:balancedsplitter}, $$|W| \leq \omega(R_1) + (k+1)b \leq 9|U|/12+(k+1)|U|/(12k+12) = 10|U|/12.$$ Therefore, it suffices to “guess” the partition $(R_1,R_2,R-(R_1\cup R_2))$ of $R$ and to use Theorem~\ref{thm:congseparator} in order to find the separator $S$. This would take a multiplicative factor of $2^{O(k)}$, by trying all 3-partitions of $R$, which is itself of size $O(k)$. 

But recall that this argument only works when the $2/3$-balanced separator $Z$ of $U$ does not intersect $R$. If it does, we need more branching. Indeed we need to guess a vertex $u \in Z \cap R$ and then to deal only with instance $U-\{u\}$, but looking for separators of size at most $k$ instead of $k+1$ (the size of the balanced separator is decreased by 1). Moreover $U-\{u\}$ might not be connected, so we need to deal with its largest component only. These branchings lead to an overall multiplicative factor of $k^{\bigo{k}}$, since we have $\bigo{k}$ branching choices at each step and the depth of the branching is at most $k+1$. This leads to the $\biglo{k^{O(k)}D}$ round complexity.

Although the whole description has been made on a unique connected subset $U \subseteq V$, since we only used partwise aggregation tools the same holds for the computation of 10/12-balanced separators of pairwise disjoint connected subsets $U_1\dots, U_{\ell} \subseteq V$.
\end{proof}

\subsection{Odd Steps}\label{subsec:odd}

We need to keep as invariant of the algorithm the fact that even steps receive as input instances $(U,X)$ such that $|X|\leq 6k+4$. Recall that during even steps we add to the new bag some separator of size up to $k+1$, so when we recurse on odd steps the input $(U',X')$ only guarantees that $|X'| \leq 7k+5$. We need to reduce the size of the separator to at most $6k+4$. Note that if the set $X'$ is already of size at most $6k+4$, we simply ignore this odd step. 

\begin{lemma}\label{lemma:sepofX}
    Let $G$ be a graph of treewidth at most $k$ and diameter $D$.
    There exists an $\biglo{2^{O(k)}D}$-rounds deterministic algorithm in the \CONGEST\ model such that, given a connected subset $U'$ of $G$ and $X' \subseteq N(U')$ such that $6k+4 \leq |X'| \leq 7k+5$, computes a 2/3 separator of $X'$ in $G[U' \cup X']$ in $\biglo{2^{O(k)}D}$ rounds. 

    Moreover, the same computation can be preformed within the same number of rounds in parallel, on disjoint connected sets $U'_1, \dots, U'_{\ell}$ and sets $X'_1,\dots,X'_{\ell}$, with $X'_i \subseteq N(U'_i)$ and the sise constraints as above, for all $i \in [\ell]$.
\end{lemma}
\begin{proof}
We “guess” how the $2/3$-balanced $S$ separator of $X'$ in $G[U' \cup X']$ splits the set $X'$, that is we guess the parts $X'_1$ and $X'_2$ of size at most $2|X'|/3$ that will be separated, and $Y' = X' - (X'_1 \cup X'_2)$ belongs the separator $S$ of size at most $k+1$ that exists by Proposition~\ref{prop:sepbt}. Assume that the three-partition of $X$ was guessed correctly, and observe that $X'_1$ and $X'_2$ cannot be empty because of the sizes of each part. Therefore, when using~\Cref{thm:congseparator} to find a separator of size at most $k+1 - |Y'|$ separating $X'_1$ and $X'_2$ in $G[U \cup X'] - Y']$, we obtain $S$. For any component $U''$ of $G[U' \cup X'] - S$, its neighborhood $X''$ is contained in $S \cup X'_1$ or in $S \cup X'_2$. Since $|X'_1| \leq 2|X'|/3 \leq 2(7k+5)/3$ and $|S|\leq k+1$, we deduce that $|X''| \leq (14k+10+3k+3)/3 \leq 6k+4$ for all $k>0$. 

Therefore one can guess the correct partition and then the separator $S$ in $\biglo{2^{\bigo{k}} D}$ rounds. 

 When running in parallel all different instances $(U'_1,X'_1),\dots,(U'_{\ell},X'_{\ell})$ of the odd step of a same phase, sets $U'_i$ are pairwise disjoint, but sets $X'_i$ are not necessarily disjoint. To cope with this issue, we must compute the separators of $G[U'_i \cup X'_i]$ making sure that vertices of $U'_i$ “take in charge” the $\bigo{k}$ vertices of $X'_i$. This is possible because all vertices of $U'_i$ know the whole set $X'_i$, as indicated in Lemma~\ref{lemma:mabseparator}.
\end{proof}

\subsection{Wrap-up}
\begin{proof}[Proof of~\Cref{th:treedec} and~\Cref{le:twtech}]
  Recall that we start the process with the instance $(U,X) = (V,\emptyset)$. The tree decomposition has a root $r$ node whose bag $B_r$ is the empty set.
  
  Then we alternate even and odd steps, in phases $i$ starting with $i=0$. 
  
  In general, the even step of phase $i$ works on instances $(U_1,X_1),\dots, (U_{\ell},X_{\ell})$ with $G[U_j]$ connected, and $X_j \subseteq U_i$ of size at most $6k+4$ for all $j \in [\ell]$. As an invariant (see notations of~\Cref{le:twtech}) the vertices of $U_j$ correspond to the set $V_u \setminus B_u$ for some node $u$ of the decomposition tree. Also, all vertices of $U_j$ already know the list of bags $(B_u, B_{p_u},\dots,B_r)$ computed during the previous steps.
  
  Whenever an instance satisfies $|X_j \cup U_j| \leq 7k+5$ then we simply create a new leaf in the decomposition tree, whose parent bag is the bag that “created” this instance. Technically, all vertices of $U_j$ append this new bag to the list of bags, in the beginning of the list. 

  Otherwise, using~\Cref{lemma:sepofU}, we find a $10/12$-balanced separator $S_j$ for every $U_i$ with $|S_j| \leq k+1$. In particular this guarantees that we can stop after $\bigo{\log n}$ phases.
  We create a new node in the decomposition, with bag $X_j \cup S_j$, for every instance, in the sense that every vertex of $U_j$ learns by partwise aggregation  the new bag and the subgraph of $G$ induced by it. 
 Observe that each vertex of $S_i$ can locally determine that it entered a bag for the first time, so it learns its depth in the decomposition tree as the length of its list of bags. Then, by part-wise aggregation, as each node knows if it belongs to $U_j$ and/or $S_j$, by an analogous algorithm as the one presented for \Cref{lemm:detectbridges}, we compute the components of $G[U_j] - S_j$, and for each such component $U'$ we create a new instance $(U',X')$ with $X' = N_G(U')$. By construction $|X'| \leq |X_j| + |S| \leq 7k+5$.

  We then perform the odd step on all instances $(U'_1,X'_1),\dots,(U'_{\ell'},X'_{\ell'})$, obtained from all instances $(U_1,X_1),\dots, (U_{\ell},X_{\ell})$.

  The aim of the odd step is to shrink the size of each $X'_j$ from at most $7k+5$ to at most $6k+4$. If this is already the case, there is nothing to do. We apply~\Cref{lemma:sepofX} only on the instances having $|X'_j|>6k+4$, obtaining 2/3-separators $S'_j$ of $X'_j$ in $G[U'_j \cup X'_j]$. We create as before a new bag $X'_j \cup S'_j$ and all vertices of $U'_j$ learn it and the induced subgraph, as required by~\Cref{le:twtech}. As observed in~\Cref{lemma:sepofX}, for each component $U''$ of $G[U'_j] - S'_j$, its neighborhood $X'' = N_G[U'']$ is contained in $S'_j$ and one of the parts of $X'_j$ separated by $S'_j$, so $|X''| \leq 2|X'_j|/3 + |S'_j| \leq 6k+4$. The process continues with all instances of type $(U'',X'')$.

  Each of the $\bigo{\log n}$ phases takes $\biglo{k^{O(k)}D}$ rounds by~\Cref{lemma:sepofU} and~\Cref{lemma:sepofX}.
\end{proof}

\section{Decision and optimization for MSO properties}\label{se:MSO}

Consider an MSO graph property $\cP$, i.e., a property expressible by an MSO formula $\varphi$. Recall that properties such as 3-colorability (and non-3-colorability) or hamiltonicity are in this case. Courcelle's theorem~\cite{Courcelle90} states that any such property there in an $O(n)$ time sequential algorithm deciding the property on classes of graphs of bounded treewidth, if a tree decomposition is also part of the input. Informally, the algorithm proceeds by dynamic programming on the tree decomposition, which is considered as rooted. At every node of the decomposition, only an information of constant size needs to be computed, from the information retrieved at the children nodes, and eventually at the root node one can decide property $\cP$. All constants, in particular the constant of the big-Oh, depend on formula $\varphi$ and the width of the decompositions. 

Among the several existing proofs of Courcelle's theorem~\cite{Courcelle90}, we shall use the algorithm and vocabulary of Borie, Parker and Tovey~\cite{BoPaTo92}. Firstly, we introduce the notation and key concepts, secondly, we present the sequential algorithm, and thirdly its adaptation to the \CONGEST\ model. Eventually, we discuss how to extend the decision algorithm to optimization problems.
 
\subsection{Tree Decompositions and $w$-terminal Recursive Graphs}\label{subsec:twterm}

Graphs of treewidth smaller than $w$ can alternatively defined as \emph{$w$-terminal recursive graph}. A $w$-terminal graph is a pair $(G,W)$ formed from a graph $G=(V,E)$, and a set $W \subseteq V$ of size at most $w$, called the \emph{terminals} of $G$. Moreover, $W$ is a totally ordered set: we can speak of the first, second, up to the $w$th terminal of $(G,W)$. For simplicity, here we will consider our vertices ordered by their identifiers.

The class of $w$-terminal recursive graphs is defined as the set of $w$-terminal graphs obtained from \emph{base} graphs by sequences of \emph{gluings}. A \emph{base} graph is a $w$-terminal graph where all vertices are terminals. In particular, $w$-terminal recursive base graphs have at most $w$ vertices.

\begin{figure}[H]
    \centering
    \includegraphics[scale=0.25]{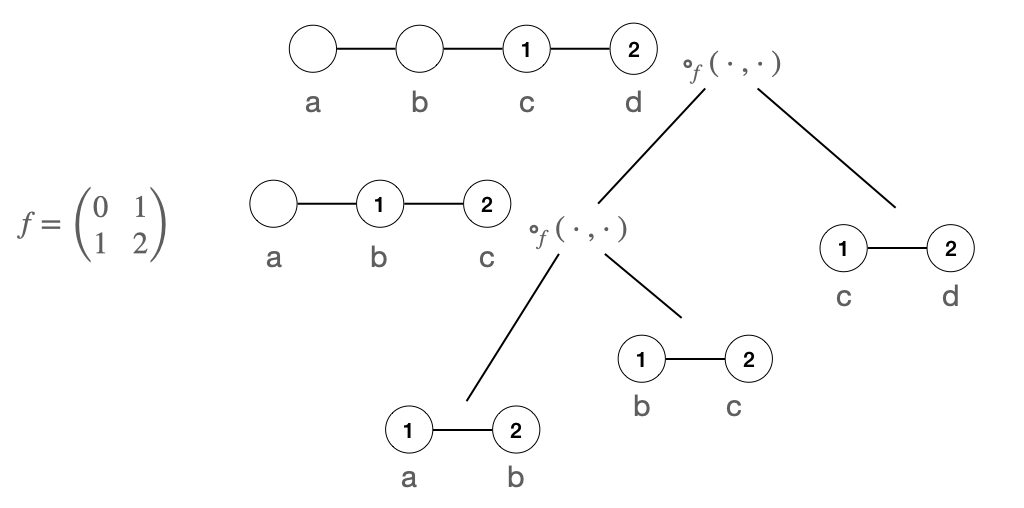}
    \caption{A path as a 2-terminal recursive graph.}
    \label{fig:TwG}
\end{figure}

A gluing $\circ_f$ of two $w$-terminal recursive graphs $G_1$ and $G_2$ produces a new $w$-terminal recursive graph $G = \circ_f(G_1,G_2)$. The gluing operation is described by the $w \times 2$ matrix $f$ with values in $0,1\dots, w$, as follows. First, we make two disjoint copies of $G_1$ and $G_2$. Each terminal of $G_1$ is identified with at most one terminal of $G_2$. More precisely the $i$th terminal of $G_1$ (resp. $G_2$) becomes the terminal vertex of $G$ with number $f[i,1]$ (resp. the $f[i,2]$), if $f[i,1] \neq 0$ (resp. $f[i,2] \neq 0)$. If $f[i,1]=0$, then the $i$th terminal of $G_1$ becomes a non-terminal vertex of $G$; in particular if $G_1$ has $t_1 < w$  terminals, then for each $i \in \{t_1+1,\dots,w\}$ we must have $f[i,1]=0$. The same holds for graph $G_2$;  actually, in all our constructions, the terminals of $G$ will correspond exactly to the terminals of $G_2$. We emphasis that in each column of matrix $f$, every non-empty value appears at most once. Also, the number of possible matrices $f$ only depends on $w$, because these matrices have $2w$ elements in the set $0,\dots,w$. Therefore for any fixed $w$ there is a constant number of gluing operations. An example of a representation of paths as 2-terminal recursive graphs, inspired from~\cite{FFMRT24}, is given in Figure~\ref{fig:TwG}.

By Theorem~40 in~\cite{Bodlaender98arb}, The class of $w$-terminal recursive graphs is exactly the class of graphs of treewidth strictly smaller that $w$. We describe here how a tree decomposition of width $<w$ of a graph $G$ can be transformed into a representation of $G$ as a $w$-terminal recursive graph, with the root bag as set of terminals.

\begin{figure}[H]
    \centering
    \includegraphics[scale=0.2]{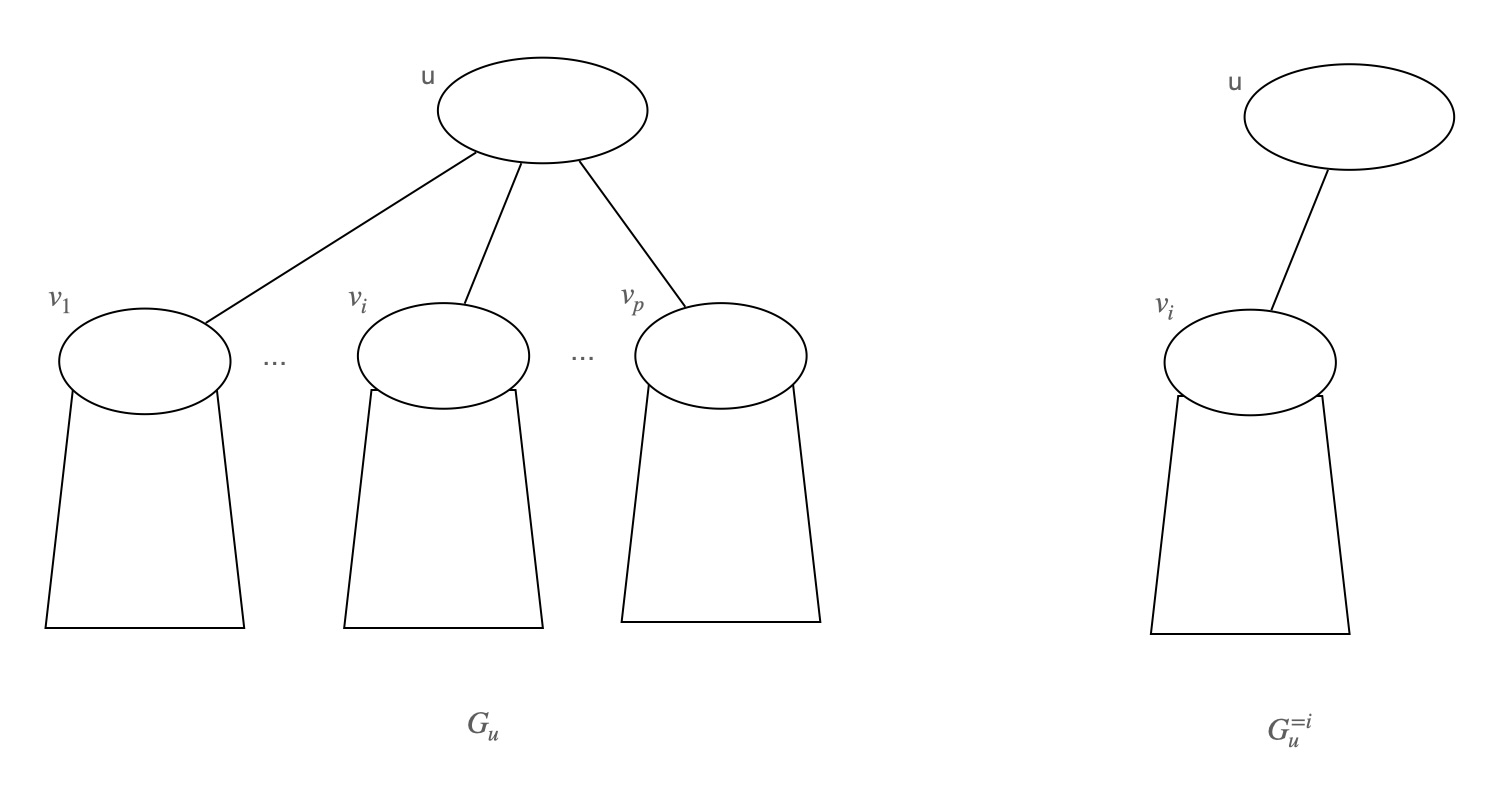}
    \caption{Tree-decompositions: graphs $G_u$ and $G^{=i}_u$.}
    \label{fig:Gu}
\end{figure}

Assume that we are given a graph $G=(V,E)$ together with a rooted tree decomposition $(T=(I,F),\{B_u\}{u \in I})$ with bags of size at most $w$. We introduce the following notations, at the node $u$ of the tree decomposition (see also Figure~\ref{fig:Gu}, inspired from~\cite{FFMRT24}).

\begin{itemize}
    \item $T_u$ is the subtree of $T$ rooted at $u$.
    \item $B_u$ is the bag of node $u$ and $G_u^{base} = (G[B_u],B_u)$ is the base graph induced by the bag in $G$.
    \item $V_u$ is the union of all bags of nodes in $T_u$, and $G_u = (G[V_u],B_u)$ is the $w$-terminal graph induced by $V_u$ in $G$, with $B_u$ as set of terminals.
    \item Denote by $v_1,\dots, v_p $ the children of $u$ in $T$. For $i \in \{1,\dots, p\}$, let $G^{=i}_u = (G[V_{v_i} \cup B_u],B_u)$.
\end{itemize}

We then make the following simple but crucial observations.

\begin{lemma}\label{le:equ}
For all $i \in \{1,\dots,p\}$,
$$G^{=i}_u = \circ_{f(B_{v_i},B_u)}(G_{v_i},G^{base}_u)$$
Where matrix $f(B_{v_i},B_u)$ glues each vertex of $B_{v_i}$ on the corresponding vertex of $B_u$.
\end{lemma}
\begin{proof}
Let us simply describe matrix $f(B_{v_i},B_u)$. Since the terminals of the glued graphs are exactly those of the second operand, the second column of $f$ is the identity, padded with $0$s. That is, $f[j,2]=j$ for all $j \in \{1 \dots |B_u|\}$, and $f[j,2]=0$ for $j \in \{1 +|B_u|\dots, w\}$.

For any $j \in \{1,\dots,|B_{v_i}|\}$, we set $f[j,1]=0$ if the $j$th vertex of $B_{v_i}$ does not belong to $B_u$, otherwise we assigne to $f[j,1]$ the rank of that vertex in $B_u$. For $j \in \{1+|B_{v_i}|,\dots,w\}$, we also pad with $f[j,1]=0$.
\end{proof}

\begin{lemma}\label{le:allu}
$G_u$ is obtained by successively gluing $G^{=1}_u, G^{=2}_u, \dots G^{=p}_u$ using the same gluing operation $\circ_{f(B_u,B_u)}$ which identifies corresponding vertices of $B_u$. 

Moreover, operation $\circ_{f(B_u,B_u)}$ is commutative and associative, and $G^{base}_u$ is a neutral element. 
\end{lemma}
\begin{proof}
    The proof is immediate. Observe that now matrix $f(B_u,B_u)$ is defined as $f[i,1]=f[i,2]=i$ for all $i \in \{1 \dots |B_u|\}$, and $f[i,1]=f[i,2]=0$ for $i \in \{1 +|B_u|\dots, w\}$. This entails the commutativity and associativity of the operation, plus the fact that $G^{base}_u$ is neutral. 
\end{proof}

\subsection{Regularity of MSO Properties and the Sequential Algorithm}

Borie, Parker and Tovey~\cite{BoPaTo92} define de notion of regular properties over $w$-terminal recursive graphs, and then show that properties expressible by MSO formulae are regular. This immediately provides an efficient algorithm for deciding such properties for graphs of bounded treewidth.

\begin{definition}[regular property]\label{de:reg}
A graph property $\cP(G)$ is \emph{regular} if, for any value~$w$, we can associate to~$w$  and $\cP$:
\begin{itemize}
\item a finite set $\cC$ of \emph{homomorphism classes},
\item an \emph{homomorphism function} $h$, assigning to each $w$-terminal recursive graph $G$ a \emph{class} $h(G) \in \cC$, and 
\item an \emph{update function} $\odot_f : \cC \times \cC \to \cC$ for each gluing operation $f$, 
\end{itemize}
such that:
\begin{enumerate}
\item If $h(G_1) = h(G_2)$ then $\cP(G_1) = \cP(G_2)$;
\item For any two $w$-terminal recursive graphs $G_1$ and $G_2$, 
 $$h\Big(\circ_f\big(G_1,G_2)\big)\Big) = \odot_f\Big(h(G_1),h(G_2)\Big).$$

\end{enumerate}
\end{definition}

A homomorphism class $c \in \cC$ is called \emph{accepting} if graphs $G$ such that $h(G)=c$ satisfy property $\cP$; otherwise, class $c$ is called \emph{rejecting}. 

To give an example of these concepts, let us consider the 3-coloring property, which is MSO expressible as explained in the Introduction, on $w$-terminal recursive graphs $(G,W)$. 
Each homomorphism class will correspond to a 3-partition of the set $W$ of terminals, and a boolean value. Given the partition $(R_W,G_W,B_W)$ of $W$, the boolean is set to $\mathrm{true}$ if there is a 3-coloring $(R,G,B)$ of the whole vertex set, such that $(R_W,G_W,B_W)$ corresponds to intersection of these sets and $W$, otherwise the boolean is $\mathrm{false}$. 
Since terminals are numbered from $1$ to $w$, each 3-partition of $W$ is encoded as a partition of set $[w]$. In particular, the number of homomorphism classes is $2^{\bigo{w}}$. Accepting classes are classes with the boolean value at $\mathrm{true}$. It is a matter of exercise to compute functions $\odot_f$, for any gluing operation $f$. This extends to any property expressible in MSO.

\begin{theorem}[\cite{BoPaTo92}]\label{th:reg}
Any property $\cP(G)$ expressible by an MSO formula $\varphi$ is regular. Moreover, given the formula $\varphi$ and a parameter $w$, one can compute in constant time
\begin{itemize}
 \item the set of classes~$\cC$,  
 \item the update functions $\odot_f$ over all possible gluing matrices~$f$,
 \item the homomorphism classes $h(G)$ for all base graphs $G$.
\end{itemize}
  
\end{theorem}

Thanks to this powerful theorem, we can now decide any MSO property $\cP$ over graphs of treewidth at smaller than $w$ by dynamic programming. Assuming that a tree decomposition of width $<w$ is part of the input, the algorithm takes $O(n)$ time, for fixed $w$ and $\cP$. With the notations of the previous subsection, at each node $u$ of the tree decomposition we compute the base graph $G^{base}_u$ (which is of constant size), and then its homomorphism class $h(G_u)$ of $G_u$ using the base case of Theorem~\ref{th:reg}.

Then, from the bottom to the top, at each vertex $u$ we compute using Theorem~\ref{th:reg}:
\begin{itemize}
    \item For each child $v_i$ of $u$, the homomorphism class $h(G^{=i}_u) = \odot_{f(B_{v_i},B_u)}\Big(h(G_{v_i}),h(G^{base}_u)\Big)$, see Lemma~\ref{le:equ}.
    \item The homomorphism class $h(G_u)$ is obtained from the classes $h(G^{=1}_u), h(G^{=2}_u),\dots, h(G^{=p}_u)$ by iteratively appying operation $\odot_{f(B_u,B_u)}$, cf. Lemma~\ref{le:allu}.
\end{itemize}
Observe that, when dealing with node $u$, the homomorphism classes $h(G_{v_i})$ have already been computed at previous steps of the dynamic programming.
Eventually, at the root $r$ of the tree decomposition, we return $\mathrm{true}$ if $h(G_r)$ is accepting, otherwise we return $\mathrm{false}$.

\subsection{The \CONGEST\ Algorithm Deciding
MSO Properties}\label{subsec:MSOdec} 

The distributed algorithms mimics the sequential one, in order to prove our main result:

\mainresult*
\begin{proof}

Here we use a tree decomposition of constant width $7k+5$, and of depth $\bigo{\log n}$ obtained by Theorem~\ref{th:treedec}. The algorithm proceeds in phases, for decreasing values $j \in \{\mathrm{depth}(T),\dots, 0\}$. At phase $j$, we compute in parallel the homomorphism classes $h(G_u)$ for all nodes $u$ of depth $j$, in  $\tilde{O}(D)$ \CONGEST{} rounds. 
For any node $u$ of the decomposition, let $S_u = B_u \cap B_{p(u)}$ and $C_u = V_u \setminus S_u$ (at the root $r$, we have $B_r = S_r = \emptyset$ and $C_r = V$). We rely on the connectivity property of our tree decomposition (Section~\ref{se:treewidth}, Lemma~\ref{le:twtech}): $C_u$ is a connected subgraph of $G$, and $S_u = N(C_u)$. Observe that, by definition of tree decompositions, two nodes $u,u'$ such that none is ancestor of the other satisfy that $C_u$ and $C_{u'}$ are disjoint, in particular the sets $C_u$ corresponding to nodes $u$ of a same depth $j$ are pairwise disjoint. For any $u$ different from the root, let $\ell(u)$ the vertex of minimum id in $B_u \setminus S_u$ (by construction of our tree decomposition, this set is not empty). At the root $r$, we take as $\ell(r)$ the vertex of minimum id in $V$. This vertex $\ell(u)$ is called the \emph{leader of node $u$}, and at the end of phase $j$ it will know the homomorphism class $h(G_u)$. Also by construction, all vertices of $C_u$ know (with no communication) the bag $B_u$ of depth $j$ and the parent bag $B_{p(u)}$, so they can locally compute the leader $\ell(u)$; when $u$ is the root, computing the leader only requires $\tilde{O}(D)$ rounds for the whole graph.

Let us describe how leaders $\ell(u)$ at depth $j$ are able to compute the homomorphism classes of $G_u$. We focus here on a single node $u$, but observe that all following steps can be carried out in parallel in $\tilde{O}(D)$ rounds. The leader $\ell(u)$ knows the bag $B_u$ and the induced subgraph $G[B_u]$, in particular it knows the $w$-terminal recursive graph $G^{base}_u=(G[B_u],B_u)$.

Denote as in the sequential case by $v_1,\dots,v_p$ the children of $u$ in the decomposition tree, all of depth $j+1$. All leaders $\ell(v_1),\dots,\ell(v_p)$ have already computed the homomorphism classes $h(G_{v_i})$ during the previous phase $j+1$. Actually, these leaders $\ell(v_i)$ also know the bag $B_u$ (and all bags from $v_i$ to the root), and the graph $G[B_u]$. Therefore each $\ell(v_i)$ can compute locally the homomorphism class $h(G^{=i}_u)$ of $G^{=i}_u = (G[B_u \cup V_{v_i}],B_u)$, obtained by gueing of $G_{v_i}$ and $G^{base}_u$, as in the sequential case (see also Lemma~\ref{le:equ}). Observe that the homomorphism class $h(G^{base}_u)$ of $G^{base}_u=(G[B_u],B_u)$ can be computed locally by every vertex of $C_u$.
By construction these leaders are pairwise distinct (actually they are also distinct from $\ell(u)$ except maybe when $u$ is the root of the tree decomposition).

 We associate to each vertex of $C_u$ a $w$-terminal recursive graph, and its homomorphism class, as follows. Leader $\ell(v_i)$ is associated to $G^{=i}_u$ and its class $h(G^{=i}_u)$. All vertices in $C_u$ that are not leaders of depth $j+1$ are associated to the base graph $G^{base}_u$ and its class $h(G^{base}_u)$. 
Observe (see also Lemma~\ref{le:allu}) that $G_u$ is obtained by the gluing of all terminal recursive graphs associated to vertices of $C_u$, using the same gluing $\circ_{f(B_u,B_u)}$ which identifies vertices of $B_u$ from the two copies. 
We deduce from Theorem~\ref{th:reg} that the homomorphism class $h(G_u)$ is obtained by the homomorphism classes associated to all vertices of $C_u$, through operation $\odot_{f(B_u,B_u)}$ acting on homomorphism classes. By Lemma~\ref{le:allu}, operation $\circ_{f(B_u,B_u)}$ over $w$-terminal recursive graphs is commutative and associative, thus operation $\odot_{f(B_u,B_u)}$ is also associative and commutative. Therefore, we can use~\Cref{broadcasts} on operation $\bigoplus = \odot_{f(B_u,B_u)}$ and homomorphism classes associated to each vertex of $C_u$, and obtain at  $\ell(u)$ the homomorphism class $h(G_u)$ in $\biglo{D}$ rounds. Again, this overall number of $\biglo{D}$ rounds holds for the parallel computing of all homomorphism classes $h(G_u)$, over all nodes $u$ of depth $j$.

After phase $j=0$, the root leader $\ell(r)$ accepts of the homomorphism class $h(G_r)$ is accepting, otherwise it rejects. All other vertices accept.

Altogether, this decides property $\cP$ in $\biglo{D}$ \CONGEST{} rounds.
\end{proof}

\subsection{\CONGEST\ Algorithms for Optimization and Counting}

Courcelle's theorem extends to optimization and counting~\ref{prop:Courcelle}. Consider an MSO formula $\varphi(X)$ with a free set variable, corresponding to a vertex subset or edge subset. Moreover, assume that the input graph $G=(V,E)$ as weighted, i.e. we have as input a weight function $\mathrm{w} : V \cup E \to \mathbb{Z}$ associating to each vertex or edge an integer weight. The only constraint is that weights are polynomial, thus represented on $\bigo{\log n}$ bits. 

The optimization problem $max\varphi$ consists in finding the maximum weigh set $S$ such that $G \models \varphi(S)$. Typical problems in this framework are finding a maximum independent set, a minimum dominating set, etc. 
We show that this problem can also be solved in $\biglo{D}$ \CONGEST\ rounds.

We can also consider the counting problem $count\varphi$, i.e., counting the number of different sets $S$ such that $G \models \varphi(S)$. As examples, we can count the maximal independent sets, or the minimal dominating sets or the triangles of the input graph $G$. Let $\nbsol(n)$ denote the maximum number of possible solutions for $n$ vertex graphs. Note that $\nbsol(n)$ can be exponential in $n$,  even $2^{kn}$ for edge sets, and we need to communicate such numbers between vertices. Therefore the total number of rounds will become $\biglo{D \log(\nbsol(n))}$, which is still $\biglo{D}$ for problems such as counting triangles.

We obtain:

\begin{theorem}\label{thm:optcount}
  Fix a positive integer $k$ and and an MSO formula $\varphi(X)$ with a free set variable $X$. 
  
  Given as input a communication network $G=(V,E)$ with $n$ vertices, of treewidth at most $k$, of diameter $D$, and a integer weight function $\mathrm{w}: V \cup E \to \mathbb{Z}$ with polynomial weights, 
  
  \begin{itemize}
      \item there is a \CONGEST\ algorithm computing the maximum weight set $S$ such that $G \models \varphi(S)$ in $\biglo{D}$ rounds;
      \item there is a \CONGEST\ algorithm counting the number of different sets $S$ such that $G \models \varphi(S)$ in $\biglo{D \log(\nbsol(n))}$, where $\nbsol(n)$ is an upper bound on the number of solutions.
  \end{itemize}
\end{theorem}
\begin{proof}
Let us discuss the differences with the decision algorithms in the sequential setting, then the implementation in \CONGEST\ will follow directly. 

First we need to extend the notion of gluing of graphs to gluing of pairs formed by a graph and an vertex/edge subset,

$$(G,X) = \odot_f((G_1,X_1),(G_2,X_2)),$$

nevertheless the gluing is only valid under specific constraints. Consider the case when $X_1, X_2$ are vertex subsets. Let $t$ be a terminal vertex of $G$, obtained by the gluing of terminals of $G_1$ and $G_2$, say $t_1$ and $t_2$. Then we must either have $t_1 \in X_1, t_2 \in X_2$, or  $t_1 \notin X_1, t_2 \notin X_2$. Set $X$ is obtained from the union of $X_1$ and $X_2$ by identifying pairs of vertices that have been mapped on a same terminal of $G$. The situation is similar for edge subsets, we refer to~\cite{BoPaTo92} for full details.

The notion of regularity, introduced in Definition~\ref{de:reg} on graph properties, extends to properties $\cP(G,X)$, depending on the graph $G$ and the vertex or edge set $X$ of $G$. We simply replace the two conditions of the definition by:
\begin{enumerate}
\item If $h(G_1,X_1) = h(G_2,X_2)$ then $\cP(G_1,X_1) = \cP(G_2,X_2)$;
\item For any two $w$-terminal recursive graphs $G_1$ and $G_2$, 
 $$h\Big(\circ_f\big((G_1,X_1),(G_2,X_2))\big)\Big) = \odot_f\Big(h(G_1,X_1),h(G_2,X_2)\Big).$$ 
\end{enumerate}

Let  $G$ be a $w$-terminal recursive graph and let $W$ denote its set of terminals. Given a set $X$, we may assume w.l.o.g. that the homomorphism class $c=h(G,X)$ encodes the intersection of $X$ with $W$. Concretely, if $X$ is a vertex subset than we encode in $c$ the list of terminals contained in $X$, as a list of numbers from $1$ to $w$. If $X$ is an edge subset, we encode similarly the terminal vertices incident to $X$, and the edges among those vertices. Both information are of constant size. Therefore, given a class $c$ and the set $W$, we may assume that the function $\selected(c,W)$ retrieves the unique  intersection of $X$ with $G[W]$. By ‘unique’ we mean that different sets $X_1, X_2$ such that $h(G,X_1)=h(G,X_2)=c$ must have the same intersection with $G[W]$. 

Theorem~\ref{th:reg} holds for any property $\cP(G,X)$ over graphs and vertex sets, expressible by some MSO formula $\varphi(X)$ (actually, the result of Borie, Parker and Tovey~\cite{BoPaTo92} is stated on formulae with arbitrary number of variables). 

In order to solve the optimization problem $max\varphi$ for formula $\varphi(X)$ (see also~\cite{BoPaTo92,FMRT24}) let $\cC$ the set of homomorphism classes corresponding to the formula as in Theorem~\ref{th:reg}. We associate to each $w$-terminal recursive graph $G$ a table $\OPT(G)$ with $|\cC|$ entries, where for each $c \in \cC$, $\OPT(G)[c]$ is the maximum weight of a set $S$ such that $G \models \varphi(S)$, or $-\infty$ if no such $S$ exists.

Then if graph $G$ is obtained as a gluing $\circ_f(G_1,G_2)$, we have that (see also~\cite{FMRT24})

\begin{equation}
    \label{eq:optcomp}
   \begin{split}  
    \OPT(G)[c]  = \max_{c_1,c_2 \in \cC \text{~s.t.~} c = \odot_f (c_1,c_2)}  \OPT(G_1)[c_1] &+ \OPT(G_2)[c_2] - \\
      & -\w(\selected(c_1,W_1) \cap \selected(c_2,W_2))
    \end{split}
\end{equation}

For base graphs $G=(V,E)$ with $W=V$ as set of terminals, note that and class $c \in \cC$,

\begin{equation}\label{eq:optbase}
\OPT(G)[c] = \w(\selected(c,V)),
\end{equation}

In both equations we set $\OPT(G)[c]$ to $-\infty$ if the right-hand side is not defined, (e.g., for the later, if there is no vertex/edge subset $X$ such that $c=h(G,X)$). 

The sequential algorithm proceeds bottom-up and computes, at each node $u$ of the decomposition tree, the table $\OPT(G_u)$. Observe that the table is of size $\bigo{\log n}$, thanks to the fact that we deal with $|\cC| = \bigo{1}$ entries of logarithmic size, thanks to the fact that weights are polynomial in $n$.

The \CONGEST\ algorithm proceeds as for the decision case, see Subsection~\ref{subsec:MSOdec}. Again we use our tree decomposition of constant width and of logarithmic depth, and we proceed in phases $j$ from $\mathrm{depth}(T)$ to $0$, and at phase $j$ all nodes $u$ of this depth in $T$ compute in parallel $\OPT(G_u)$. The computation is performed by the leader vertex $\ell(u)$ associated to node $u$ of the decomposition, based on information from the leaders $\ell(v_1), \dots, \ell(v_p)$ where $v_1, \dots, v_p$ are the children on $u$ in $T$. 
If $u$ is a leaf, this computation is simply performed by brute force on the base graph $G^{base}_u =(G[B_u],B_u)$. 

Otherwise, at the previous phase j+1, leaders $\ell(v_i)$ have computed $\OPT(G_{v_i})$. With no extra communication, they can even compute $ \OPT(G^{=i}_u)$, obtained by the gluing of $G(v_i)$ with the base graph $G^{base}_u$. Since $G_u$ is obtained by the gluing of all $G^{=i}_u$, through the gluing ‘identity’ operation $\circ_f(B_u,B_u)$ (Lemma~\ref{le:allu}), which is associative and commutative, we need to observe that the this gluing $\circ_{f(B_u,B_u)}$ can be translated into an associative and commutative function over tables $\OPT$ of size $\bigo{\log n}$. Indeed if $G = \circ_{f(B_u,B_u)}(G_1, G_2)$ then 
$OPT(G) = \oplus_{f(B_u,B_u)}(\OPT(G_1), \OPT(G_2))$ for some function $\oplus_{f(B_u,B_u)}$ described by Equation~\ref{eq:optcomp}. By commutativity and associativity of $\circ_{f(B_u,B_u)}$, function $\oplus_{f(B_u,B_u)}$ is also commutative and associative. We
could use again~\cref{broadcasts} to perform a part-wise aggregation on sets $C_u$ on operation $\bigoplus = \oplus_{f(B_u,B_u)}$ from the local information 
by associating
 to each vertex $x$ of $C_u$ the information $\OPT_x$ defined as $\OPT(G^{=i}_u)$ if $x = \ell(v_i)$ for some $v_i$, or as $\OPT(G^{base}_u)$ otherwise. Indeed since $G_u$ is obtained by the gluing of all $G_{v_i}$ through operation $\odot_{f(B_u,B_u)}$, to which we can glue an arbitrary number of time the base graph $G^{base}_u$ (as this operation does not modify the graph), we obtain $\OPT(G_u)$ as the aggregate of $\oplus_{f(B_u,B_u)}$ over all sets $\OPT_x$, for all vertices $x \in C_u$, in particular the leader $\ell(u)$ learns $\OPT(G_u)$.
 
 This would suffice to obtain the maximum weight of a set $S$ satisfying $\varphi(S)$, indeed the root leader $\ell(r)$ would return (transmit to all vertices) the maximum value $OPT_r[c]$ over all accepting classes $c$. 

 Nevertheless we need to be more careful as we also want to mark an optimal set $S$, i.e. mark all vertices of $S$ if $S$ is a vertex set, or if $S$ is an edge set compute, at each vertex $x$ of $G$, its incident edges belonging to $S$. For this we need a bit more details about the low congestion shortcuts machinery. By~\cite{haeupler2018round}, graphs of treewidth at most $k$ do not only admit, for any partition into connected subsets, shortcuts of quality $\bigo{k\cdot D}$ that can be computed in $\biglo{k\cdot D}$ rounds, but they admit shorcuts of dilation $\biglo{k\cdot D}$ and congestion $\bigo{k \log n}$, that can be computed in $\biglo{k\cdot D}$ rounds.

 Therefore, when dealing with all connected subgraphs $C_u$ for all nodes $u$ of the decomposition tree of a given depth, we firstly compute such shortcuts. We can then retrieve for each $C_u$ a subtree $A_u$ of $G$, spanning $C_u$, of diameter $\biglo{k\cdot D}$. These trees $A_u$ are not pairwise disjoint, nevertheless each edge of $G$ is used by $\biglo{1}$ such trees. Again we associate to each vertex $x$ of the tree the corresponding information $\OPT_x$ as described before, if $x \in C_u$; when $x \not\in C_u$ we also set $\OPT_x = \OPT(G_u^{base})$.
 In $\biglo{k\cdot D}$ rounds we can do a bottom-up computation at every vertex $x$ of $A_u$, to make it learn $\bigoplus_{y} \OPT_y$ over all descendants $y$ of $x$ in $A_u$, with $x \in C_u$. In particular the root $r_u$ of $A_u$ will learn $\OPT(G_u)$ and will communicate it to $\ell(u)$. To be more precise, let $x_1,x_2,\dots,x_{\alpha}$ be the children of vertex $x \in V(A_u)$ in the tree $A_u$. Recall that the table $OPT_x$ is obtained from the tables $\OPT_{x_1}, \dots, \OPT_{x_{\alpha}}$ by repeatedly applying Equation~\ref{eq:optcomp}, $\alpha-1$ times. Therefore, whenever $\OPT_x[c]$ is computed for some homomorphism class $c$ as a maximum, vertex $x$ will memorize the “argmax”, i.e., the classes $c_1,\dots, c_{\alpha}$ on which this maximum was attained, by choosing the corresponding values in $\OPT_{x_1}[c_1],\dots, \OPT_{x_{\alpha}}[c_{\alpha}]$. Each bottom-up phase takes $\biglo{k \cdot D \cdot |\OPT|}$ time since we deal with families of trees of diameter $\biglo{k \cdot D}$, whose non-disjointness causes a congestion $\biglo{1}$ and we transmit information of  $|OPT|$ bits, i.e. the number of homomorphism classes of $\varphi$, times $\bigo{\log n}$ bits.  
 
 As previously, at the end of this bottom-up part, at the root $r$ of the decomposition tree the leader $\ell(r)$ will choose the accepting class $c(r)$ maximizing $\OPT_r[c]$. Now this leader needs to propagate to each of the children $w_1,\dots,w_{\beta}$ of the root in the decomposition tree the classes $c(w_1),\dots,c(w_{\beta})$ that produced class $c(r)$ at the root, by gluing. This is propagated top-down in the tree $A_r$, from its root $r_r$ to the leaders $\ell(w_1),\dots,\ell(w_{\beta})$. Indeed recall that each vertex $x$ of $A_r$ stored, for each homomorphism class $c'$, the “argmax” from its children in $A_r$ that produced this class at $x$, by gluing, i.e. a tuple of classes $(c'_1,\dots,c'_{\alpha})$. In the top-down phase, if $x$ has received class $c'$ from its parent, then it transmits to each child its corresponding class. In particular, at the end of this top-down phase, each leader $\ell(w_1),\dots, \ell(w_\beta)$ will learn the corresponding class $c(w_1),\dots, c(w_\beta)$ corresponding to the optimal solution.

 This process is repeated top-down on the decomposition tree, in phases $j$ from $0$ to $\bigo{\log n}$, on all nodes $u$ of the corresponding depth $j$ in the tree. Whenever a leader $\ell(u)$ learns its optimal class $c(u)$, it can mark the vertices or edges covered by bag $B_u$, since this information is encoded in the class, and transmits to the leaders $\ell(v_1),\dots,\ell(v_p)$ their optimal classes $c(v_1),\dots,c(v_p)$. Clearly each phase takes $\biglo{k \cdot D}$ rounds, since we treat in parallel trees $A_u$ of depth $\biglo{k \cdot D}$, non-disjoint but with polylogarithmic congestion on each edge.

This ends the proof of the optimization part of Theorem~\ref{thm:optcount}. 

The counting part is very similar (see also Theorem~5 in~\cite{BoPaTo92}), we mainly need to replace table $\OPT$ with table $\COUNT$. At each node $u$ we compute $\COUNT(G_u)$, a table with $|\cC|$ entries, where for each $c \in \cC$, $\COUNT(G_u)[c]$ denotes the number of partial solutions $S_u$ such that $h(G_u,S_u) = c$. It remains to replace Equation~\ref{eq:optbase} and~\ref{eq:optcomp} for with their correspondents for solution counting. This is quite straightforward for base graphs $G_u^{base}$: $\COUNT(G^{base}_u)[c]$ is either $1$ if a solution exists for class $c$, or 0 otherwise. In order to replace Equation~\ref{eq:optcomp} for graph gluing, when $G = \circ_f(G_1,G_2)$, 

$$ \COUNT(G)[c]  = \sum_{c_1,c_2 \in \cC \text{~s.t.~} c = \odot_f (c_1,c_2)}  \COUNT(G_1)[c_1] \cdot \COUNT(G_2)[c_2],$$
 or $0$ if the right-hand side has no terms. At the root we obtain the total number of solutions, as the sum of $\COUNT(G)[c]$ over all accepting classes $c$.
 The counting distributed algorithm follows the same steps as for optimization one, the main difference being that one needs to use these new calculations, and since we deal with numbers up to $\sharp\textrm{sol}(n)$, the round complexity is multiplied by $\log(\sharp\textrm{sol}(n))$. 
\end{proof}

\section{Conclusion}\label{sec:conclusion}

We presented in this paper distributed version of Courcelle's theorem, deciding MSO properties on bounded treewidth graphs in $\biglo{D}$ rounds in the \CONGEST{} model. Our algorithm is based on an approximation of treewidth, namely goven $k$ we decide that $\tw(G) >k$ or compute a tree decomposition of width $\bigo{k}$ in $\biglo{k^{O(k)}D}$ rounds in \CONGEST. Let us mention that we could replace the exponential dependency on $k$ in the number of bounds with a polynomial function, using the approach of~\cite{fomin2015} for approximating treewidh in fully polynomial time, instead of the approximation of~\cite{Reed92, Lagergren96}; on the negative side, we would only obtain a decomposition of width $\bigo{k^2}$.

The first natural question is whether our distributed tree decomposition can be used for deciding other properties in \CONGEST{}, beyond MSO, on bounded treewidth graphs. For example, is there an $\biglo{D}$-rounds algorithm deciding if the input graph has a non-trivial automorphism? The problem admits a sequential $2^{\mathrm{poly}(k)} \mathrm{poly}(n)$-time algorithm~\cite{GNSW20}.

It would be even more interesting to know whether we could combine locality and bounded bounded treewidth, in the following sense. A family of graphs that is closed by subgraphs is said to have \emph{bounded local treewidth} if the treewidth of such gaphs is upper bounded by a function of the diameter. E.g., graphs of bounded genus and in particular planar graphs  are of bounded local treewidth~\cite{Eppstein00}. 
We can say that an MSO formula $\varphi(x)$, depending on a vertex $x$, is \emph{local} if there is some constant $r$ such that $\varphi(x)$ depends only on the ball of radius $r$ around $x$. 
In the spirit of~\cite{NesetrilM16}, we ask: given an MSO local formula $\varphi(x)$ and a class of graphs $\mathcal G$ of bounded local treewidth, is there an algorithm that takes as input graph $G \in \mathcal G$ and marks all vertices satisfying $\varphi(x)$  in $\biglo{1}$-rounds in the \CONGEST{} model. E.g., can we mark all vertices such that their $d$-neighborhood, for some fixed $d$, contains a fixed graph $H$ as a minor?
Eventually let us note that graphs of bounded local treewidth have been introduced as a generalization of planar graphs on which one can apply Baker's technique~\cite{Baker94} to obtain (sequential) polynomial time approximation schemes for a variety of optimization problems. We believe it is unlikely that this approach can be implemented in CONGEST in a constant number of rounds, but it is legitimate to consider whether it could be used as inspiration for certain optimization problems.

\appendix

\bibliographystyle{plain}
\bibliography{bibliography.bib}
\end{document}